\documentclass{article}

\usepackage{microtype}
\usepackage{graphicx}
\usepackage{subcaption}
\usepackage{booktabs} % for professional tables
\usepackage{tabularx}
\usepackage{multirow}
\usepackage{caption}
\usepackage{makecell}
\usepackage{longtable}
\usepackage[normalem]{ulem}
\usepackage{adjustbox}
\usepackage{colortbl}
\usepackage{xcolor}
\usepackage{array}
\usepackage{siunitx}
\usepackage{bbm}
\usepackage{hyperref}
\usepackage{enumitem}

\usepackage[accepted]{icml2026}
\usepackage{algorithmic} % Use 'algorithmic', NOT 'algpseudocode'
\usepackage{amsmath}
\usepackage{amssymb}
\usepackage{mathtools}
\usepackage{amsthm}
\usepackage{fontawesome5}
\usepackage[T1]{fontenc}
\usepackage[capitalize,noabbrev]{cleveref}

\theoremstyle{plain}

\theoremstyle{definition}

\theoremstyle{remark}

\newcommand{\symbolimg}[2][0.35cm]{%
  \ensuremath{\vcenter{\hbox{\includegraphics[height=#1]{#2}}}}%
}

\usepackage[textsize=tiny]{todonotes}

\newcommand{\crcommentResolved}[1]{}

\icmltitlerunning{Quantifying the Salience of Geo-Cultural Values for Pluralistic Safety Alignment}

\begin{document}

\twocolumn[
  % \icmltitle{Salience of Geo-Cultural Values and Demographics for Pluralistic AI Alignment}
  % \icmltitle{Culture Shock: Salience of Geo-Cultural Values Beyond Demographics for Pluralistic Alignment}
    \icmltitle{Quantifying the Salience of Geo-Cultural Values for Pluralistic Safety Alignment}
  
  % It is OKAY to include author information, even for blind submissions: the
  % style file will automatically remove it for you unless you've provided
  % the [accepted] option to the icml2026 package.

  % List of affiliations: The first argument should be a (short) identifier you
  % will use later to specify author affiliations Academic affiliations
  % should list Department, University, City, Region, Country Industry
  % affiliations should list Company, City, Region, Country

  % You can specify symbols, otherwise they are numbered in order. Ideally, you
  % should not use this facility. Affiliations will be numbered in order of
  % appearance and this is the preferred way.
  \icmlsetsymbol{equal}{*}

   \begin{icmlauthorlist}
    \icmlauthor{Arkadiy Saakyan}{CU}
    \icmlauthor{Charvi Rastogi}{GDM}
    \icmlauthor{Lora Aroyo}{GDM}
  \end{icmlauthorlist}

  \icmlaffiliation{CU}{Columbia University, New York, NY, USA. Work done during an internship at Google DeepMind.}
  \icmlaffiliation{GDM}{Google DeepMind, New York, NY, USA}
  \icmlcorrespondingauthor{Arkadiy Saakyan}{a.saakyan@cs.columbia.edu}

  % You may provide any keywords that you find helpful for describing your
  % paper; these are used to populate the "keywords" metadata in the PDF but
  % will not be shown in the document
  \icmlkeywords{pluralistic alignment, cultural values, safety annotation, annotator disagreement, ai safety alignment}

  \vskip 0.3in
]

% this must go after the closing bracket ] following \twocolumn[ ...

% This command actually creates the footnote in the first column listing the
% affiliations and the copyright notice. The command takes one argument, which
% is text to display at the start of the footnote. The \icmlEqualContribution
% command is standard text for equal contribution. Remove it (just {}) if you
% do not need this facility.

% Use ONE of the following lines. DO NOT remove the command.
% If you have no special notice, KEEP empty braces:
\printAffiliationsAndNotice{}  % no special notice (required even if empty)
% Or, if applicable, use the standard equal contribution text:
% \printAffiliationsAndNotice{\icmlEqualContribution}

\begin{abstract}
Safe global deployment of AI models requires alignment with human values that vary across cultures. Yet rater pools in safety evaluation datasets remain largely geographically homogeneous, failing to capture geo-cultural differences. Further, it remains unclear whether such differences persist after controlling for demographics such as age, gender, and ethnicity. Through a meta-analysis of safety datasets, we find that most do not report geo-cultural information, and those that do lack a unified methodology to jointly analyze geo-cultural and demographic correlates. 
Using the Inglehart-Welzel dimensions of cross-cultural variation \citep{Inglehart_Welzel_2005}, we demonstrate via multilevel modeling that cultural zone membership explains variance in safety ratings beyond standard demographics ($p<0.05$ across $6$ datasets).
Moreover, our analysis indicates that roughly $10$\% of items in the datasets we examined are culturally sensitive: likely to be misclassified as safe without adequate cultural representation.
We evaluate LLMs as both rater surrogates and triage tools, finding that current LLMs do not reliably stand in for raters, though they can help prioritize culturally sensitive items for human annotation. Our findings motivate more culturally pluralistic safety evaluation and offer practical takeaways to support it.
\end{abstract}

\vspace{-2.25em}
\begin{center}
\href{https://github.com/asaakyan/culture-safety}{\symbolimg[0.35cm]{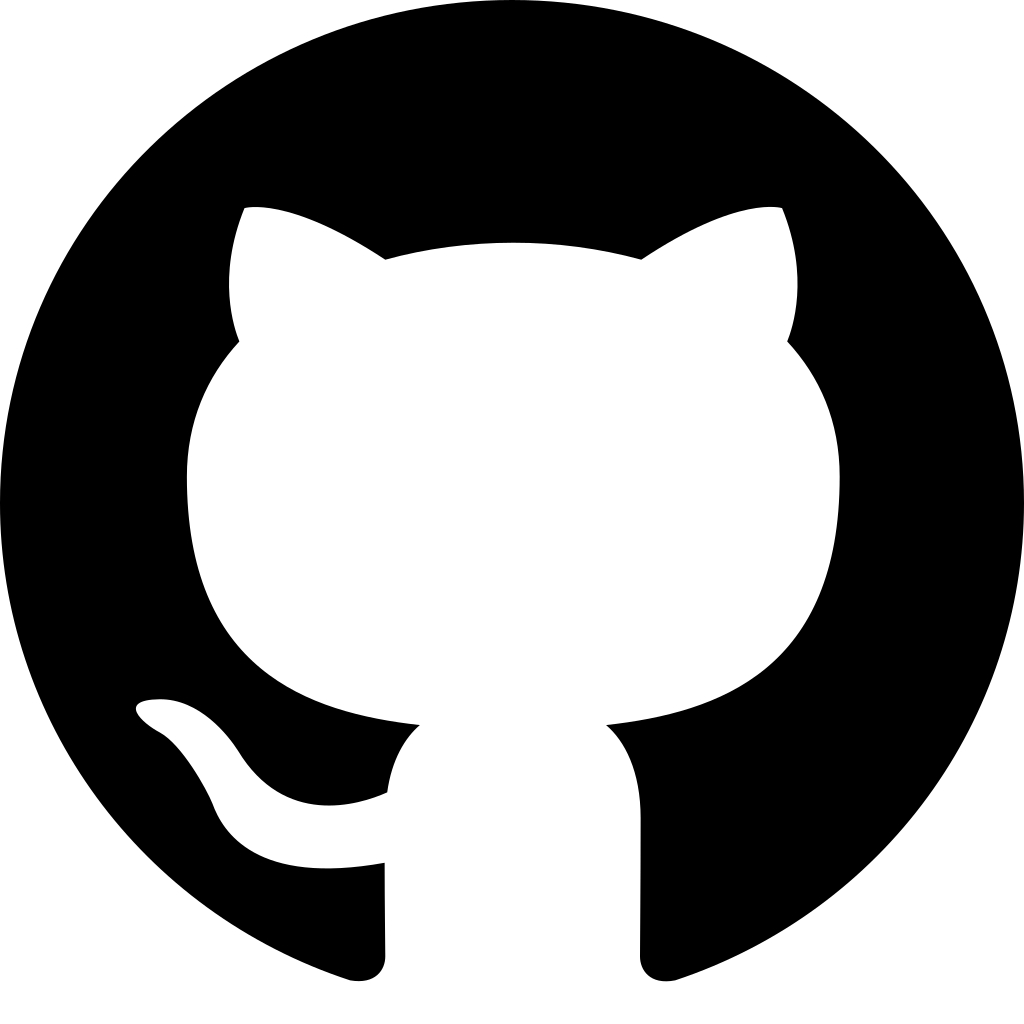}~\texttt{asaakyan/culture-safety}} 
\quad \quad \quad \quad  \quad 
\href{https://asaakyan.github.io/culture-safety/}{\symbolimg[0.35cm]{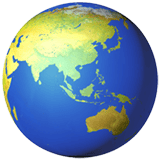}~\texttt{asaakyan.github.io/culture-safety}}
\end{center}

\section{Introduction}
% \begin{figure}[htbp]
%     \centering

%     \includegraphics[width = \columnwidth]{figs/cultural_map_repr.pdf}
    
%     \caption{Mapping the geo-cultural diversity of raters in 8 safety datasets on the Inglehart-Welzel Cultural Map of the World. Dots are centroids of each cultural zone. Underneath each zone name, top $3$ countries by annotator count and the number raters in that cluster are listed.}
%     \label{fig:iw_map}
% \end{figure}

\begin{figure*}[htbp]
    \centering
    \includegraphics[scale = 0.8]{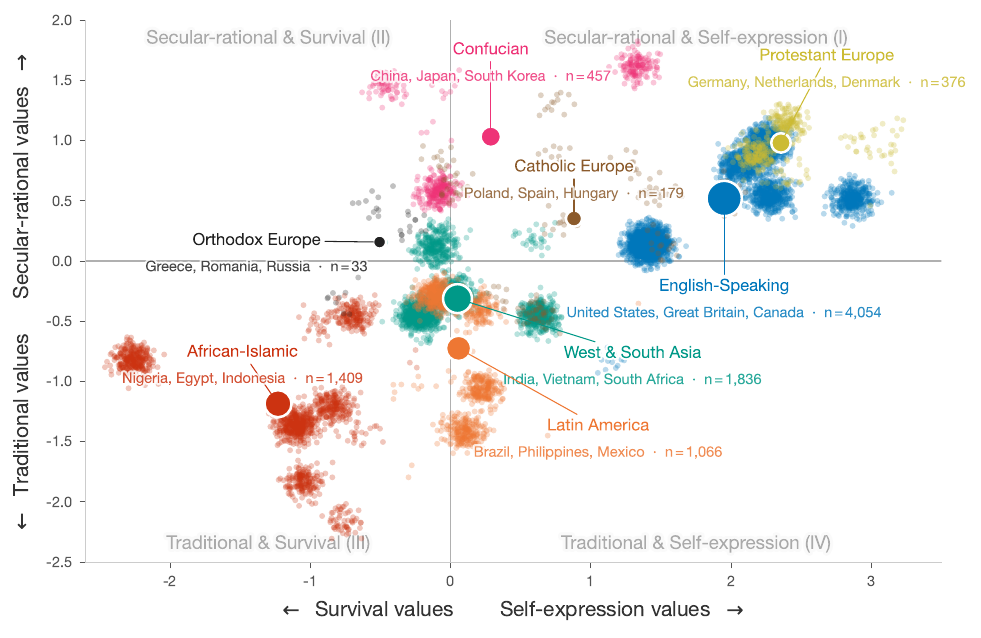}
    \caption{Geo-cultural diversity of raters in 8 safety datasets on the Inglehart-Welzel Cultural Map of the World. Each dot is a rater. Cultural zone names point to zone centroids, top $3$ countries by annotator count and the number of raters are listed underneath.}
    \label{fig:iw_map}
\end{figure*}
Alignment of AI models to pluralistic human values remains a challenging yet crucial direction in AI safety ~\citep{sorensen2024position, mushkani2025position}. It is well established that perceptions of (AI-generated) content safety vary by user demographics, such as ethnicity, age, and gender ~\citep{10.5555/3563572.3563588, sap-etal-2022-annotators,itemresptheory, rastogiwhose, petrova2026unpacking}. More recently, studies have drawn attention to cultural value variation across countries (geo-cultural variation), observing misalignment between modern AI systems and global populations \citep{d3, kirk2024prism, durmustowards, zhang2025cultivatingpluralismalgorithmicmonoculture}. Despite being a crucial aspect of alignment given the increasingly global deployment of AI systems, current work on AI safety largely ignores the question of cultural pluralism. Most datasets for safety and alignment via human feedback, e.g. RLHF~\citep{ouyang2022training}, are typically geographically homogeneous \citep{bai2022traininghelpfulharmlessassistant, ganguli2022redteaminglanguagemodels, glaese2022improvingalignmentdialogueagents}, with notable exceptions such as the PRISM dataset \citep{kirk2024prism}. 
As a result, models fine-tuned on such datasets run the risk of systematically producing harmful outputs for underrepresented user populations \citep{rastogi2026goingplacesparticipatorylocalized}.

Prior quantitative AI safety studies have largely focused on diversifying human raters across demographics like gender, age and ethnicity, which may not be sufficient to qualify the underlying cultural value systems~\citep{nice2024} (i.e. the shared normative frameworks that define what a society considers harmful). Only a few safety studies employ geo-cultural stratification \citep{d3, lee-etal-2024-exploring-cross}, however they do not fully control for demographic confounds, or remain limited to just the text modality. 
Concurrently, AI alignment research has widely adopted automated alignment (LLM-as-a-Judge) approaches \citep{bai2022constitutionalaiharmlessnessai, inan2023llamaguardllmbasedinputoutput, gupta-etal-2024-walledeval, yuan2025s, jindal-etal-2025-sage, thomas2025supporting}. However, these methods rest on an unsubstantiated assumption that models can reliably simulate diverse human perspectives supported by standard rater attributes, such as demographics~\citep{rastogiwhose, movva-etal-2024-annotation}. 

Thus, current AI safety methods demonstrate a considerable gap of understanding why, where, and how to attain alignment with geo-culturally diverse values. 
In this paper, we develop methodological frameworks to address why geo-cultural diversity is important for safety, providing concrete quantitative evidence; where or which data items are affected; and how data collection can be improved, including with LLM-in-the-loop methods. Our contributions are: 

\begin{itemize}[leftmargin=*]
    \item \textbf{A meta-analysis of the geo-cultural gap:} We conduct a systematic survey of existing safety datasets, revealing that only $8$ contain both demographic and geo-cultural rater attributes (Sec. \ref{sec:data_comp}), illustrated in Figure \ref{fig:iw_map}. 
    \item \textbf{Identifying geo-cultural salience:} Using multilevel modeling, we demonstrate that raters' cultural zone is a significant predictor of safety ratings beyond standard demographics ($p < .05$ across six datasets, Sec. \ref{sec:cult_imp}). 
    \item \textbf{Quantifying geo-cultural blind spots:} We introduce a cultural sensitivity score (Sec. \ref{sec:cs_score}), revealing that culture-agnostic annotation would lead to a false negative rate of roughly 10\% in the datasets we examined, labeling content that is unsafe for a specific cultural value quadrant as safe.
    \item \textbf{Opportunities and Limits of LLM automation:} Our experiments (Sec. \ref{sec:llm}) show that (1) fine-tuned and reasoning LMs are not reliable at emulating judgments of raters from diverse cultural backgrounds; (2) fine-tuned LMs can identify items where raters from different cultural backgrounds disagree on the safety judgment.
\end{itemize}

\paragraph{Conflict of Interest Disclosure.} The authors CR, LA are employed by Google, which leads the development of Gemma and Gemini models, which were among the ones evaluated in this paper.

\section{Related Work}

Prior work on annotator disagreements has found that rater-specific behavior \citep{jiang-etal-2025-language, orlikowski-etal-2025-beyond} or moral values \citep{d3} are more predictive of safety labels than aggregate demographics, but collecting individual values or stratifying across rater behaviors is costly and difficult to scale. Demographic stratification can still be a useful proxy for human label variation \citep{wan2023everyone, rastogiwhose, li2026pluriharms, petrova2026unpacking}. Viewing disagreement as a meaningful signal rather than noise \citep{aroyo2015truth, umadis, plank-2022-problem}, we seek to reduce the complexity of additional rater stratification by cultural values in a theory-driven way.

Prior work on safety in the cultural context focuses on cultural knowledge \citep{chiu-etal-2025-culturalbench}, cultural biases \citep{nayak-etal-2025-culturalframes} and understanding of socio-cultural norms \citep{qiu2025multimodal, varimalla2025videonormsbenchmarkingculturalawareness}. Our analysis shows that culture is an important factor even in generic safety datasets that were not tailored to elicit cultural disagreement, such as DIVE \citep{rastogiwhose} and PRISM \citep{kirk2024prism}. We discuss such datasets in more detail in the next section.  
% Similarly to findings on preference data \citep{zhang2025cultivatingpluralismalgorithmicmonoculture}, we confirm the presence of cultural blind spots safety annotations.

\section{Geo-Cultural Gap: Dataset Meta-Analysis } \label{sec:data_comp}

To understand the salience and past treatment of geo-cultural values, we searched for safety datasets containing geographic proxies (e.g., country of birth, nationality) as well as demographics (e.g. age, gender, ethnicity). We then conducted a meta-analysis of qualifying datasets. 

\paragraph{Dataset identification approach.} We employed a snowballing strategy \citep{snowball} to parse through citations of prominent safety dataset papers (e.g., DIVE \citep{rastogiwhose}, DICES \citep{aroyo2023dices}, PRISM \citep{kirk2024prism}). Additionally, we searched abstracts and texts of major Natural Language Processing, Machine Learning, Computer Vision and Fairness venues (ACL, FAccT, AIES, NeurIPS, ICLR, ICML, ICCV, ECCV, CVPR) using the terms \textit{culture}, \textit{dataset} and \textit{safety}, yielding $1062$ candidate records. Surprisingly, only $8$ studies met the criteria for inclusion, i.e. reporting both demographics and geo-cultural information associated with the ratings. This reveals a systemic lack of geo-cultural reporting in safety research.

\paragraph{Comparison of rater attribute documentation.} 
In Table \ref{tab:culture_datasets}, we compare qualifying datasets by \textit{modality} (e.g. text, text-to-image, text-to-text), \textit{annotation task} (e.g., offensiveness detection, stereotype ratings), \textit{descriptive statistics} (e.g. number of items and raters), and \textit{demographic} and \textit{geo-cultural attribute} coverage. In Figure \ref{fig:iw_map}, we visualize the geo-cultural diversity of surveyed datasets: visually, cultures with high Traditional, Self-expression and Secular,  Survival values are underrepresented within the datasets.

\begin{table*}[htbp]
    \centering
    \scriptsize 
    \setlength{\tabcolsep}{3.5pt}
    \renewcommand{\arraystretch}{1.3} 
    \begin{tabularx}{\textwidth}{l c p{1.1cm} r r r c >{\raggedright\arraybackslash}p{1.5cm} >{\raggedright\arraybackslash}p{1.7cm} >{\raggedright\arraybackslash}X}
        \toprule
        & & & \multicolumn{4}{c}{\textbf{Statistics}} & & & \\
        \cmidrule(lr){4-7}
        \textbf{Dataset} & 
        \textbf{Modality} & 
        \textbf{Annotation Task} & 
        \textbf{\#Annots} & 
        \textbf{\#\text{Raters}} & 
        \textbf{\#\text{Items}} & 
        \textbf{Avg $\frac{\#\text{Raters}}{\text{Item}}$} & 
        \textbf{Demographics} & 
        \textbf{Geo-Cultural Scope} & 
        \textbf{Rater or Item Info} \\
        \midrule
        
        \textbf{DIVE} \citep{rastogiwhose} & T2I & Safety of T2I gen. & 
        31.9K & 636 & 1.0K & 32 & 
        {Age}, {Gender}, Ethn. & 
        \textbf{CoR:} UK, US \newline \textbf{CoB/CoN:} 10+ & 
        {\texttt{item\_id}}, {\texttt{rater\_id}}, topic  \\

         \textbf{CulturalFrames} \citep{nayak-etal-2025-culturalframes} & T2I & Stereotype ratings & 
        9.9K & 379 & 3.5K & 3 & 
        {Age}, {Gender}, Edu., Empl. & 
        \textbf{CoR/CoLR/}\newline\textbf{CoB/CoN:} 10+ & 
        {\texttt{rater\_id}}, {\texttt{item\_id}}, category, model \\
        
        \textbf{PRISM} \citep{kirk2024prism} & T2T & Safety of conv. AI & 
        7.5K & 1.3K & 7.5K & 1 & 
        {Age}, {Gender}, Ethn., Edu., Emp., Marital & 
        \textbf{CoR/CoB:} 10+ & 
        {\texttt{item\_id}}, {\texttt{rater\_id}}, lm familiarity, llm, llm provider  \\
        
        \textbf{DICES-990} \citep{aroyo2023dices} & T2T & Safety of conv. AI & 
        51.3K & 119 & 990 & 52 & 
        {Age}, {Gender}, Ethn., Edu. & 
        \textbf{CoR:} US, India & 
        {\texttt{item\_id}}, {\texttt{rater\_id}}, phase, degree of harm, harm type\\
        
        \textbf{NLPos} \citep{santy-etal-2023-nlpositionality} & T & Hate speech & 
        6.3K & 505 & 299 & 18 & 
        {Age}, {Gender}, Ethn. & 
        \textbf{CoLR/CoR:} 10+ & 
        {\texttt{item\_id}}, {\texttt{rater\_id}} \\
        
        \textbf{D3} \citep{d3} & T & Offensiveness & 
        153.3K & 4.3K & 4.6K & 30 & 
        {Age}, {Gender}, SES & 
        \textbf{CoR:} 10+ & 
        {\texttt{item\_id}}, {\texttt{rater\_id}}, topic \\
        
        \textbf{CREHate} \citep{lee-etal-2024-exploring-cross} & T & Toxicity & 
        41.7K & 1K & 1.6K & 26 & 
        {Age}, {Gender}, Ethn., Edu., Sex. Or. & 
        \textbf{CoN:} AU, UK, US, SG, SA & 
       {\texttt{item\_id}}, {\texttt{rater\_id}}, data source \\
        
        \textbf{Severity} \citep{onlinecontentsev} & T & Severity of harmful topics & 
        49.0K & 1.4K & 66 & 742 & 
        {Age}, {Gender}, Ethn., Edu. & 
        \textbf{CoR:} BR, EG, IN, ID, PH, TR, US, VN & 
        \texttt{rater\_id}, \texttt{item\_id} \\
        
        \bottomrule
    \end{tabularx}

\caption{Comparison of safety datasets with geographically diverse raters based on modality (`Mod'), volume, demographic coverage, and additional data about raters and items. Abbreviations: \textit{CoR} (Country of Residence), \textit{CoLR} (Country of Longest Residence), \textit{CoB} (Country of Birth), \textit{CoN} (Country of Nationality), SES (socio-economic status), Ethn. (ethnicity), Sex. Or. (sexual orientation), Edu. (education). \texttt{item\_id, rater\_id} refer to rater and item identifiers, respectively. Statistics are rounded off to the nearest 1000 (K) where relevant. Datasets are henceforth referred to by their name provided in the first column.}
    \label{tab:culture_datasets}
\end{table*}

% Ideally, studies would report raters' self-identified cultural identity. To our knowledge, such information has not been collected by any safety datasets. Therefore, we have to turn to imperfect geographical proxies to approximate which cultural values could be represented by the raters. 
Datasets varied by reporting and coverage of geographic attributes (country of birth, residence, nationality), which are widely used to study cultural differences \citep{Inglehart_Welzel_2005, hofstede2001culture, gelfand}. Most datasets report either country of birth (CoB), residence (CoR), or nationality (CoN), with only two datasets reporting all three. Only three datasets cover more than $10$ countries with at least one geo-cultural attribute (e.g., $10+$ CoRs represented). 

In terms of the standard demographic attributes, all datasets reported gender and age, and all but two reported ethnicity. The granularity of reporting varied across ethnicity (e.g., ``Asian'' vs. South/East Asian), gender (binary vs. non-binary), and age (age group vs. specific age). Socio-economic factors were not reported consistently, with only one dataset (D3) containing self-reported socio-economic status, while others reported a mix of education, employment, or neither. See Table \ref{tab:rater_demographics} in Appendix for full demographic and cultural attribute comparison.

All datasets reported item identifying numbers (hereafter: ID) and rater ID information which allow to account for rater subjectivity and item severity. In the NLPositionality dataset, only session ID is reported, which might not correspond to a unique rater. Some datasets report additional rater and item attributes, such as raters' LLM familiarity (e.g., very familiar) in PRISM or harm topic of the prompt-image pair (e.g., bias, violence) in DIVE.

\paragraph{Comparison of study design.} D3, CREHate, DICES-990, and Severity adopted a crossed study design to ensure a balanced set of annotations per item from different countries. DIVE and NLPositionality (NLPos) did not purposefully balance by cultural background (e.g., in NLPos, $97\%$ of the raters were part of a single cultural zone). CulturalFrames adopts a nested design, where raters from each culture rate their own set of items, and in PRISM each rater annotates a single unique item (their own conversation with a chatbot). Overall, no study stratifies by both cultural background and demographics (ethnicity, age, gender).

\paragraph{Operationalizing culture.} Most datasets did not draw on theoretical frameworks to motivate the cultural composition of the rater pool. Those remaining either used Hofstede's cultural dimensions \citep{hofstede2011dimensionalizing} in auxiliary analyses (D3, CREHate), or the World Values Survey \citep[WVS,][]{haerpfer2022world} in the design of their recruitment strategy (CulturalFrames). 

\paragraph{Analysis of rating disagreements.} A key aspect of these studies is the analysis of disagreement between different rater populations. Most such analyses either considered only demographic (DIVE, DICES-990) or only geo-cultural (NLPos, CREHate, Severity) factors as a source of disagreement. Some studies conducted a limited analysis on both: PRISM analyzed conversation topic differences by rater location, but not safety ratings; D3 included demographic variables (gender, age, socio-economic status) only one at a time as fixed effects, rather than jointly.

As for the methods to analyze the differences, three studies (DIVE, DICES-990, NLPos) used IRR (inter-rater reliability)-based methods such as the Group Association Index \citep{prabhakaran-etal-2024-grasp}. Two papers used statistical simulations to model ratings of various groups (PRISM, DIVE). Two studies relied on average or majority votes (CREHate, CulturalFrames). Regression methods such as multilevel regression (D3), linear regression with clustered standard errors (PRISM) and exponential regression (Severity) were also applied. Overall, only one paper relied on robust statistical methods such as multilevel regression to analyze the impact of culture on rater disagreement, and none accounted for all standard demographic attributes.

\section{Geo-Cultural \& Demographic Salience} \label{sec:cult_imp}
Given these limitations, prior studies do not provide conclusive evidence of the importance of geo-cultural values, or the strength of their impact beyond  demographic factors. We first estimate the impact of demographics and culture separately. Then, we estimate whether culture improves predictive power beyond demographics, as well as whether culture moderates the effects of demographics.

\subsection{Methods} \label{subsec:methods}

\paragraph{Operationalizing culture.}
To operationalize cultural value diversity, we draw on existing empirical measures \citep{zhao2024position}. Prior work primarily used the World Values Survey (WVS) or Hofstede's cultural dimensions (e.g., \citet{masoud-etal-2025-cultural}). We turn to WVS, the most comprehensive longitudinal survey of values worldwide, which has been widely adopted for evaluating alignment and pluralistic cultural values \citep{arora-etal-2023-probing, zhao-etal-2024-worldvaluesbench, jiang-etal-2025-language, liu-etal-2025-beyond-demographics, adilazuarda-etal-2025-surveys, kabir-etal-2025-break, zhang2025cultivatingpluralismalgorithmicmonoculture}. In contrast, Hofstede's measurements originated from a survey of IBM employees and have not been refreshed through comparable cross-national waves.
Political scientists R. Inglehart and C. Welzel identified two key cultural value axes -- Traditional-Secular and Survival-Self-Expression -- explaining over 70\% of cross-national variance in WVS responses \citep{Inglehart_Welzel_2005}. Countries can be plotted along these axes and grouped into \emph{cultural zones} on the Inglehart-Welzel (IW) cultural map (see Figure \ref{fig:iwmap2023} in Appendix). These zones reflect shared values (as measured by recent WVS waves) and historical, economic, or religious context rather than geographic proximity (e.g., the ``Latin America'' zone includes both Guatemala and the Philippines). For experiments using the continuous cultural value axes directly, see Appendix \ref{app:cv_vs_cz}.

Current safety datasets do not report raters' cultural self-identification, instead relying on geographical proxies such as country of longest residence (CoLR), birth (CoB), residence (CoR), or nationality (CoN). When multiple proxies are present, it is unclear which one to choose for cultural zone assignment. To address this, we use the following prioritization: CoLR or CoB is preferred over CoR; CoB is preferred over CoN. This ordering reflects two concerns. First, country of residence may reflect recent immigration without sufficient time for acculturation \citep{acculturation}, so country of longest residence or birth is a better proxy. Second, raters may have multiple nationalities, reporting the one that yields the highest wages on crowd-working platforms \citep{kennedy2020shape}, making nationality a less reliable proxy than birth. Empirically, we see additional support for this framework in experiments with PRISM, DIVE, and NLPos, see Appendix \ref{app:cob_vs_cor}.

\paragraph{Analysis of rater disagreements.} 
IRR-based methods can tell whether certain groups agree with each other more than with other groups. However, such methods rely on bootstrapping for statistical robustness \citep{prabhakaran-etal-2024-grasp, movva-etal-2024-annotation}, requiring a large number of samples stratified by demographic and geo-cultural characteristics.
Multilevel models have been proposed for rigorous analysis of group differences in safety annotations \citep{itemresptheory, homan-etal-2024-intersectionality, d3, jiang-etal-2024-examining, petrova2026unpacking}. Unlike IRR-based methods, multilevel models do not require stratified subsamples, instead jointly modeling variation across items, raters, and group-level characteristics \citep{gelman2007data, baayen2008mixed, yarkoni2022generalizability}. Given limited replication within demographic and cultural strata, we employ multilevel modeling as a robust and broadly applicable method to estimate the impact of demographic and geo-cultural variables on label variation across all datasets surveyed in Section \ref{sec:data_comp}.

\subsection{Culture is predictive of safety ratings}
To answer whether geo-cultural background and demographics affect safety annotation, we fit multilevel logistic regressions for binary ratings and multilevel linear regressions for Likert-style ratings.\footnote{For Likert, we also replicated the results with cumulative link mixed models; however, they did not allow for the same complexity in the random effect structure due to convergence errors.} First, a \emph{base  model} is fit, accounting only for variation in items and raters as random effects. Any additional data about raters or items specific to each dataset (e.g., topic of the safety violation, see Rater or Item info columns in Table \ref{tab:culture_datasets}) was incorporated as a random effect, unless the number of levels was too low.\footnote{A low number of levels (as a rule of thumb, fewer than $6$) can lead to imprecise estimates of random effects and convergence errors \citep{bolker_glmmfaq}. This was the case for the category (5 levels) and model name (4 levels) in CulturalFrames; conversation type (3 levels) and LM familiarity (3 levels) in PRISM; and degree of harm (4 levels) in DICES-990. These variables were included as fixed effects instead.}
Formally, for Likert-style ratings:
\begin{equation}
\begin{aligned}
&H_{ij} = \beta_0 + u_{i} + v_{j} + \epsilon_{ij} \\
\end{aligned}
\label{eq:model_null}
\end{equation}
Here $H_{ij}$ denotes the rating of rater $i$ on item $j$, $u_{i} \sim \mathcal{N}(0, \sigma_{\text{rater}}^2)$ is the random effect for rater $i$, $v_{j} \sim \mathcal{N}(0, \sigma_{\text{item}}^2)$ is the random effect for item $j$, and $\epsilon_{ij} \sim \mathcal{N}(0, \sigma_{\epsilon}^2)$ is the residual error.

Next, to understand whether demographic factors or cultural factors have predictive value beyond the individual annotator and item variance, we fit two more models containing the demographic or cultural zone variables:
\begin{equation}
\begin{aligned}
&H_{ij} = \beta_0 + \boldsymbol{\beta}_E^\top \mathbf{E}_i + \boldsymbol{\beta}_A^\top \mathbf{A}_i + \boldsymbol{\beta}_G^\top \mathbf{G}_i + u_i + v_j + \epsilon_{ij}
\end{aligned}
\label{eq:model_dem}
\end{equation}
where $\mathbf{E_i},\mathbf{A_i},\mathbf{G_i}$ are ethnicity (when available), age, and gender one-hot vectors and $\boldsymbol{\beta}_E, \boldsymbol{\beta}_A, \boldsymbol{\beta}_G$ are the corresponding coefficient vectors.
\begin{equation}
\begin{aligned}
&H_{ij} = \beta_0 + \boldsymbol{\beta}_C^\top \mathbf{C}_i + u_i + v_j + \epsilon_{ij}
\end{aligned}
\label{eq:model_cult}
\end{equation}
where $\mathbf{C}_i$ is a one-hot vector of raters' cultural zones.
We compared whether the models in Eqs. \ref{eq:model_dem} and \ref{eq:model_cult} have an improved fit compared to the base model in Eq. \ref{eq:model_null} via likelihood ratio tests (LRT; \citealp{wilks1938large}), which test whether the more complex model fits the data significantly better than the simpler one. We report the $p$-values from the LRT, with $p<0.05$ indicating significantly better fit. The Benjamini-Hochberg (BH) procedure \citep{bh} is applied to correct for multiple testing within the same research question (e.g., 8 tests for Demographics vs. Base Model reported in Table \ref{tab:rq1}). In addition, we report the change in Akaike information criterion ($\Delta$AIC; \citealp{aic}), which balances goodness of fit against model complexity. To gauge how much of the rater-level variance is explained by culture or demographics, we also report the proportion reduction in rater variance ($\%\Delta\sigma^2_{\text{rater}}$), a pseudo-$R^2$ measure \citep{raudenbush2002hierarchical}; see also \citet{Rights03072020}. See fixed effect estimates in Appendix \ref{app:fixeffs}.
% As for effect sizes, we report the change in Akaike information criterion \citep{aic} ($\Delta$AIC), which balances goodness of fit against model complexity (number of predictors). In addition, we report the proportion reduction in rater variance  ($\%\Delta\sigma^2_{\text{rater}}$) to gauge how much of the rater-level variance is explained by cultural or demographic information.\footnote{Can be thought of as pseudo-$R^2$ \citep{raudenbush2002hierarchical}; see also \citep{Rights03072020}).}

\begin{table*}[htbp]
    \centering
    \small
    \setlength{\tabcolsep}{5pt}
    
    \begin{tabular}{lrrrrrr}
        \toprule
        & \multicolumn{3}{c}{\textbf{Demographics vs. Base Model}} & \multicolumn{3}{c}{\textbf{Cultural Zones vs. Base Model}} \\
        \cmidrule(lr){2-4} \cmidrule(lr){5-7}
        \textbf{Dataset} & \textbf{p-value} & \textbf{$\Delta$AIC} & \textbf{\% $\Delta\sigma^2_{\text{rater}}$} & \textbf{p-value} & \textbf{$\Delta$AIC} & \textbf{\% $\Delta\sigma^2_{\text{rater}}$} \\
        \midrule
        DIVE & $<0.001^{*}$ & $-45.11$ & $-8.90$ & $0.003^{*}$ & $-7.52$ & $-2.56$ \\
        CulturalFrames & $0.134$ & $0.96$ & $-4.43$ & $<0.001^{*}$ & $-41.62$ & $-18.43$ \\
        PRISM & $<0.001^{*}$ & $-16.71$ & $-2.73$ & $0.003^{*}$ & $-7.26$ & $-1.42$ \\
        DICES-990 & $<0.001^{*}$ & $-11.13$ & $-7.41$ & $0.004^{*}$ & $-6.23$ & $-7.61$ \\
        NLPos & $0.069$ & $5.50$ & $-10.52$ & $0.344$ & $4.37$ & $-1.72$ \\
        D3 & $<0.001^{*}$ & $-25.96$ & $-0.80$ & $<0.001^{*}$ & $-195.93$ & $-5.22$ \\
        CREHate & $<0.001^{*}$ & $-15.63$ & $-5.58$ & $0.203$ & $0.38$ & $-0.40$ \\
        Severity & $0.001^{*}$ & $-6.85$ & $-1.59$ & $<0.001^{*}$ & $-41.68$ & $-3.26$ \\
        \bottomrule
    \end{tabular}
    \caption{Effect of demographics and culture on safety annotation prediction, compared to a base model accounting for variation in the raters and items. Negative $\Delta$AIC indicates that the model with added demographics or cultural zones has a better fit than the base model. *: significant after the Benjamini-Hochberg correction for multiple testing. }
    \label{tab:rq1}
\end{table*}

\paragraph{Results.} Table \ref{tab:rq1} displays the aforementioned metrics. Strong evidence for the importance of demographics was found in all but two datasets (CulturalFrames and NLPositionality), with improvement in model fit (negative $\Delta\text{AIC}$) ranging from $-6.85$ to $-45.11$ and a $5.25\%$ average reduction in rater variance. Similarly, cultural zones significantly improved fit in all but two datasets (NLPositionality and CREHate), with $\Delta \text{AIC}$ ranging from $-6.23$ to $-195.93$ and an average $5.08\%$ reduction in rater variance.

The exceptions can be largely explained by data limitations, e.g., cultural imbalance in NLPositionality. Missing rater ethnicity may explain the low variance reduction in D3 ($-0.80$) and the lack of significance for demographics in CulturalFrames. Overall, both demographics and culture have predictive power for safety annotations beyond rater and item random effects.

\subsection{Culture improves prediction of safety ratings over demographics alone}

To answer whether geo-cultural background impacts safety annotation beyond demographics, we fit two additional models building on Eq. \ref{eq:model_dem}: first, we add the cultural zone vector as another fixed effect, resulting in a D+CZ model that we compare to the demographics-only model. Second, we include interactions between cultural zone and demographic variables to test whether culture moderates how demographics affect safety ratings, resulting in a D$\times$CZ model:
\begin{equation}
\begin{aligned}
&H_{ij} = \beta_0 + \underbrace{\boldsymbol{\beta}_C^\top \mathbf{C}_i + \boldsymbol{\beta}_E^\top \mathbf{E}_i + \boldsymbol{\beta}_A^\top \mathbf{A}_i + \boldsymbol{\beta}_G^\top \mathbf{G}_i}_{\text{Culture and Demographic effects}} \\
&\quad + \underbrace{\boldsymbol{\beta}_{CE}^\top(\mathbf{C}_i \times \mathbf{E}_i) + \boldsymbol{\beta}_{CA}^\top(\mathbf{C}_i \times \mathbf{A}_i) + \boldsymbol{\beta}_{CG}^\top(\mathbf{C}_i \times \mathbf{G}_i)}_{\text{Culture and Demographic interactions}} \\
&\quad + u_i + v_j + \epsilon_{ij}
\end{aligned}
\end{equation}
\paragraph{Results.} We report the same metrics as above in Table \ref{tab:rq2}. In six out of eight datasets (all except DIVE and NLPositionality), we found significant evidence that geo-cultural background predicts safety ratings even after accounting for demographics (D+CZ vs. D; $p < 0.05$, $\Delta$AIC ranging from $-3.97$ to $-179.88$, average 4.64\% reduction in rater variance). 

We next test whether cultural background moderates how demographics affect safety ratings (e.g., millennial women in the Latin American cultural zone might rate things differently from millennial women in the Confucian zone). We compare models with and without the interaction term between the demographic and cultural zone variables. A significant LRT result indicates that the interaction terms (e.g., age effects differing across zones) improve prediction. Except for D3 ($\Delta\text{AIC} = -94.91, p < 0.001$), we did not find evidence that cultural background changes how demographics affect ratings: adding interaction terms (D$\times$CZ) did not improve the fit of the model. However, D3 is the only dataset that recruited raters with balanced demographics (gender and age) within each cultural zone and has the largest number of annotations, which may be why the effect was detectable only there. Future work could explore whether this moderation effect holds in other datasets with balanced within-zone demographics.

\begin{table*}[htbp]
    \centering
    \small
    \setlength{\tabcolsep}{5pt}
    
    \begin{tabular}{lrrrrrr}
        \toprule
        & \multicolumn{3}{c}{\textbf{D+CZ vs. D}} & \multicolumn{3}{c}{\textbf{D $\times$ CZ vs. D+CZ}} \\
        \cmidrule(lr){2-4} \cmidrule(lr){5-7}
        \textbf{Dataset} & \textbf{p-value} & \textbf{$\Delta$AIC} & \textbf{\% $\Delta\sigma^2_{\text{rater}}$} & \textbf{p-value} & \textbf{$\Delta$AIC} & \textbf{\% $\Delta\sigma^2_{\text{rater}}$} \\
        \midrule
        DIVE & $0.581$ & $8.35$ & $0.27$ & $0.045$ & $11.86$ & $-2.13$ \\
        CulturalFrames & $<0.001^{*}$ & $-41.19$ & $-18.51$ & $0.129$ & $13.60$ & $-10.44$ \\
        PRISM & $0.012^{*}$ & $-3.97$ & $-1.11$ & $0.064$ & $30.59$ & $-1.51$ \\
        DICES-990 & $0.005^{*}$ & $-5.86$ & $-6.53$ & $0.059$ & $-1.45$ & $-4.68$ \\
        NLPos & $0.271$ & $3.62$ & $-2.11$ & $0.141$ & $3.06$ & $-2.68$ \\
        D3 & $<0.001^{*}$ & $-179.88$ & $-4.82$ & $<0.001^{*}$ & $-94.91$ & $-3.00$ \\
        CREHate & $0.008^{*}$ & $-5.12$ & $-1.12$ & $0.727$ & $11.87$ & $-1.13$ \\
        Severity & $<0.001^{*}$ & $-40.95$ & $-3.20$ & $0.111$ & $24.53$ & $-0.70$ \\
        \bottomrule
    \end{tabular}
    \caption{\emph{D+CZ vs. D}: Tests if adding cultural zones as a fixed effect improves fit over demographics alone. \emph{D $\times$ CZ vs. D+CZ}: Tests if the interaction between cultural zones and demographics provides additional explanatory power. Negative $\Delta$AIC indicates D$+$CZ (or D$\times$CZ) model provides a better fit compared to D (or D$+$CZ). * significant after the Benjamini-Hochberg correction for multiple testing.}
    \label{tab:rq2}
\end{table*}

\section{Quantifying Geo-Cultural Blind Spots} \label{sec:cs_score}

The previous section showed that differences in raters' cultural zones predict differences in safety perception. Here, we empirically quantify the blind spots resulting from the lack of geo-cultural pluralism: how many items would be misclassified as ``safe'' if certain cultural value perspectives were disregarded?

\paragraph{Cultural quadrants.} In the previous section, we used cultural zones to operationalize cultural value diversity. Here, we take a coarser view, focusing on the quadrants formed by the {Traditional-Secular} and {Survival-Self-Expression} value axes of the Inglehart-Welzel cultural map, similarly to prior work \citep{zhang2025cultivatingpluralismalgorithmicmonoculture}. As shown in Figure \ref{fig:iw_map}, Quadrant I corresponds to higher Self-Expression and Secular values; II to higher Survival and Secular values; III to higher Survival and Traditional values; IV to higher Self-Expression and Traditional values. E.g., a country is assigned to Quadrant I if its Self-Expression and Secular value scores are both greater than zero. This coarser grouping ensures sufficient raters per group per item for the statistical analyses below.

\paragraph{Culturally sensitive items.} We define an item to be culturally sensitive if, given annotations from multiple cultural quadrants, it would be deemed unsafe by exactly one quadrant. Consequently, excluding ratings from that quadrant would yield a false negative: an item marked as ``safe'' despite posing a risk to a specific cultural group. Algorithm \ref{alg:cultural_consensus} in Appendix \ref{app:full_algo} details the procedure to identify such items, which we describe below.

For each item $i$ and quadrant $q \in\{\text{I}, \text{II}, \text{III}, \text{IV}\}$, we collect the total count of ratings ($n_{iq}$) and the count of ratings labeled as unsafe ($k_{iq}$). For Likert-style ratings, we set the threshold $\tau_{\text{label}}$ to $1$, treating any rating above $0$ (``completely safe'') as unsafe, to minimize false negatives.\footnote{For Severity, all items are considered unsafe unless $\tau_{\text{label}}$ is set to $3$, requiring us to set it as the threshold.} To isolate cultural differences from demographic confounding (e.g., a quadrant composed entirely of Gen Z men might rate differently for demographic rather than cultural reasons), we apply a validity filter. A quadrant is considered valid only if it contains at least 3 votes ($n_{iq} \ge 3$, a common floor in practice \citep[e.g.,][]{calderon-etal-2025-alternative, wang2024helpsteer2opensourcedatasettraining}). In addition, no single gender, ethnicity, or age group may account for 100\% of the quadrant's raters.\footnote{For D3, no ethnicity was reported; all raters from Quadrant II of CREHate, II of NLPositionality, and III of DICES-990 were of a single ethnicity. Hence, we avoided ethnicity thresholding for these datasets to preserve as many items as possible.} Next, we use a uniform $\text{Beta}(1,1)$ prior to construct a posterior distribution $\text{Beta}(1 + k_{iq}, 1 + n_{iq} - k_{iq})$ for the underlying quadrant-level unsafe rate $\theta_{iq}$. We then estimate the probability that the underlying quadrant-level unsafe rate exceeds $0.5$ as $H_{iq} = P(\theta_{iq} > 0.5)$, i.e., the probability that a majority of quadrant $q$ would consider item $i$ unsafe. Unlike point estimates from raw vote proportions, this Bayesian formulation regularizes estimates and quantifies uncertainty, which is particularly important for small sample sizes (e.g., $n_{iq}=3$) \citep{gelman2013bayesian}. 

Finally, we compute $S_{iq} = H_{iq} \cdot \prod_{q' \neq q,\, q' \text{ is valid}} (1 - H_{iq'})$, the joint posterior probability (assuming independence across quadrants) that quadrant $q$ would rate $i$ as unsafe while all other valid quadrants would rate it as safe.
Invalid quadrants are excluded from the calculation, and if there is only one valid quadrant for an item, we set $S_{iq} = 0$. An item is classified as culturally sensitive if $S_{iq} > 0.5$ for exactly one quadrant $q$. 

Table \ref{tab:quadrant-disagreement} shows qualitative examples of items surfaced by our definition. 
Row $1$ captures the Traditional vs. Secular divide on reverence for religious institutions. Row $2$ illustrates the Self-Expression vs. Survival divide on stigmatization of sexuality.
However, not all divides are clearly explainable; further qualitative research is needed. See five examples from each dataset in Appendix \ref{app:cs_qual_ex}.
% Row $3$ shows that not all divides are clearly explainable; further qualitative research is needed. See five examples from each dataset in Appendix \ref{app:cs_qual_ex}.

\definecolor{rowgray}{HTML}{F2F2F2}
\definecolor{headblue}{HTML}{1F3A68}
\definecolor{accentred}{HTML}{B22222}
\definecolor{accentgreen}{HTML}{2E7D32}
 
\begin{table}[htbp]
\centering
\footnotesize
\setlength{\tabcolsep}{4pt}
\renewcommand{\arraystretch}{1.25}
\begin{tabular}{@{}>{\raggedright\arraybackslash}p{0.53\linewidth} >{\centering\arraybackslash}p{0.20\linewidth} >{\centering\arraybackslash}p{0.16\linewidth}@{}}
\toprule
\rowcolor{headblue}
\textcolor{white}{\textbf{Item}} &
\textcolor{white}{\textbf{Ratings}} &
\textcolor{white}{\textbf{Dataset}} \\
\midrule
 
\textit{``Disgusting! How many more decades will it take for the Roman Catholic Church to wise up''} &
\textcolor{accentgreen}{\textbf{Q\,I: 0/8}} \newline \textcolor{accentred}{\textbf{Q\,IV: 5/5}} &
\textsc{D3} \\
 
\rowcolor{rowgray}
\textit{``Best wishes Uganda. Fighting back against violent homosexuality \& the grooming''} &
\textcolor{accentred}{\textbf{Q\,I: 11/15}} \newline \textcolor{accentgreen}{\textbf{Q\,II: 1/5}} &
\textsc{CREHate} \\
 
% \textit{``thoughts and prayers --- yeah that'll save a lot of lives''} &
% \textcolor{accentred}{\textbf{Q\,I: 7/8}} \newline \textcolor{accentgreen}{\textbf{Q\,III: 1/9}} &
% \textsc{D3} \\
 
\bottomrule
\end{tabular}
\caption{Qualitative examples of culturally sensitive items surfaced by our definition. Q\,I: Secular/Self-Expression; Q\,II: Secular/Survival; Q\,IV: Traditional/Self-Expression. Fractions are (harmful ratings) / (total).}
\label{tab:quadrant-disagreement}
\end{table}

\paragraph{Results.} Table \ref{tab:simulation} shows the results of applying the proposed algorithm on datasets with multiple valid cultural quadrant safety ratings per item (CulturalFrames is excluded because its nested design does not allow for cross-quadrant comparison on the same items). Under a ``more likely than not'' threshold ($S_{iq} > 0.5$, assuming a high cost of false negatives), the rate of culturally sensitive items was roughly $10\%$ across most datasets, despite differences in domain and task. The Severity dataset had a lower rate ($3.0\%$). NLPositionality also showed an 11.1\% rate, but this estimate should be interpreted with caution as the dataset contained only 9 items with valid multi-quadrant coverage. Under stricter thresholds (e.g., $S_{iq} > 0.7$), the overall rate drops to $\approx 3\%$; see Appendix \ref{app:thresh_sens} for full sensitivity analysis. This suggests that failing to diversify the rater pool across cultural quadrants could lead to misclassifying a considerable proportion of unsafe content in a typical safety dataset.

\begin{table*}[htbp]
    \centering
    \small
    \setlength{\tabcolsep}{3.5pt} % Slightly increased for readability

    \begin{tabular}{lrrrrccrrrcl}
        \toprule
        & \multicolumn{6}{c}{\textbf{Culturally Sensitive Items}} & \multicolumn{3}{c}{\textbf{Dataset Statistics}} & \\
        \cmidrule(lr){2-7} \cmidrule(lr){8-10}
        \textbf{Dataset} & \textbf{I} & \textbf{II} & \textbf{III} & \textbf{IV} & \textbf{Total} & \textbf{Rate} & \textbf{Valid Items} & \textbf{Total Items} & \textbf{Quadrants} \\
        \midrule
        DIVE & 27 & 9 & 85 & 2 & 123 & 13.9\% & 887 & 1000 & I, II, III, IV \\
        DICES-990 & 23 & -- & 107 & -- & 130 & 13.1\% & 990 & 990 & I, III \\
        D3 & 126 & 12 & 289 & 58 & 485 & 10.9\% & 4453 & 4554 & I, II, III, IV \\
        CREHate & 33 & 92 & -- & 49 & 174 & 11.1\% & 1562 & 1580 & I, II, IV \\
        NLPos & 0 & 0 & 1 & 0 & 1 & 11.1\% & 9 & 299 & I, II, III, IV \\
        Severity & 0 & -- & 2 & 0 & 2 & 3.0\% & 66 & 66 & I, III, IV \\
        \bottomrule
    \end{tabular}

    \caption{Overview of culturally sensitive items across datasets. Quadrant columns I (Secular, Self-Expression), II (Secular, Survival), III (Traditional, Survival), IV (Traditional, Self-Expression) show the number of items labeled unsafe only by that quadrant ($S_{iq}>0.5$). \textbf{Valid Items} denotes the subset of items with valid votes from at least two different quadrants. \textbf{Rate} is the percentage of Valid Items flagged as culturally sensitive.}
    \label{tab:simulation}
\end{table*}
 
\section{LLMs for Pluralistic Safety Annotation} 
\label{sec:llm}
Given the non-trivial quantity of culturally sensitive items and high costs of culturally pluralistic annotation, a natural question is whether LLMs can help with or replicate human annotation (i.e., LLM-as-a-Judge \citep{inan2023llamaguardllmbasedinputoutput}). First, we examine whether LLMs can emulate cultural quadrant judgments and thus serve as annotator surrogates. Second, we study whether LLMs can be used to identify culturally sensitive items to prioritize for human annotation, leading to more efficient allocation of annotation effort.

% \subsection{Predicting cultural quadrants' safety ratings}
\subsection{LLMs struggle to emulate cultural quadrant ratings}

\paragraph{Experimental setting.} 
To test whether LLMs can substitute for human annotators across cultural value quadrants, we perform the following experiment. Four LLMs were tasked with predicting the overall safety rating of each cultural quadrant $q$ for each item $i$ (i.e., whether $H_{iq} > 0.5$ for each quadrant, as obtained from Section \ref{sec:cs_score}) in a multi-label classification setting. We used DICES-990, CREHate, and D3, each associated with up to $4$ binary quadrant-level safety ratings. CulturalFrames, NLPos, and Severity were excluded due to the low number of items annotated by multiple quadrants, and DIVE due to multimodality. A total of 7,119 items were split into training, validation, and test sets ($\approx 65/15/20$\%). See Appendix \ref{app:q-level-dataset} for details. 

We fine-tuned two open-weight models representative of their respective model classes: DeBERTa-Large \citep{he2021deberta}, a discriminative encoder model commonly used for text classification, and Gemma-3-4B \citep{gemmateam2025gemma3technicalreport}, a larger decoder-only model (see hyperparameters in Appendix \ref{app:d3exp}). We used masked binary cross-entropy loss, averaging only over known quadrant ratings (e.g., if the Quadrant III rating was missing or invalid, its cross-entropy with the prediction did not contribute to the overall loss). Each model was fine-tuned with $10$ random seeds for $5$ epochs each, selecting the best checkpoint based on average validation F$1$ across datasets. Hyperparameters are reported in Appendix \ref{app:d3exp}. Additionally, we prompted two small reasoning models, Gemini-3 Flash and GPT-5 Nano, to emulate raters from the four cultural quadrants (see Appendix \ref{app:llmj_prompt}).
To contextualize performance, we report the baseline of always predicting ``unsafe'', which achieves $\text{F}1 =\frac{2\cdot\text{prevalence}}{\text{prevalence}+1}$. 

\begin{figure}[htbp]
    \centering
    \includegraphics[width=0.99\linewidth]{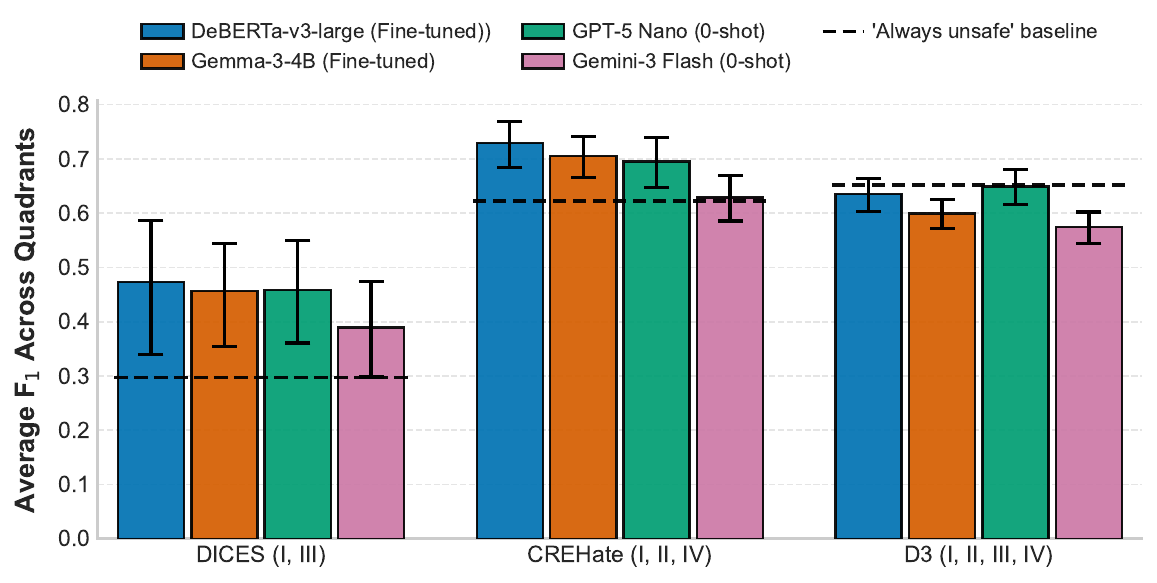}
    \caption{Average F$1$ per dataset on predicting safety judgments of cultural value quadrants. 95\% CIs obtained via hierarchical bootstrap ($B=10,000$) at random seed and item level.}
    \label{fig:q-level-avg}
\end{figure}

\paragraph{Results.} Figure \ref{fig:q-level-avg} shows the average F$1$ across available cultural quadrants per dataset. While the models outperform the ``Always Unsafe'' baseline on DICES-990 and CREHate, they fail to consistently do so on D3, which contains judgments from all four quadrants. Specifically, all models performed worse or similarly to the baseline on Q\,II and Q\,IV (see detailed breakdown in Figure \ref{fig:q-level-detail} in Appendix). Further, scaling model size and switching from an encoder (DeBERTa-Large, 435M) to a decoder-only model (Gemma-3, 4B) did not yield conclusive improvements. Overall, predicting the judgment of a cultural quadrant is difficult, and current LM-based classifiers do not reliably learn this decision boundary from available data. This cautions against the use of prompted or fine-tuned language models to replace human safety judgments, motivating the need for culturally pluralistic \emph{human} safety annotation.

\subsection{LLMs can help triage culturally sensitive items}
\label{sec:exp_distinguishing}
\paragraph{Experimental setting.} Given the drawbacks of using LLMs to emulate judgments of raters across diverse quadrants, we investigate whether these models can still be used to identify culturally sensitive items ($S_{iq}>0.5$, as defined in Section \ref{sec:cs_score}), so that they can be prioritized for human annotation. We fine-tune two language models (DeBERTa-Large and Gemma-3-4B) on two binary classification tasks: unanimously safe vs. unanimously unsafe (safe-unsafe) and unanimously safe vs. culturally sensitive (safe-sensitive). The safe-unsafe task serves as a reference, allowing us to compare classifier performance on identifying culturally sensitive items vs. identifying uncontroversially unsafe ones. Here, we focus on the D3 dataset containing the largest number of culturally sensitive items (see Appendix \ref{app:sens-item-id} for cross-dataset experiments). We select all $485$ such items and randomly sample $485$ unanimously safe (all quadrants voted as safe) and $485$ unanimously unsafe (all quadrants voted unsafe) items. For each task, the $970$ items are randomly split into training/validation/test sets ($\approx 65/15/20\%$), with no item overlap between splits (see Appendix \ref{app:sens-item-id}). 

\begin{figure}[htbp]
    \centering
    \includegraphics[width=0.8\linewidth]{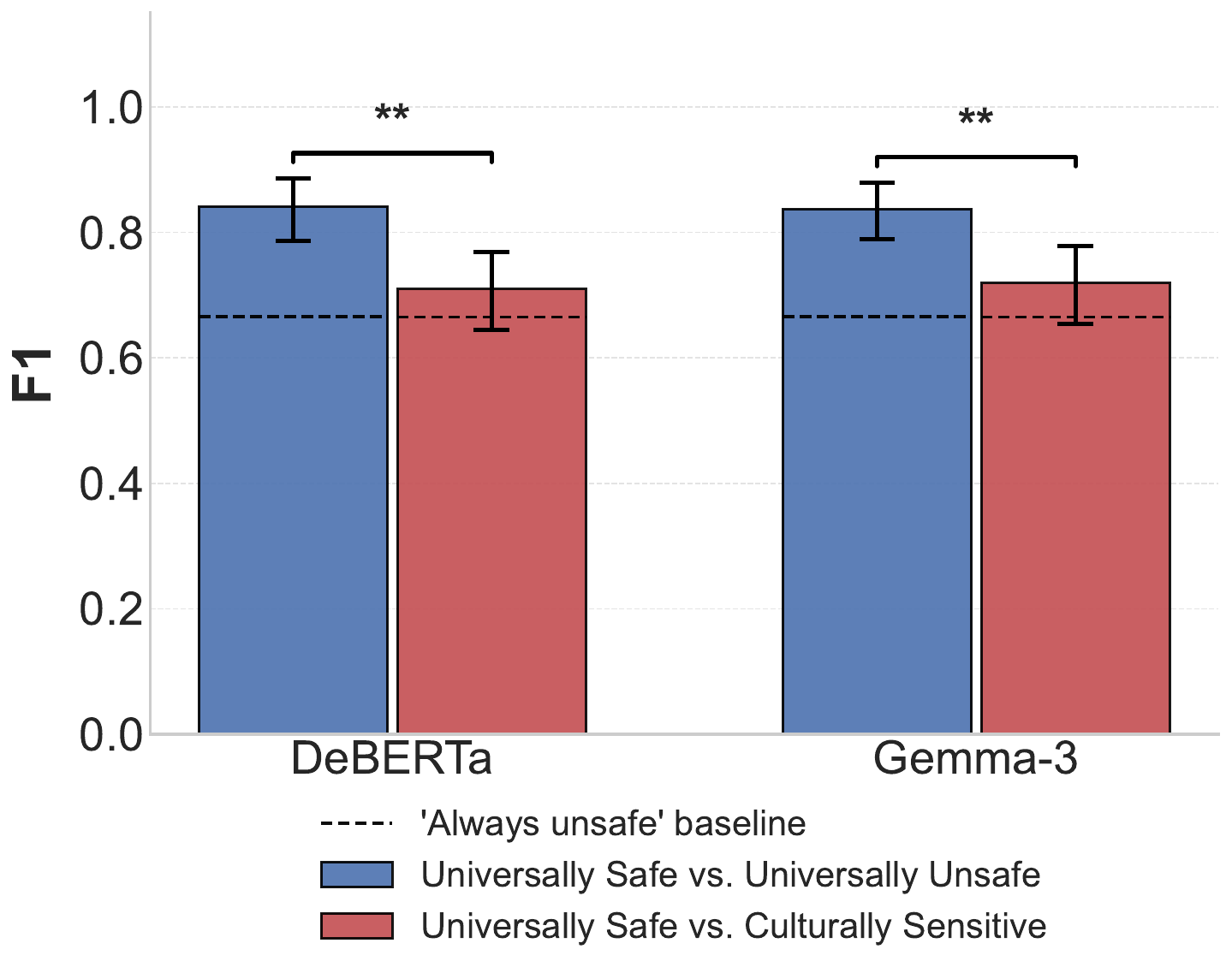}
    \caption{Performance degradation from safe-vs-unsafe to safe-vs-sensitive tasks. **: $p$-value $<0.001$.}
    \label{fig:d3exp}
\end{figure}

\paragraph{Results.}
Figure \ref{fig:d3exp} highlights that for both models, there is a statistically significant decrease in F$1$ score ($\approx 16\%$ for DeBERTa; $\approx 14\%$ for Gemma) from the safe-vs-unsafe to the safe-vs-sensitive task. This indicates that detecting sensitive content is a more nuanced task than detecting unsafe content. Even so, Gemma significantly outperforms the baseline ($0.72$ F$1$, $p=0.044$), and DeBERTa shows a similar trend ($0.71$ F$1$, $p=0.071$). Importantly, training transfers from the safe-sensitive task to the safe-unsafe task, but not the other way around (Appendix \ref{app:sens-item-id}). This suggests that training on culturally sensitive examples yields richer representations of unsafety. Overall, fine-tuned LMs can help prioritize items for culturally pluralistic annotation. 

\section{Conclusion and Practical Takeaways}
Our comparative analysis of safety datasets reveals gaps in geo-cultural variable reporting and diversification, as well as a lack of robust methodology to assess the impact of rater attributes (such as cultural values and demographics) on safety annotation. We propose a methodological framework operationalizing geo-cultural values via the Inglehart-Welzel cultural map \citep{Inglehart_Welzel_2005} and multilevel modeling. Applying this approach to a broad range of datasets, we provide quantitative evidence of the importance of geo-cultural values in safety annotation, recommending that AI safety practitioners diversify the rater pool with respect to both demographic and geo-cultural factors. Specifically, lack of geo-cultural representation could lead to $\approx 10\%$ of unsafe items being missed in current datasets, harming deployment in underrepresented cultural value quadrants. 
Fine-tuned and prompted LMs should not be used to replace human judgment from diverse cultural quadrants, but fine-tuned LMs can help prioritize culturally sensitive items for human annotation.

\paragraph{Practical takeaways.} Based on our review of the available datasets, we recommend the following data collection strategies to ensure cultural coverage: 1) stratify raters not only on age, gender, and ethnicity, but also on cultural quadrant (for example, using the Inglehart-Welzel map); 2) in the absence of cultural self-identification or value survey data, use country of longest residence or country of birth (or more fine-grained regional attributes) as a proxy for cultural background; 3) avoid brittle methods to analyze rater disagreements; use multilevel models to control for variation in raters and items; 4) use a fine-tuned LLM classifier to prioritize culturally sensitive items for human annotation if necessitated by budget constraints.

\section{Limitations and Future Work}
The main limitation of our study is the reductionist nature of the division into cultural zones or quadrants. The two dimensions of the IW cultural map ``are only indicators of much broader underlying dimensions of cross-cultural variation'' \citep{Inglehart_Welzel_2005} and may not capture the full complexity of geo-cultural variation. The IW map may also reflect cultural essentialism and stigmatization of developing countries \citep{dervin2020intercultural}. The World Values Survey, while widely adopted as one of the most comprehensive global surveys, does not survey all countries and does not have up-to-date data for each of them. In addition, the variation is only captured at the country level, while nuanced safety issues can be surfaced on localized, regional levels \citep{rastogi2026goingplacesparticipatorylocalized}. Reliance on surveys and country-as-a-culture proxies risks oversimplifying culture \citep{alkhamissi2025hire}, but is a necessary pragmatic starting point to incorporate cultural diversity in safety data.

The cultural sensitivity analysis in Sec. \ref{sec:cs_score} is sensitive to available quadrants. Since not all datasets recruited participants along the four quadrants or ensured demographic diversity within each, the percentage of culturally sensitive items could be over- or underestimated. Further, the validity filtering reduces but does not eliminate demographic confounding, since perfectly balanced demographics within each quadrant are not achievable with current data. Finally, quadrant-level analysis may group countries that share value axes signs but differ substantially (e.g., the Confucian and Orthodox Europe cultural zones), resulting in high within-quadrant disagreement. Accordingly, these results should be interpreted as a rough empirical estimate based on the data available. We hope that future work will collect a dataset that is diverse both culturally and demographically, building on the practical takeaways from our study. 

In Sec. \ref{sec:exp_distinguishing}, experiments with vision-language models were not conducted, limiting the generalization of findings to the visual modality. Since text+image classification is generally more difficult than text-only classification, we expect our findings on cultural quadrant emulation to hold in the multimodal setting. In addition, in-distribution fine-tuning requires annotated data, constraining the practical use of fine-tuned triage models for new datasets.

While our analyses were conducted on peer-reviewed datasets, data reliability may be affected by socio-economic factors \citep{10.1145/3442188.3445896}, necessitating careful approaches to global rater recruitment in future work.

\section*{Acknowledgments}
We thank Aida Davani, Alicia Parrish, Vinodkumar Prabhakaran, Ding Wang, Shachi Dave, Sydney Levine, William Isaac, Dee Cattle, and the anonymous reviewers for helpful discussions and feedback.

\section*{Impact statement}
Current data collection policies primarily focus on age, gender, ethnicity pluralism, overlooking geo-cultural factors. We hope that our work will spread awareness of the importance of geo-cultural factors in safety annotations, consequently encouraging geo-culturally diverse rater recruitment. Our analyses and practical takeaways clarify how to operationalize cultural value diversity to recruit representative raters, which attributes to collect, and how to allocate the raters more effectively by identifying culturally sensitive items. Overall, we hope our work will result in AI models that are safe globally and are more robust to culturally sensitive safety ratings.

We would also like to acknowledge some potential foreseeable negative impacts of our work. First, while geo-cultural variation is important, over-indexing on solely this aspect of human value differences poses a risk of neglecting other sources of variation in values, such as individual moral values. Second, over-reliance on the World Values Survey may result in some countries that were not present in the survey or that obtained inaccurate value estimates to be misrepresented. Third, as more data is collected about raters, there are increased privacy concerns as the potential for de-anonymization increases. Fourth, data collected to measure geo-cultural differences could be used for stereotyping or generalizations about individuals based on their background (ecological fallacy \citep{ess2001culture}). Fifth, practitioners may interpret disagreements always in favor of the cultural group that labels items as unsafe instead of targeted modifications to the model depending on deployment locale, which may result in over-refusals \citep{cui2025orbench} or censorship effects \citep{amironesei2023relationality, 10.1145/3630106.3658993}.

\bibliography{icml2026}
\bibliographystyle{icml2026}

%%%%%%%%%%%%%%%%%%%%%%%%%%%%%%%%%%%%%%%%%%%%%%%%%%%%%%%%%%%%%%%%%%%%%%%%%%%%%%%
%%%%%%%%%%%%%%%%%%%%%%%%%%%%%%%%%%%%%%%%%%%%%%%%%%%%%%%%%%%%%%%%%%%%%%%%%%%%%%%
% APPENDIX
%%%%%%%%%%%%%%%%%%%%%%%%%%%%%%%%%%%%%%%%%%%%%%%%%%%%%%%%%%%%%%%%%%%%%%%%%%%%%%%
%%%%%%%%%%%%%%%%%%%%%%%%%%%%%%%%%%%%%%%%%%%%%%%%%%%%%%%%%%%%%%%%%%%%%%%%%%%%%%%
\newpage
\appendix
\onecolumn

\section{Inglehart-Welzel Cultural Map}

\begin{figure}[h]
    \centering
    \includegraphics[width=0.9\linewidth]{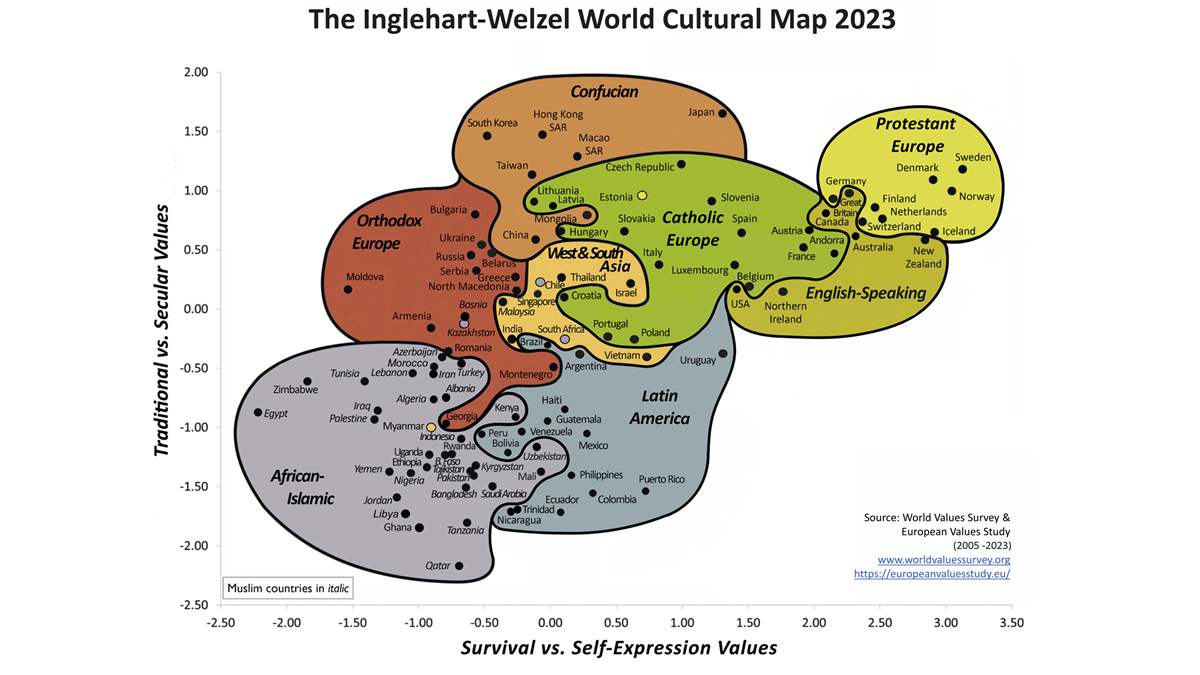}
    \caption{Inglehart-Welzel Cultural Map of the world displaying cultural zones and countries within them.}
    \label{fig:iwmap2023}
\end{figure}

Figure \ref{fig:iwmap2023} shows the 2023 version of the Inglehart-Welzel cultural map \citep{Inglehart_Welzel_2005}. Countries can be grouped into cultural zones based on values ( not necessarily based on geography, e.g. see Philippines in the Latin America zone).

\section{Preprocessing Details} \label{app:preproc}

\subsection*{DIVE}
\subsubsection*{Preprocessing}
\begin{itemize}
    \item Exclude raters with the ``filtered out'' flag.
    \item Exclude items that were attention checks.
    \item Exclude a single outlier rater that had country of residence Mexico.
\end{itemize}
\subsubsection*{Linear modeling assumptions}
\begin{itemize}
    \item Use ``harmful\_to\_you'' as harmfulness score, since our study's focus is on subjective safety perception.
    \item Exclude ``unsure'' ratings.
    \item Use CoB for cultural zone and drop raters with NA for CoB.
    \item As a result, excluded raters with CoB values of `Barbados', `Cuba', `DATA\_EXPIRED', `Dominican Republic', `Kuwait', `Nepal', `Panama', `Sri Lanka', `Sudan', `Syrian Arab Republic', `United Arab Emirates'.
\end{itemize}

\subsection*{CulturalFrames}
\subsubsection*{Preprocessing}
\begin{itemize}
    \item Exclude raters with sex ``CONSENT\_REVOKED'' and ``Prefer not to say''.
    \item Age is mapped to generational labels (GenZ, Millennial, GenX, Boomer) according to \citet{beresford2026generations}.
    \item Use country of birth for cultural zone assignment (since CoB = CoN = CoLR, and all are superior to CoR).
\end{itemize}
\subsubsection*{Linear modeling assumptions}
\begin{itemize}
    \item Use ``is\_stereotypical'' as harmfulness score.
\end{itemize}

\subsection*{PRISM}
\subsubsection*{Preprocessing}
\begin{itemize}
    \item Exclude raters with ``Prefer not to say'' or ``Other'' for ethnicity, gender, age group.
\end{itemize}
\subsubsection*{Linear modeling assumptions}
\begin{itemize}
    \item Use 100 - ``safety score'' as harmfulness score.
    \item Use CoB for cultural zone and drop raters with NA for CoB.
    \item As a result, excluded raters with CoB values of 'Afghanistan', 'Cuba', ``C\^{o}te d'Ivoire'', 'Guyana', 'Honduras', 'Jamaica', 'Malawi', 'Prefer not to say', 'Sri Lanka', 'Sudan', 'Tonga'.
\end{itemize}

\subsection*{DICES-990}
\subsubsection*{Preprocessing}
\begin{itemize}
    \item The subset of the data with expert annotation of topic of harm was used.
    \item Use fine-grained self-reported ethnicity categories, after that, exclude raters with NA, ``Other'' and ``Prefer not to answer'' as response to ethnicity.
    \item Assign self-described gender (non M, F) into ``nonbinary'' category.
\end{itemize}
\subsubsection*{Linear modeling assumptions}
\begin{itemize}
    \item Use a subset of the data with expert annotations of degrees of harm (306 items).
    \item Use ``Q\_overall'' as harmfulness score, exclude ``Unsure'' ratings.
\end{itemize}

\subsection*{NLPositionality}
\subsubsection*{Preprocessing}
\begin{itemize}
    \item Exclude raters with NA or ``No response'' as response to ethnicity.
    \item Exclude raters with ``No response'' or NA as response to gender, assign self-described gender (non M, F) into ``nonbinary'' category.
    \item Age is mapped to generational labels (GenZ, Millennial, GenX, Boomer) according to \citet{beresford2026generations}.
    \item Raters with age below GenZ (11) or NA for age are excluded.
\end{itemize}
\subsubsection*{Linear modeling assumptions}
\begin{itemize}
    \item Use ``litw'' as harmfulness score (1 harmful, -1 not harmful), exclude Unsure (0) instances.
    \item Use CoLR for cultural zone and drop raters with NA for CoLR.
    \item As a result, excluded raters with CoB values of 'Dominican Republic', 'Nepal'.
\end{itemize}

\subsection*{D3}
\subsubsection*{Preprocessing}
\begin{itemize}
    \item No filtering was needed.
\end{itemize}
\subsubsection*{Linear modeling assumptions}
\begin{itemize}
    \item Use ``rating\_raw'' as harmfulness score (1 harmful, -1 not harmful), exclude Did Not Understand (-1) instances.
    \item Use CoR for cultural zone and drop raters with NA for CoR.
    \item As a result, excluded raters with CoB values of `United Arab Emirates'.
\end{itemize}

\subsection*{CREHate}
\subsubsection*{Preprocessing}
\begin{itemize}
    \item Exclude raters with NA for age.
    \item Exclude raters with fine-grained ``Other'' ethnicities that have $\leq$ 2 raters, but keep ``Mixed'' ethnicity as a larger category to be consistent with DICES-990.
    \item Age is mapped to generational labels (GenZ, Millennial, GenX, Boomer) according to \citet{beresford2026generations}.
\end{itemize}
\subsubsection*{Linear modeling assumptions}
\begin{itemize}
    \item Use ``annotation'' as harmfulness score (1 harmful, 0 not harmful), exclude Unsure (2) instances.
\end{itemize}

\subsection*{Severity}
\subsubsection*{Preprocessing}
\begin{itemize}
    \item Exclude raters with NA or ``Prefer not to disclose'' for gender.
    \item Exclude raters with NA for age.
    \item Exclude raters with ethnicity containing ``Other'' or that have a unique ethnicity.
    \item Assign self-described gender (non M, F) into ``nonbinary'' category.
    \item Convert ``likert'' scores to numerical values in the following way:
    \begin{itemize}
        \item ``Not at all upsetting'': 0
        \item ``A little upsetting'': 1
        \item ``Somewhat upsetting'': 2
        \item ``Very upsetting'': 3
        \item ``Extremely upsetting'': 4
    \end{itemize}
\end{itemize}
\subsubsection*{Linear modeling assumptions}
\begin{itemize}
    \item Use converted ``likert'' scores as harmfulness scores.
    \item Exclude items with NA for likert score.
\end{itemize}

\subsection{Comparison of demographic and geo-cultural attribute coverage} Table \ref{tab:rater_demographics} shows the coverage of demographic and geo-cultural rater attributes across datasets. 

\begin{table}[htbp]
\centering
\resizebox{\textwidth}{!}{%
\begin{tabular}{lrrrrrrrr}
\toprule
 & \textbf{DICES-990} & \textbf{PRISM} & \textbf{D3} & \textbf{NLPos} & \textbf{DIVE} & \textbf{Severity} & \textbf{CREHate} & \textbf{CultFrames} \\
\midrule
\textbf{Total annotations} & 51,340 & 7,517 & 153,251 & 6,348 & 31,930 & 48,964 & 41,714 & 9,922 \\
\textbf{Unique raters} & 119 & 1,303 & 4,309 & 505 & 636 & 1,416 & 1,039 & 379 \\
\midrule
\multicolumn{9}{l}{\textbf{Gender}} \\
\quad Man & 57 (47.9\%) & 652 (50.0\%) & 2,149 (49.9\%) & 185 (36.6\%) & 323 (50.8\%) & 733 (51.8\%) & 515 (49.6\%) & 189 (49.9\%) \\
\quad Woman & 60 (50.4\%) & 633 (48.6\%) & 2,119 (49.2\%) & 260 (51.5\%) & 313 (49.2\%) & 673 (47.5\%) & 508 (48.9\%) & 190 (50.1\%) \\
\quad Non-binary/Other & 2 (1.7\%) & 18 (1.4\%) & 41 (1.0\%) & 60 (11.9\%) & -- & 10 (0.7\%) & 16 (1.5\%) & -- \\
\midrule
\multicolumn{9}{l}{\textbf{Ethnicity}} \\
\quad White & 27 (22.7\%) & 892 (68.5\%) & -- & 312 (61.8\%) & 128 (20.1\%) & 519 (36.7\%) & 673 (64.8\%) & -- \\
\quad Asian & 53 (44.5\%) & 91 (7.0\%) & -- & 94 (18.6\%) & 254 (39.9\%) & 803 (56.7\%) & 183 (17.6\%) & -- \\
\quad Black/African & 15 (12.6\%) & 117 (9.0\%) & -- & 38 (7.5\%) & 123 (19.3\%) & 57 (4.0\%) & 157 (15.1\%) & -- \\
\quad Latino/Hispanic & 16 (13.4\%) & 117 (9.0\%) & -- & 51 (10.1\%) & 131 (20.6\%) & -- & -- & -- \\
\quad Middle Eastern & -- & 12 (0.9\%) & -- & 2 (0.4\%) & -- & -- & 7 (0.7\%) & -- \\
\quad Indigenous/Pacific Isl. & 7 (5.9\%) & 8 (0.6\%) & -- & 6 (1.2\%) & -- & 30 (2.1\%) & -- & -- \\
\quad Mixed/Other & 1 (0.8\%) & 66 (5.1\%) & -- & 2 (0.4\%) & -- & -- & 19 (1.8\%) & -- \\
\midrule
\multicolumn{9}{l}{\textbf{Age Group}} \\
\quad Gen Z / 18--30 & 18 (15.1\%) & 253 (19.4\%) & 2,019 (46.9\%) & 374 (74.1\%) & 212 (33.3\%) & 150 (10.6\%) & 302 (29.1\%) & 147 (38.8\%) \\
\quad Millennial / 25--44 & 67 (56.3\%) & 604 (46.4\%) & 1,495 (34.7\%) & 85 (16.8\%) & 215 (33.8\%) & 445 (31.4\%) & 466 (44.9\%) & 163 (43.0\%) \\
\quad Gen X / 35--54 & 34 (28.6\%) & 355 (27.2\%) & -- & 34 (6.7\%) & 209 (32.9\%) & 662 (46.8\%) & 200 (19.2\%) & 61 (16.1\%) \\
\quad Boomer / 55+ & -- & 91 (7.0\%) & 795 (18.4\%) & 12 (2.4\%) & -- & 159 (11.2\%) & 71 (6.8\%) & 8 (2.1\%) \\
\midrule
\multicolumn{9}{l}{\textbf{Geographic Attributes}} \\
\quad Country of Birth & -- & 69 countries & -- & -- & 50 countries & -- & -- & 10 countries \\
\quad Country of Residence & 2 countries & 34 countries & 21 countries & 11 countries & 2 countries & 8 countries & -- & 22 countries \\
\quad Country of Longest Res. & -- & -- & -- & 16 countries & -- & -- & -- & -- \\
\quad Country of Nationality & -- & -- & -- & -- & 40 countries & -- & 5 countries & 10 countries \\
\bottomrule
\end{tabular}}
\caption{Demographic and geo-cultural attribute distributions across datasets. Counts (\%).}
\label{tab:rater_demographics}
\end{table}

\clearpage
\section{Multilevel Modeling}

\subsection{Comparison of Country of Birth, Residence, and Nationality as a Basis for Cultural Cluster Assignment} \label{app:cob_vs_cor}
In Table \ref{tab:cob_vs_cor}, we compare models fitted on the PRISM dataset (since it is the only dataset containing both diverse countries of residence and birth), but ablating on how we assign a cultural zone to each annotator (see cultural zone model specification for PRISM in Table \ref{tab:prism_cult}). Instead of using country of birth, we also use country of residence and their combination (e.g.,, English-Speaking country of residence with Catholic Europe country of birth). Cultural zones assigned using CoB were the most parsimonious explanation of safety ratings, as indicated by the lower AIC value for that model compared to all other configurations, including the base model.

\begin{table}[H]
\centering
\begin{tabular}{lrrrrl}
\hline
Model & npar & AIC($\downarrow$) & logLik & $\chi^2$ & $p$ \\
\hline
Null & 9 & 59505.8 & -29743.9 & --- & --- \\
Culture (CoB) & 16 & 59498.5 & -29733.3 & 21.26 & 0.003 \\
Culture (CoR) & 15 & 59501.6 & -29735.8 & 16.12 & 0.013 \\
Culture (CoB$\times$CoR) & 31 & 59517.4 & -29727.7 & 32.40 & 0.071 \\
\hline
\end{tabular}
\caption{Comparison of cultural zone predictiveness when based on country of birth (CoB), country of residence (CoR), and their combination (CoB $\times$ CoR) in PRISM.}
\label{tab:cob_vs_cor}
\end{table}

Similar results were observed in the DIVE dataset (Table \ref{tab:cob_vs_con}), comparing nationality and country of birth as a proxy (see cultural zone model specification for DIVE in Table \ref{tab:dive_cult}).

\begin{table}[H]
\centering
\begin{tabular}{lrrrrl}
\hline
Model & npar & AIC($\downarrow$) & logLik & $\chi^2$ & $p$ \\
\hline
Null & 5 & 84658.1 & -42324.1 & --- & --- \\
Culture (CoB) & 12 & 84650.6 & -42313.30 & 21.52 & 0.003 \\
Culture (CoN) & 13 & 84665.1 & -42319.6 & 8.98 & 0.344 \\
\hline
\end{tabular}
\caption{Comparison of cultural zone predictiveness when based on country of birth (CoB) and country of residence (CoN) in DIVE.}
\label{tab:cob_vs_con}
\end{table}

On the NLPositionality dataset, we were able to compare CoR and CoLR, similarly confirming the intuition that CoLR is a better predictor.

\begin{table}[H]
\centering
\begin{tabular}{lrrrrl}
\hline
Model & npar & AIC($\downarrow$) & logLik & $\chi^2$ & $p$ \\
\hline
Null & 3 & 5273.4 & -2633.7 & --- & --- \\
Culture (CoLR) & 8 & 5277.7 & -2630.9 & 5.6 & 0.3435 \\
Culture (CoR) & 7 & 5278.5 & -2632.3 & 2.86 & 0.581 \\
\hline
\end{tabular}
\caption{Comparison of cultural zone predictiveness when based on country of residence (CoR) and country of longest residence (CoLR) in NLPositionality.}
\label{tab:colr_vs_cor}
\end{table}

\subsection{Cultural Values vs. Cultural Zones vs. Cultural Quadrants as the Predictor Variable} \label{app:cv_vs_cz}

We also perform analysis by using the continuous cultural value axes scores from the WVS instead of the coarser cultural zones. As can be seen in Table \ref{tab:rq3}, cultural zones provide lower $\Delta$AIC values in the majority of the datasets compared to all possible configurations (Traditional-Secular score alone (Trad), Survival-Self-Expression score alone (SurvS), adding them separately (T+S), and adding their interaction (T$\times$S). We also find that value quadrants are similarly predictive as reported in Tables \ref{tab:rq5}, \ref{tab:rq6}, confirming the validity of quadrant-level analysis in Sections \ref{sec:cs_score}, \ref{sec:llm}. In Table \ref{tab:rq4} we also report demographics variable experiments with the best cultural value configuration instead of the cultural zone as the predictive variable for completeness.

\begin{table*}[h]
    \centering
    \small
    \setlength{\tabcolsep}{2pt}
    \resizebox{\textwidth}{!}{%
    \begin{tabular}{lrrrrrrrrrrrrrrr}
        \toprule
        & \multicolumn{3}{c}{\textbf{Cult Zones}} & \multicolumn{3}{c}{\textbf{Values (Trad)}} & \multicolumn{3}{c}{\textbf{Values (SurvS)}} & \multicolumn{3}{c}{\textbf{Values (T+S)}} & \multicolumn{3}{c}{\textbf{Values (T$\times$S)}} \\
        \cmidrule(lr){2-4} \cmidrule(lr){5-7} \cmidrule(lr){8-10} \cmidrule(lr){11-13} \cmidrule(lr){14-16}
        \textbf{Dataset} & \textbf{p} & \textbf{$\Delta$AIC} & \textbf{\%$\Delta\sigma^2$} & \textbf{p} & \textbf{$\Delta$AIC} & \textbf{\%$\Delta\sigma^2$} & \textbf{p} & \textbf{$\Delta$AIC} & \textbf{\%$\Delta\sigma^2$} & \textbf{p} & \textbf{$\Delta$AIC} & \textbf{\%$\Delta\sigma^2$} & \textbf{p} & \textbf{$\Delta$AIC} & \textbf{\%$\Delta\sigma^2$} \\
        \midrule
        DIVE & $0.003^{*}$ & $-7.52$ & $-2.56$ & $<0.001^{*}$ & $-9.75$ & $-1.91$ & $<0.001^{*}$ & $-10.18$ & $-1.98$ & $0.001^{*}$ & $-9.75$ & $-2.08$ & $<0.001^{*}$ & \underline{$-13.98$} & $-2.99$ \\
        CulturalFrames & $<0.001^{*}$ & \underline{$-41.62$} & $-18.43$ & $0.160$ & $0.03$ & $0.10$ & $0.030^{*}$ & $-2.71$ & $-2.97$ & $<0.001^{*}$ & $-20.38$ & $-8.75$ & $<0.001^{*}$ & $-29.88$ & $-13.63$ \\
        PRISM & $0.003^{*}$ & \underline{$-7.26$} & $-1.42$ & $0.654$ & $1.80$ & $0.08$ & $0.738$ & $1.89$ & $0.09$ & $0.903$ & $3.80$ & $0.18$ & $0.698$ & $4.57$ & $0.13$ \\
        DICES-990 & $0.004^{*}$ & \underline{$-6.23$} & $-7.61$ & $0.004^{*}$ & $-6.23$ & $-7.61$ & $0.004^{*}$ & $-6.23$ & $-7.61$ & $0.004^{*}$ & $-6.23$ & $-7.61$ & $0.004^{*}$ & $-6.23$ & $-7.61$ \\
        NLPos & $0.344$ & $4.37$ & $-1.72$ & $0.041^{*}$ & \underline{$-2.19$} & $-1.30$ & $0.689$ & $1.84$ & $-0.08$ & $0.056$ & $-1.78$ & $-1.82$ & $0.115$ & $0.06$ & $-1.92$ \\
        D3 & $<0.001^{*}$ & \underline{$-195.93$} & $-5.22$ & $<0.001^{*}$ & $-86.91$ & $-2.31$ & $<0.001^{*}$ & $-104.25$ & $-2.77$ & $<0.001^{*}$ & $-107.02$ & $-2.87$ & $<0.001^{*}$ & $-119.96$ & $-3.23$ \\
        CREHate & $0.203$ & $0.38$ & $-0.40$ & $0.042^{*}$ & $-2.15$ & $-0.51$ & $0.743$ & $1.89$ & $0.03$ & $0.001^{*}$ & \underline{$-9.13$} & $-2.35$ & $0.004^{*}$ & $-7.14$ & $-2.35$ \\
        Severity & $<0.001^{*}$ & $-41.68$ & $-3.26$ & $<0.001^{*}$ & $-33.74$ & $-2.54$ & $<0.001^{*}$ & $-23.60$ & $-1.79$ & $<0.001^{*}$ & $-37.40$ & $-2.87$ & $<0.001^{*}$ & \underline{$-46.43$} & $-3.59$ \\
        \bottomrule
    \end{tabular}%
    }
    \caption{Comparison of using cultural zones vs. cultural axes values directly as the predictive variable. Highest negative $\Delta$AIC (best fit) for each dataset is underlined. * $p < 0.05$.}
    \label{tab:rq3}
\end{table*}

\begin{table*}[h]
    \centering
    \small
    \setlength{\tabcolsep}{4pt}
    
    \begin{tabular}{llrrrrrr}
        \toprule
        & & \multicolumn{3}{c}{\textbf{D+CV vs. D}} & \multicolumn{3}{c}{\textbf{D $\times$ CV vs. D+CV}} \\
        \cmidrule(lr){3-5} \cmidrule(lr){6-8}
        \textbf{Dataset} & \textbf{Best CV Config} & \textbf{p-value} & \textbf{$\Delta$AIC} & \textbf{\% $\Delta\sigma^2_{\text{rater}}$} & \textbf{p-value} & \textbf{$\Delta$AIC} & \textbf{\% $\Delta\sigma^2_{\text{rater}}$} \\
        \midrule
        DIVE & Trad$\times$SurvS & $0.369$ & $1.19$ & $0.03$ & $0.248$ & $4.93$ & $-0.35$ \\
        CulturalFrames & Trad$\times$SurvS & $<0.001^{*}$ & $-29.23$ & $-13.54$ & $0.007^{*}$ & $-3.18$ & $-9.09$ \\
        PRISM & Trad & $0.828$ & $1.95$ & $0.10$ & $0.324$ & $9.45$ & $-0.15$ \\
        DICES-990 & Trad & $0.005^{*}$ & $-5.86$ & $-6.53$ & $0.059$ & $-1.45$ & $-4.68$ \\
        NLPos & Trad & $0.025^{*}$ & $-3.06$ & $-1.64$ & $0.765$ & $8.66$ & $-1.16$ \\
        D3 & Trad$\times$SurvS & $<0.001^{*}$ & $-112.07$ & $-3.02$ & $<0.001^{*}$ & $-92.62$ & $-2.76$ \\
        CREHate & Trad+SurvS & $0.022^{*}$ & $-3.62$ & $-1.27$ & $0.614$ & $20.30$ & $-2.79$ \\
        Severity & Trad$\times$SurvS & $<0.001^{*}$ & $-41.90$ & $-3.26$ & $0.280$ & $33.40$ & $-0.25$ \\
        \bottomrule
    \end{tabular}
    \caption{\emph{D+CV vs. D}: Tests if adding continuous cultural values as a fixed effect improves fit over demographics alone. \emph{D $\times$ CV vs. D+CV}: Tests if the interaction between cultural values and demographics provides additional explanatory power. Best CV Config indicates the cultural value configuration with the lowest AIC compared to null.}
    \label{tab:rq4}
\end{table*}

\begin{table*}[h]
    \centering
    \small
    \setlength{\tabcolsep}{5pt}

    \begin{tabular}{lrrrrrr}
        \toprule
        & \multicolumn{3}{c}{\textbf{Cultural Zones}} & \multicolumn{3}{c}{\textbf{Value Quadrants}} \\
        \cmidrule(lr){2-4} \cmidrule(lr){5-7}
        \textbf{Dataset} & \textbf{p-value} & \textbf{$\Delta$AIC} & \textbf{\% $\Delta\sigma^2_{\text{rater}}$} & \textbf{p-value} & \textbf{$\Delta$AIC} & \textbf{\% $\Delta\sigma^2_{\text{rater}}$} \\
        \midrule
        DIVE & $0.003^{*}$ & $-7.52$ & $-2.56$ & $<0.001^{*}$ & \underline{$-10.69$} & $-2.42$ \\
        CulturalFrames & $<0.001^{*}$ & \underline{$-41.62$} & $-18.43$ & $0.004^{*}$ & $-7.05$ & $-1.92$ \\
        PRISM & $0.003^{*}$ & \underline{$-7.26$} & $-1.42$ & $0.734$ & $4.72$ & $0.17$ \\
        DICES-990 & $0.004^{*}$ & \underline{$-6.23$} & $-7.61$ & $0.004^{*}$ & $-6.23$ & $-7.61$ \\
        NLPos & $0.344$ & $4.37$ & $-1.72$ & $0.186$ & \underline{$1.19$} & $-1.68$ \\
        D3 & $<0.001^{*}$ & $-195.93$ & $-5.22$ & $<0.001^{*}$ & \underline{$-208.59$} & $-5.52$ \\
        CREHate & $0.203$ & $0.38$ & $-0.40$ & $0.002^{*}$ & \underline{$-8.08$} & $-2.34$ \\
        Severity & $<0.001^{*}$ & \underline{$-41.68$} & $-3.26$ & $<0.001^{*}$ & $-40.53$ & $-3.12$ \\
        \bottomrule
    \end{tabular}
    \caption{Comparison of using cultural zones (categorical clusters) vs. cultural value quadrants (discretised from Trad/SurvS axes) as predictors. Highest negative $\Delta$AIC (best fit) for each dataset is underlined. * $p < 0.05$.}
    \label{tab:rq5}
\end{table*}

\begin{table*}[htbp]
    \centering
    \small
    \setlength{\tabcolsep}{5pt}

    \begin{tabular}{lrrrrrr}
        \toprule
        & \multicolumn{3}{c}{\textbf{D+Q vs. D}} & \multicolumn{3}{c}{\textbf{D $\times$ Q vs. D+Q}} \\
        \cmidrule(lr){2-4} \cmidrule(lr){5-7}
        \textbf{Dataset} & \textbf{p-value} & \textbf{$\Delta$AIC} & \textbf{\% $\Delta\sigma^2_{\text{rater}}$} & \textbf{p-value} & \textbf{$\Delta$AIC} & \textbf{\% $\Delta\sigma^2_{\text{rater}}$} \\
        \midrule
        DIVE & $0.164$ & $0.88$ & $-0.37$ & $0.161$ & $13.02$ & $-0.95$ \\
        CulturalFrames & $0.001^{*}$ & $-10.12$ & $-3.27$ & $0.207$ & $6.70$ & $-5.02$ \\
        PRISM & $0.942$ & $5.61$ & $0.26$ & $0.365$ & $25.09$ & $-0.20$ \\
        DICES-990 & $0.005^{*}$ & $-5.86$ & $-6.53$ & $0.059$ & $-1.45$ & $-4.68$ \\
        NLPos & $0.151$ & $0.69$ & $-2.02$ & $0.186$ & $4.72$ & $-2.68$ \\
        D3 & $<0.001^{*}$ & $-194.06$ & $-5.15$ & $<0.001^{*}$ & $-72.85$ & $-2.25$ \\
        CREHate & $0.012^{*}$ & $-4.83$ & $-1.36$ & $0.975$ & $22.38$ & $-1.13$ \\
        Severity & $<0.001^{*}$ & $-38.54$ & $-2.97$ & $0.511$ & $23.85$ & $0.11$ \\
        \bottomrule
    \end{tabular}
    \caption{\emph{D+Q vs. D}: Tests if adding cultural value quadrants as a fixed effect improves fit over demographics alone. \emph{D $\times$ Q vs. D+Q}: Tests if the interaction between quadrants and demographics provides additional explanatory power. Quadrants are discretised from the Trad and SurvS cultural value axes. * $p < 0.05$.}
    \label{tab:rq6}
\end{table*}

\subsection{Fixed Effect Estimates} \label{app:fixeffs}
\begin{figure*}[h]
    \centering
    % ── DIVE ──
    \begin{subfigure}[t]{0.32\textwidth}
        \includegraphics[width=\linewidth]{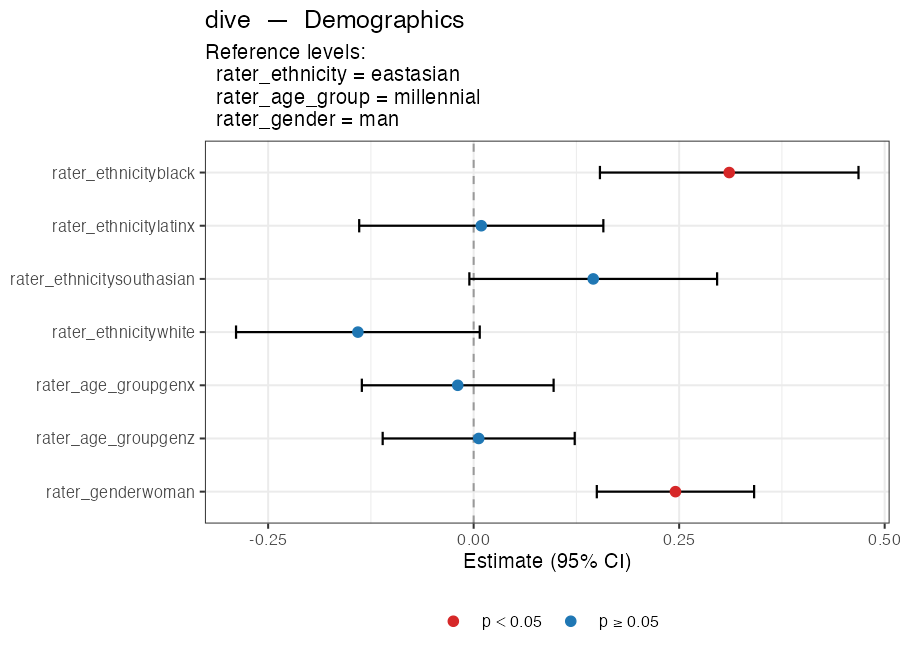}
        \caption{DIVE -- Demographics}
    \end{subfigure}\hfill
    \begin{subfigure}[t]{0.32\textwidth}
        \includegraphics[width=\linewidth]{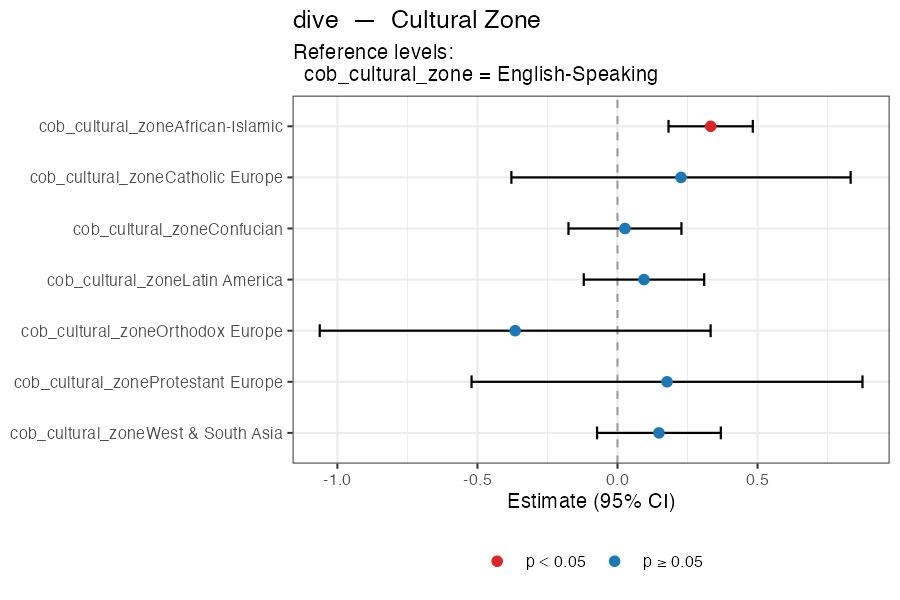}
        \caption{DIVE -- Cultural Zone}
    \end{subfigure}\hfill
    \begin{subfigure}[t]{0.32\textwidth}
        \includegraphics[width=\linewidth]{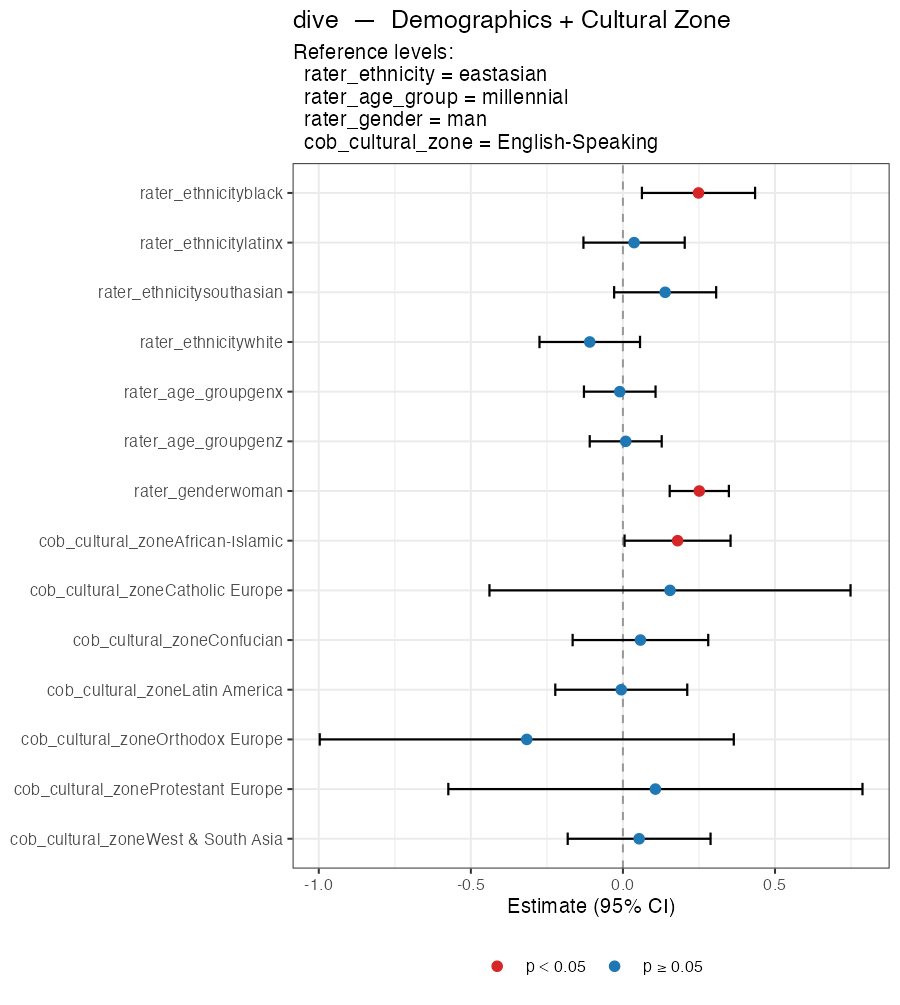}
        \caption{DIVE -- Demographics + Cultural Zone}
    \end{subfigure}
    \\[2ex]
    % ── CulturalFrames ──
    \begin{subfigure}[t]{0.32\textwidth}
        \includegraphics[width=\linewidth]{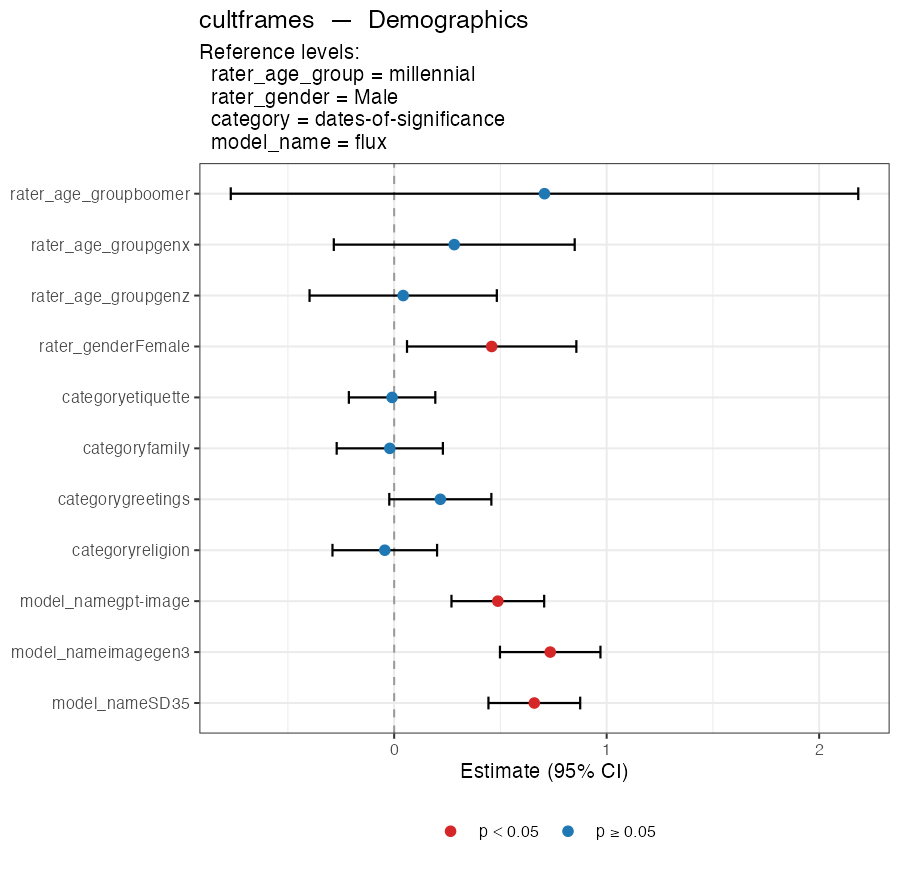}
        \caption{CulturalFrames -- Demographics}
    \end{subfigure}\hfill
    \begin{subfigure}[t]{0.32\textwidth}
        \includegraphics[width=\linewidth]{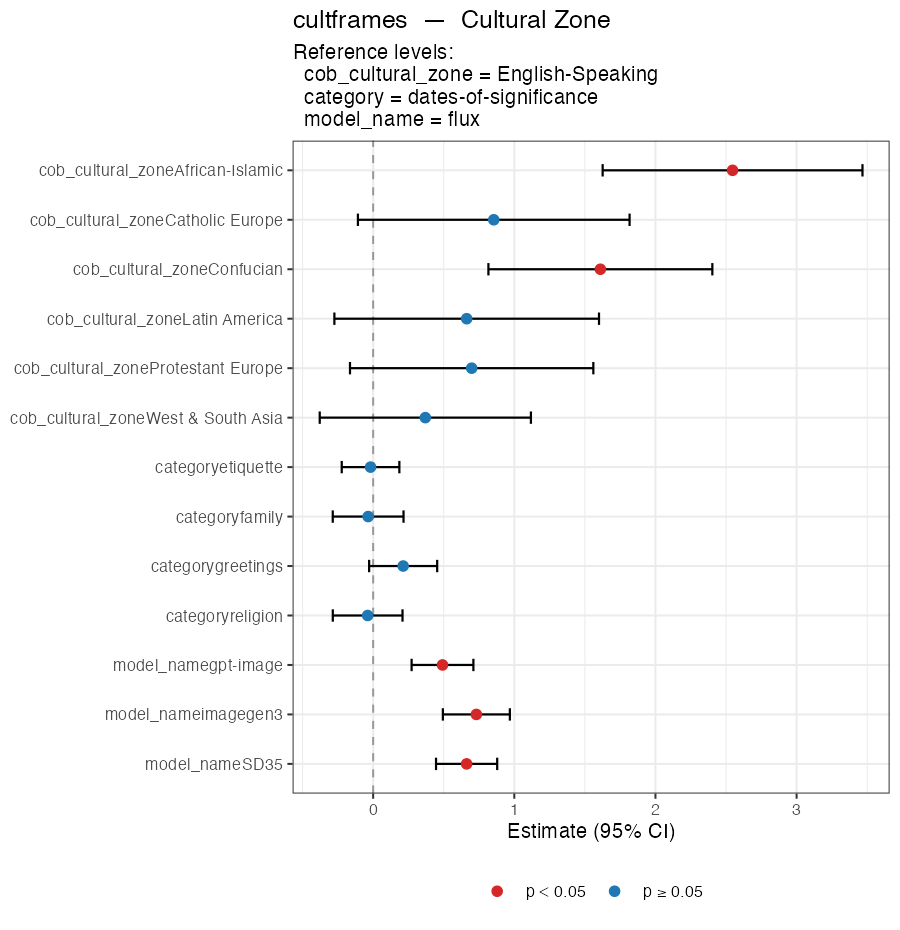}
        \caption{CulturalFrames -- Cultural Zone}
    \end{subfigure}\hfill
    \begin{subfigure}[t]{0.32\textwidth}
        \includegraphics[width=\linewidth]{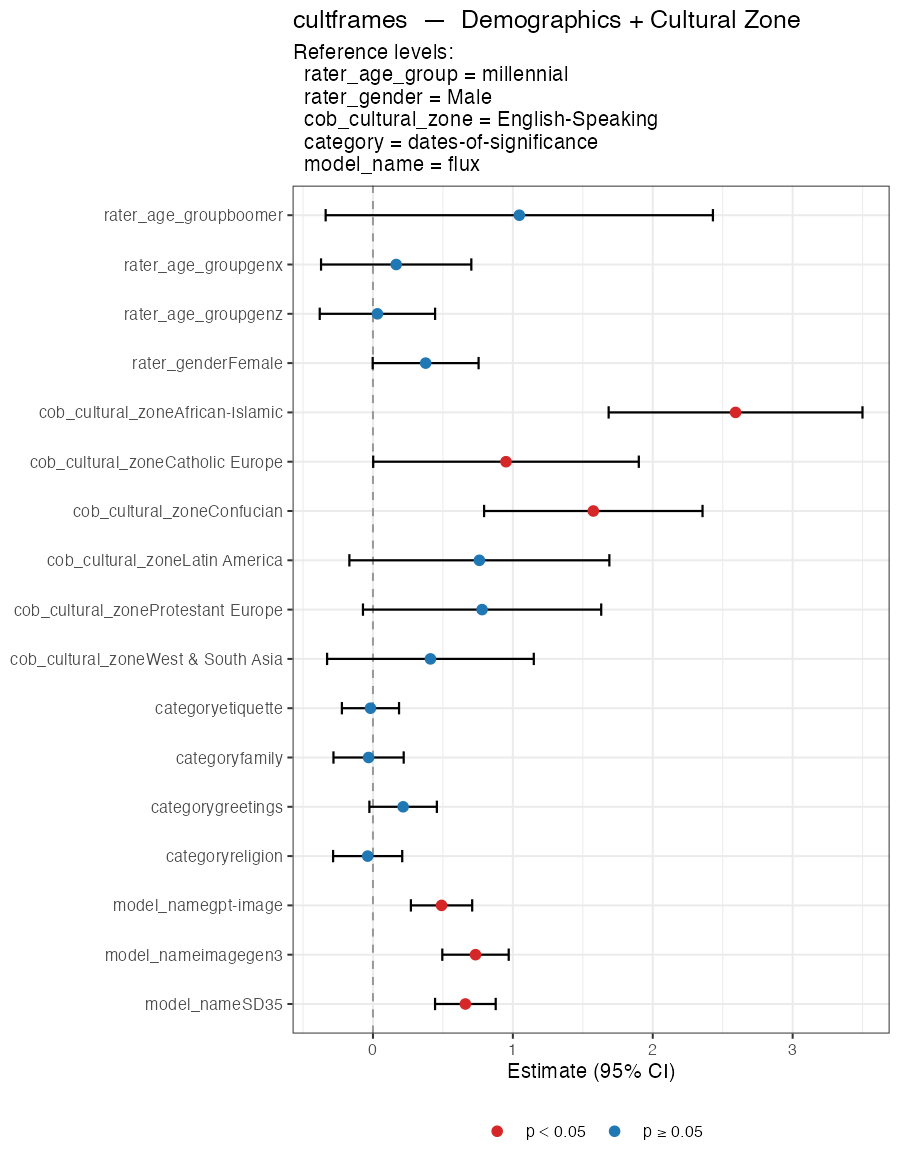}
        \caption{CulturalFrames -- Demographics + Cultural Zone}
    \end{subfigure}
    \\[2ex]
    \caption{Forest plots of fixed-effect coefficients (point estimates with 95\% CI) for the Demographics, Cultural Zone, and Demographics + Cultural Zone models (1/4). Red points: $p < 0.05$; blue points: $p \geq 0.05$. Reference levels are shown in each panel subtitle.}
    \label{fig:forest_plots_1}
\end{figure*}

\begin{figure*}[h]
    \centering
    % ── PRISM ──
    \begin{subfigure}[t]{0.32\textwidth}
        \includegraphics[width=\linewidth]{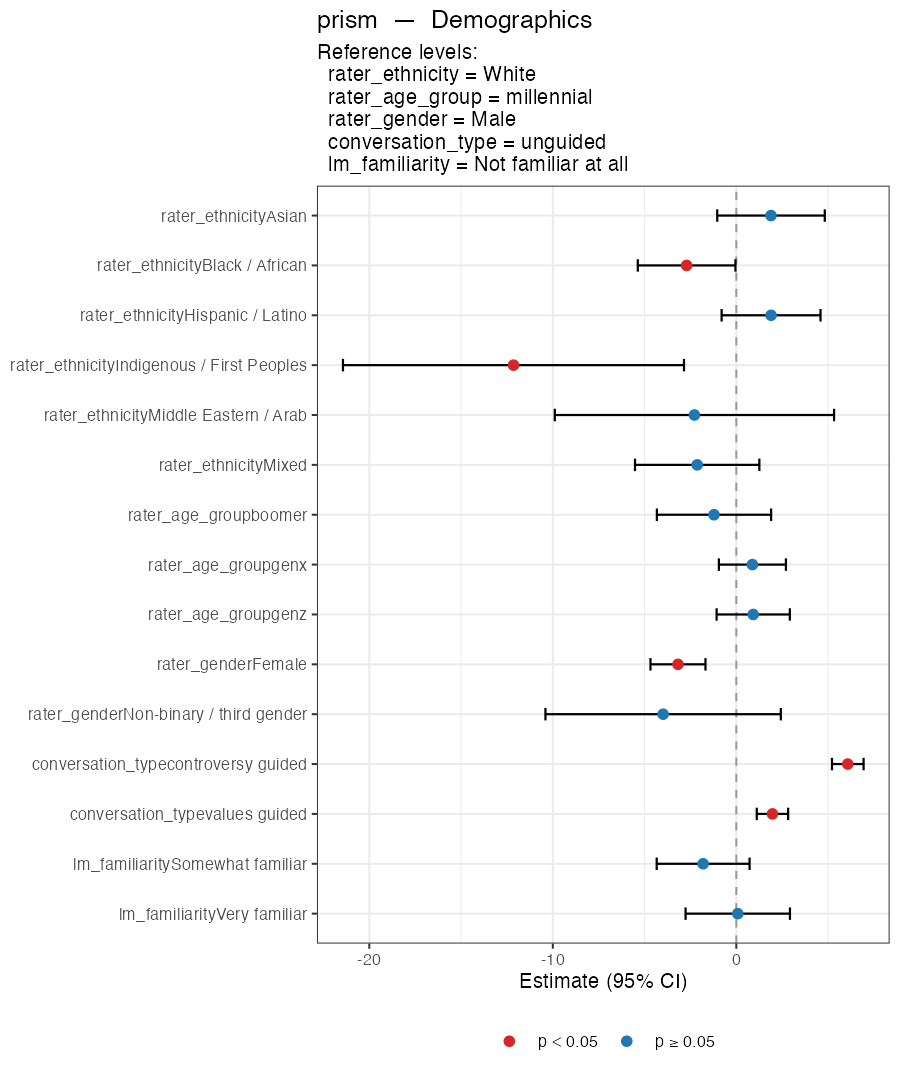}
        \caption{PRISM -- Demographics}
    \end{subfigure}\hfill
    \begin{subfigure}[t]{0.32\textwidth}
        \includegraphics[width=\linewidth]{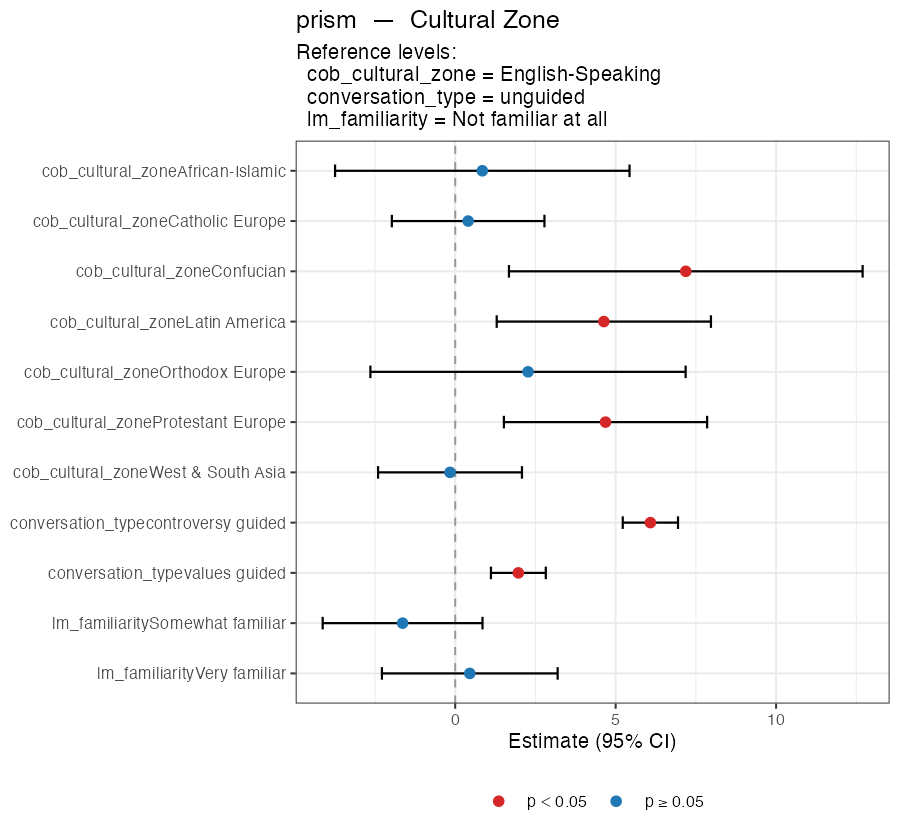}
        \caption{PRISM -- Cultural Zone}
    \end{subfigure}\hfill
    \begin{subfigure}[t]{0.32\textwidth}
        \includegraphics[width=\linewidth]{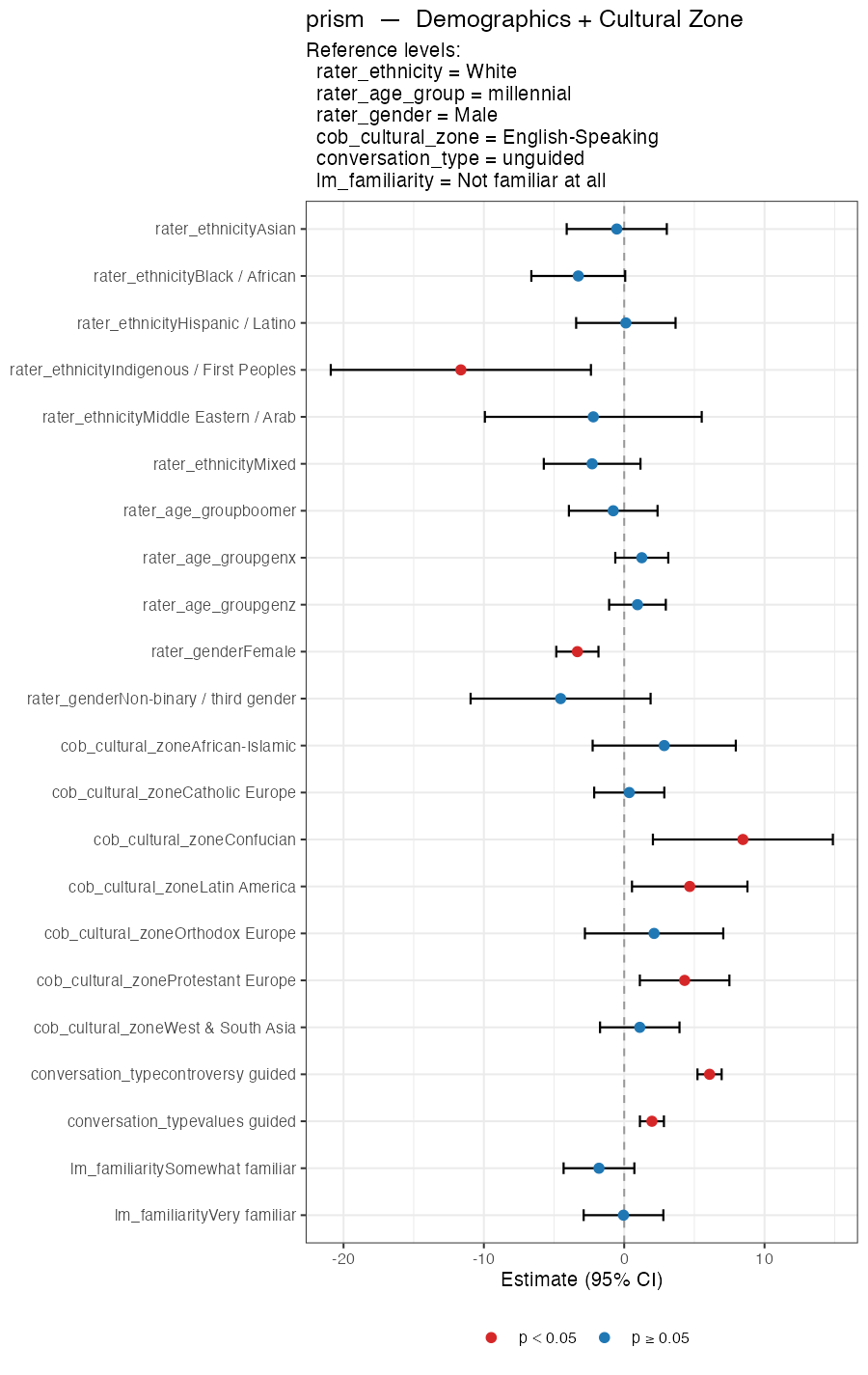}
        \caption{PRISM -- Demographics + Cultural Zone}
    \end{subfigure}
    \\[2ex]
    % ── DICES-990 ──
    \begin{subfigure}[t]{0.32\textwidth}
        \includegraphics[width=\linewidth]{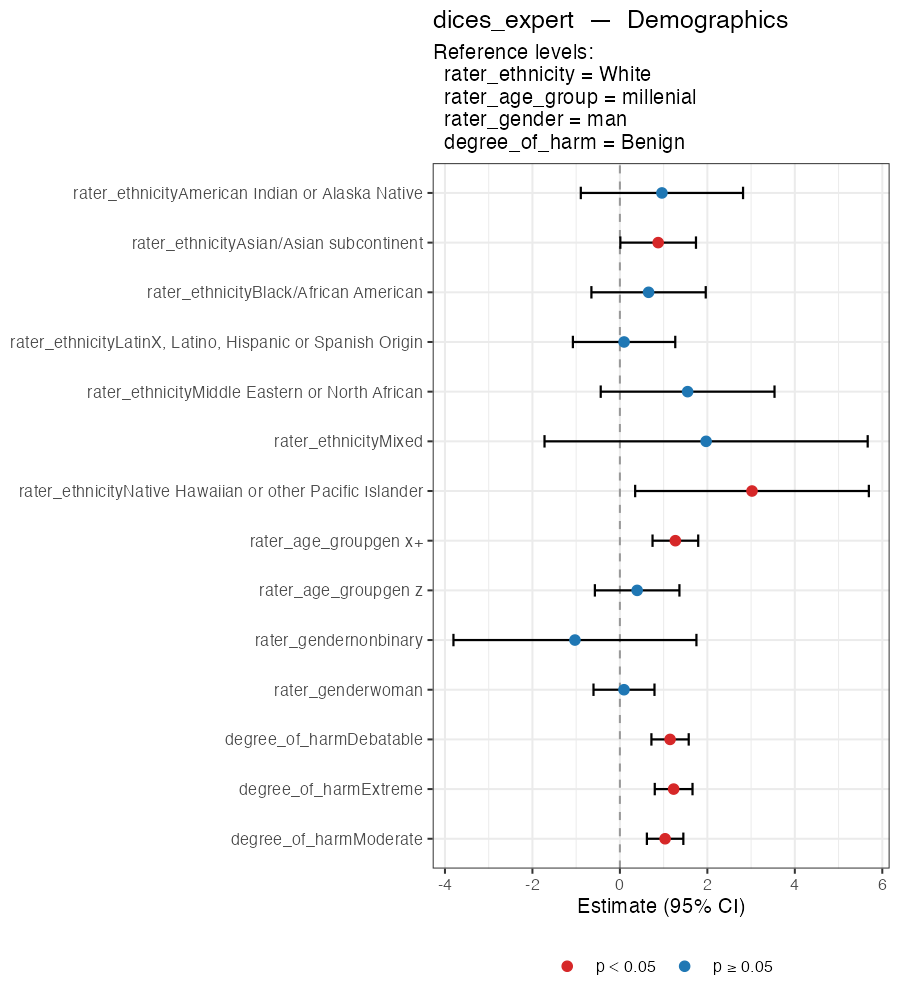}
        \caption{DICES-990 -- Demographics}
    \end{subfigure}\hfill
    \begin{subfigure}[t]{0.32\textwidth}
        \includegraphics[width=\linewidth]{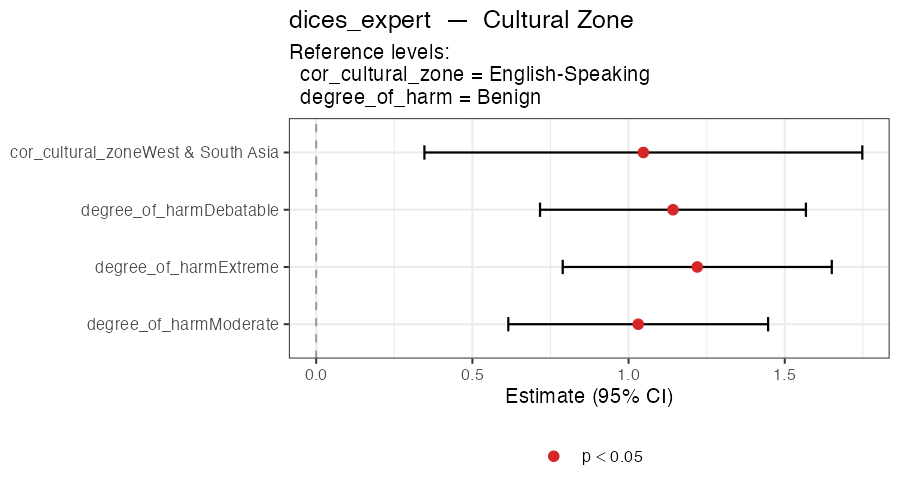}
        \caption{DICES-990 -- Cultural Zone}
    \end{subfigure}\hfill
    \begin{subfigure}[t]{0.32\textwidth}
        \includegraphics[width=\linewidth]{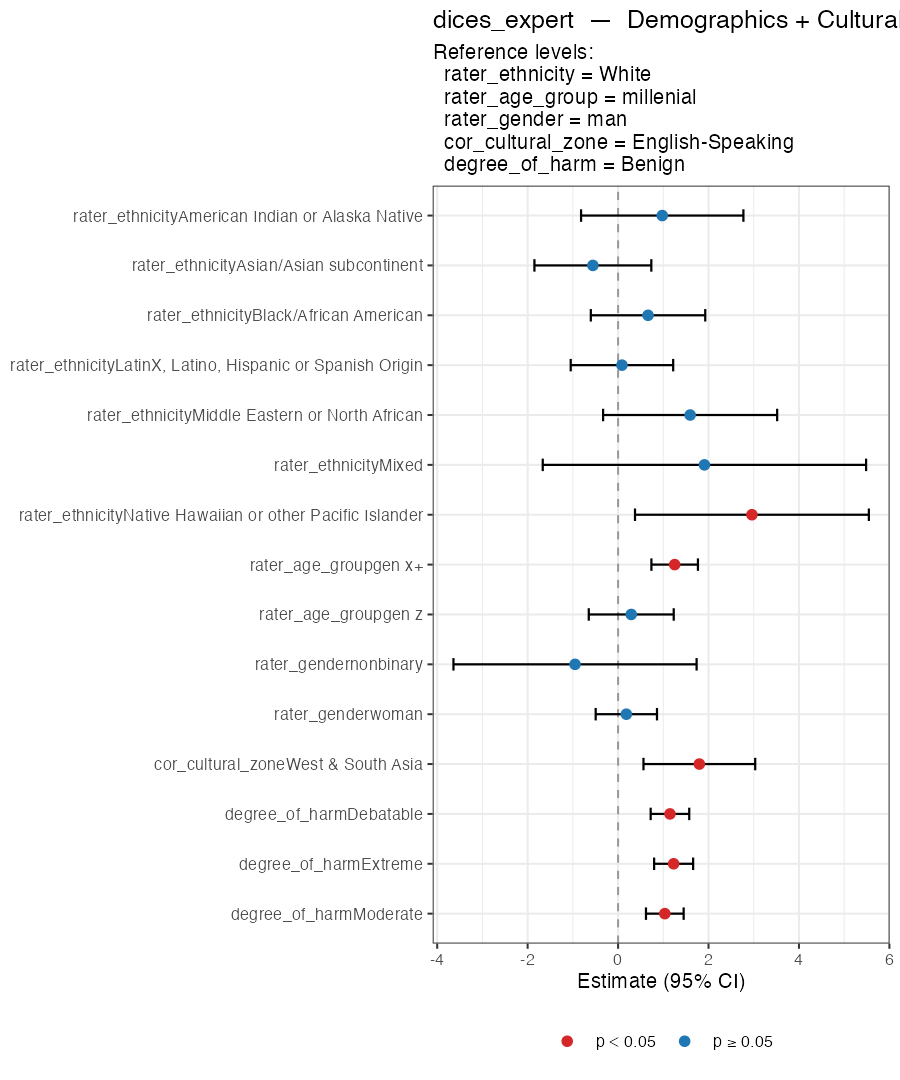}
        \caption{DICES-990 -- Demographics + Cultural Zone}
    \end{subfigure}
    \\[2ex]
    \caption{Forest plots of fixed-effect coefficients (point estimates with 95\% CI) for the Demographics, Cultural Zone, and Demographics + Cultural Zone models (2/4). Red points: $p < 0.05$; blue points: $p \geq 0.05$. Reference levels are shown in each panel subtitle.}
    \label{fig:forest_plots_2}
\end{figure*}

\begin{figure*}[h]
    \centering
    % ── NLPos ──
    \begin{subfigure}[t]{0.32\textwidth}
        \includegraphics[width=\linewidth]{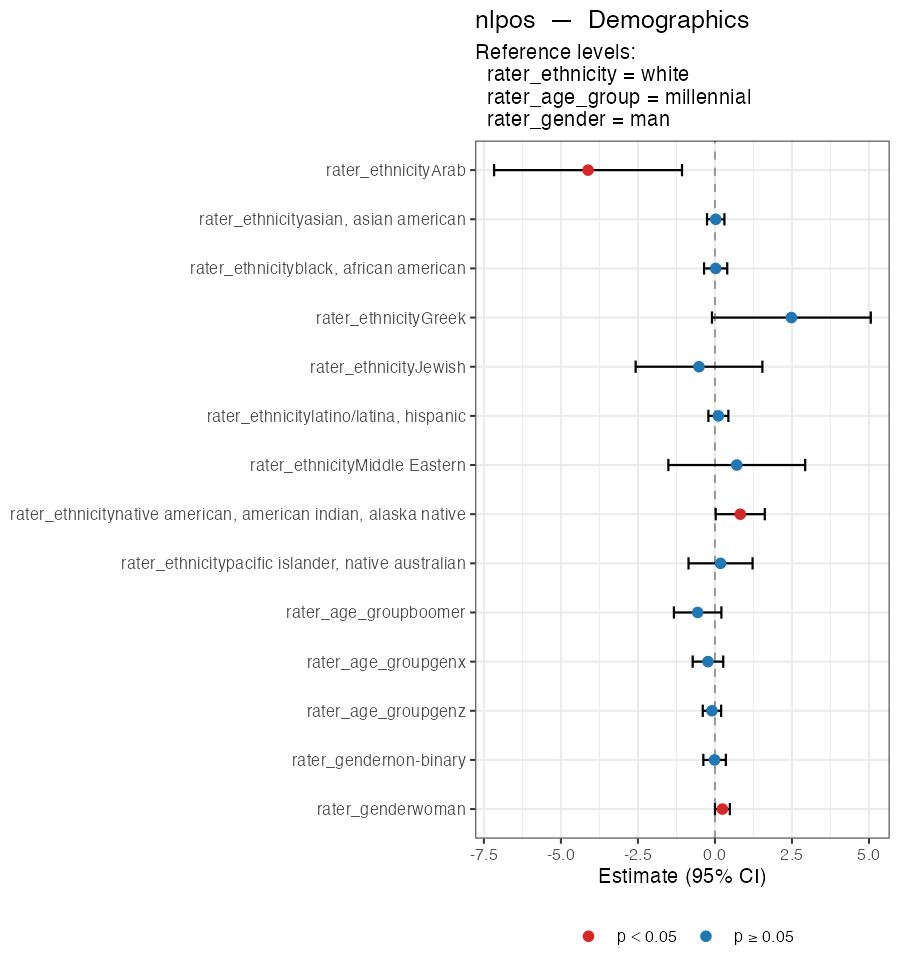}
        \caption{NLPos -- Demographics}
    \end{subfigure}\hfill
    \begin{subfigure}[t]{0.32\textwidth}
        \includegraphics[width=\linewidth]{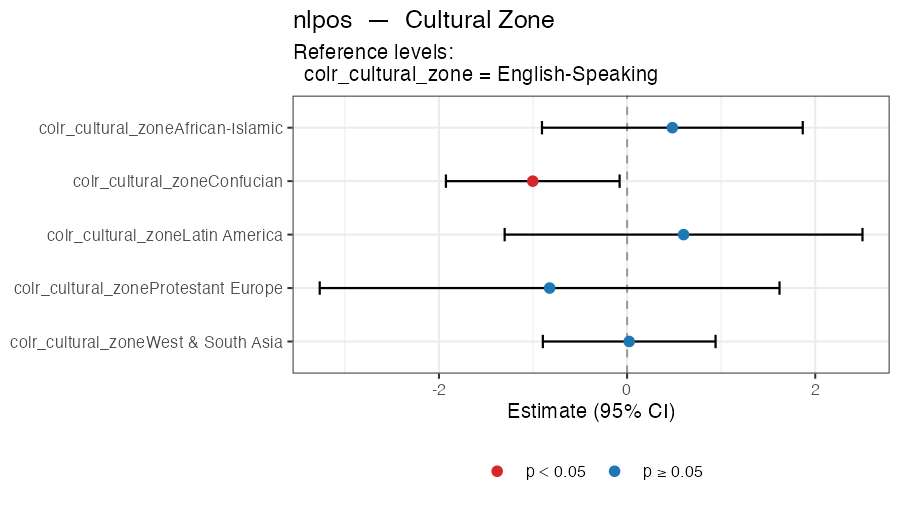}
        \caption{NLPos -- Cultural Zone}
    \end{subfigure}\hfill
    \begin{subfigure}[t]{0.32\textwidth}
        \includegraphics[width=\linewidth]{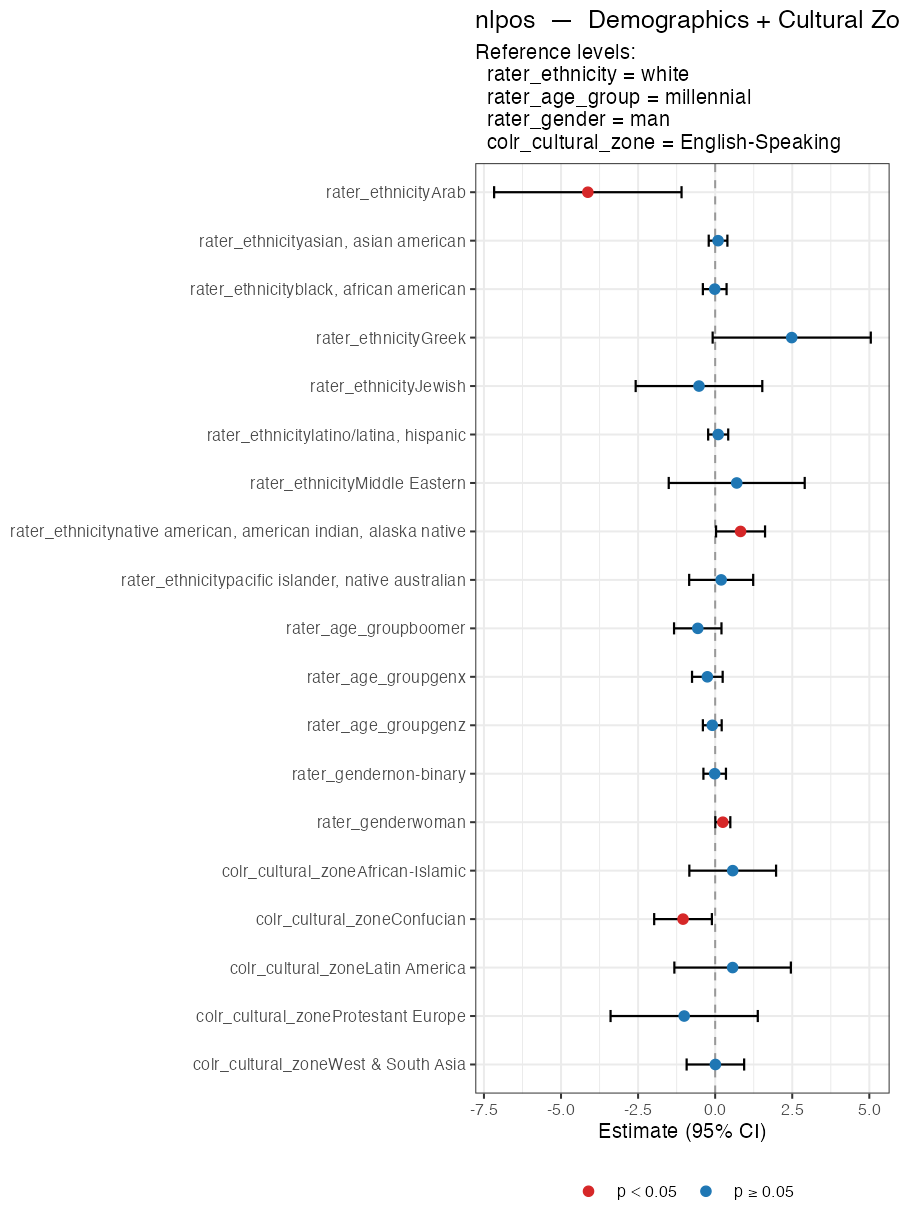}
        \caption{NLPos -- Demographics + Cultural Zone}
    \end{subfigure}
    \\[2ex]
    % ── D3 ──
    \begin{subfigure}[t]{0.32\textwidth}
        \includegraphics[width=\linewidth]{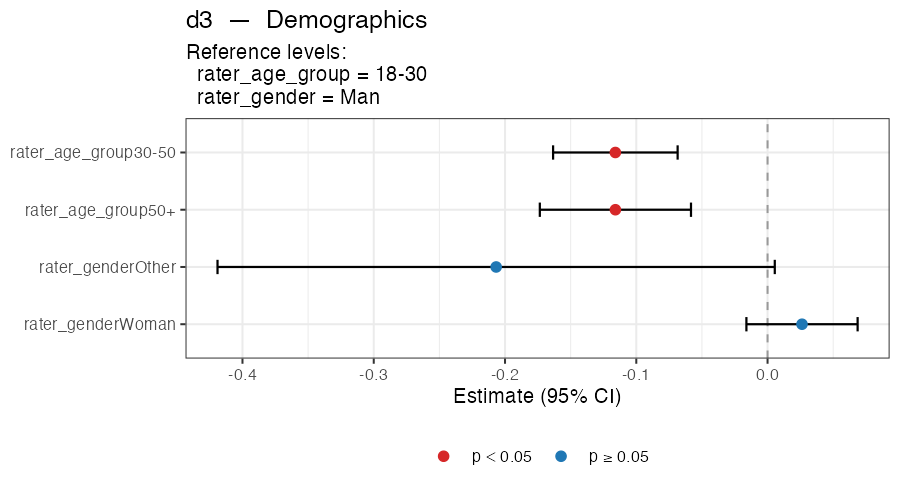}
        \caption{D3 -- Demographics}
    \end{subfigure}\hfill
    \begin{subfigure}[t]{0.32\textwidth}
        \includegraphics[width=\linewidth]{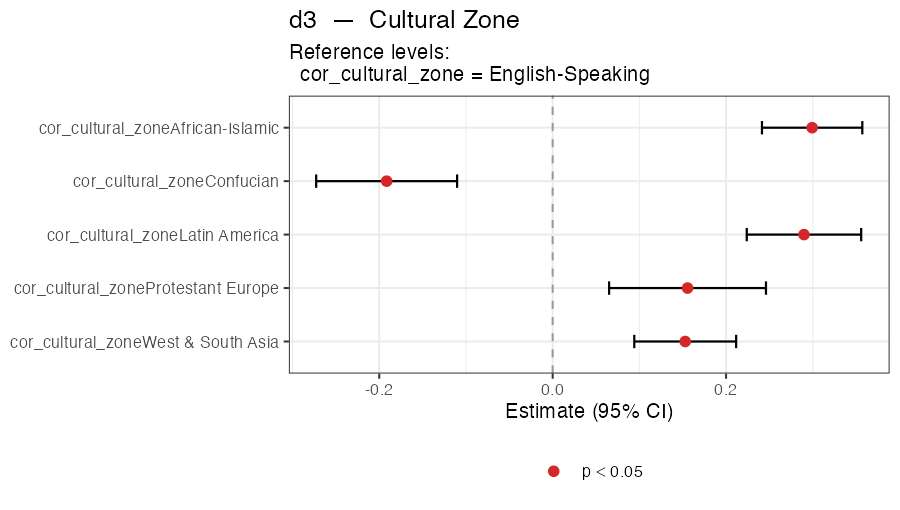}
        \caption{D3 -- Cultural Zone}
    \end{subfigure}\hfill
    \begin{subfigure}[t]{0.32\textwidth}
        \includegraphics[width=\linewidth]{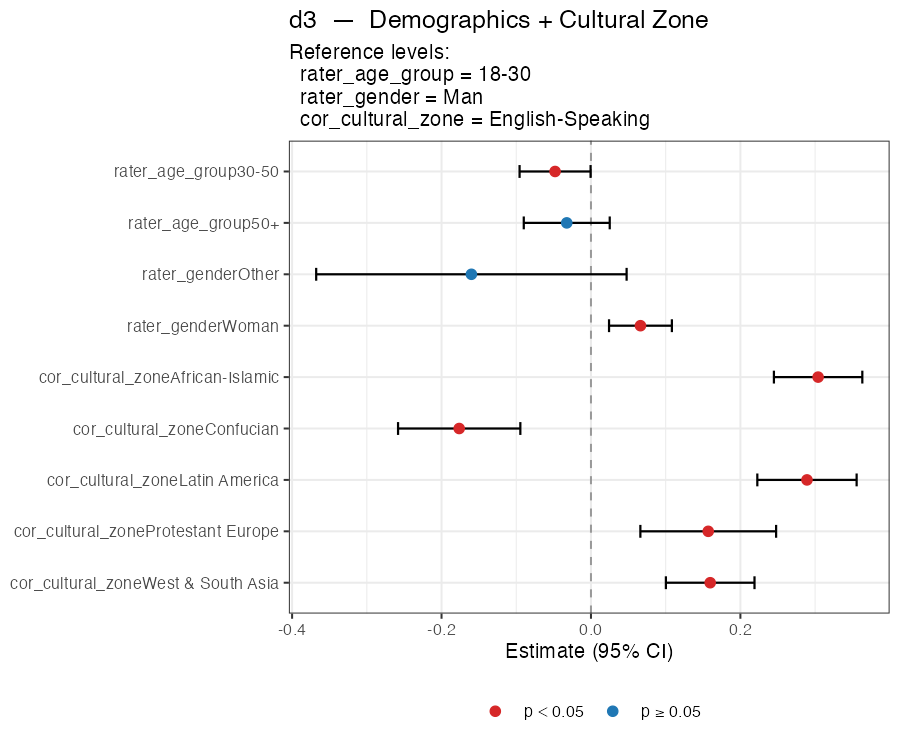}
        \caption{D3 -- Demographics + Cultural Zone}
    \end{subfigure}
    \\[2ex]
    \caption{Forest plots of fixed-effect coefficients (point estimates with 95\% CI) for the Demographics, Cultural Zone, and Demographics + Cultural Zone models (3/4). Red points: $p < 0.05$; blue points: $p \geq 0.05$. Reference levels are shown in each panel subtitle.}
    \label{fig:forest_plots_3}
\end{figure*}

\begin{figure*}[h]
    \centering
    % ── CREHate ──
    \begin{subfigure}[t]{0.32\textwidth}
        \includegraphics[width=\linewidth]{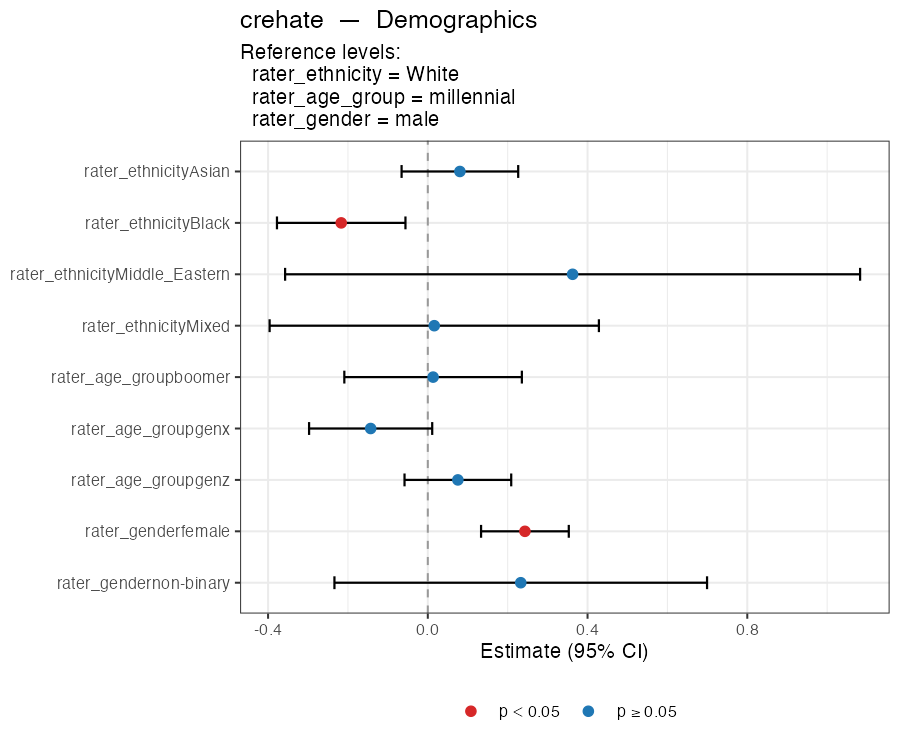}
        \caption{CREHate -- Demographics}
    \end{subfigure}\hfill
    \begin{subfigure}[t]{0.32\textwidth}
        \includegraphics[width=\linewidth]{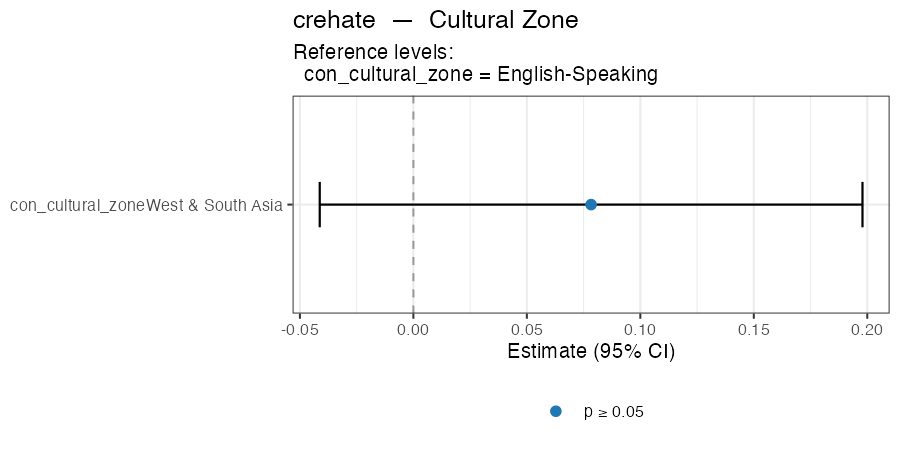}
        \caption{CREHate -- Cultural Zone}
    \end{subfigure}\hfill
    \begin{subfigure}[t]{0.32\textwidth}
        \includegraphics[width=\linewidth]{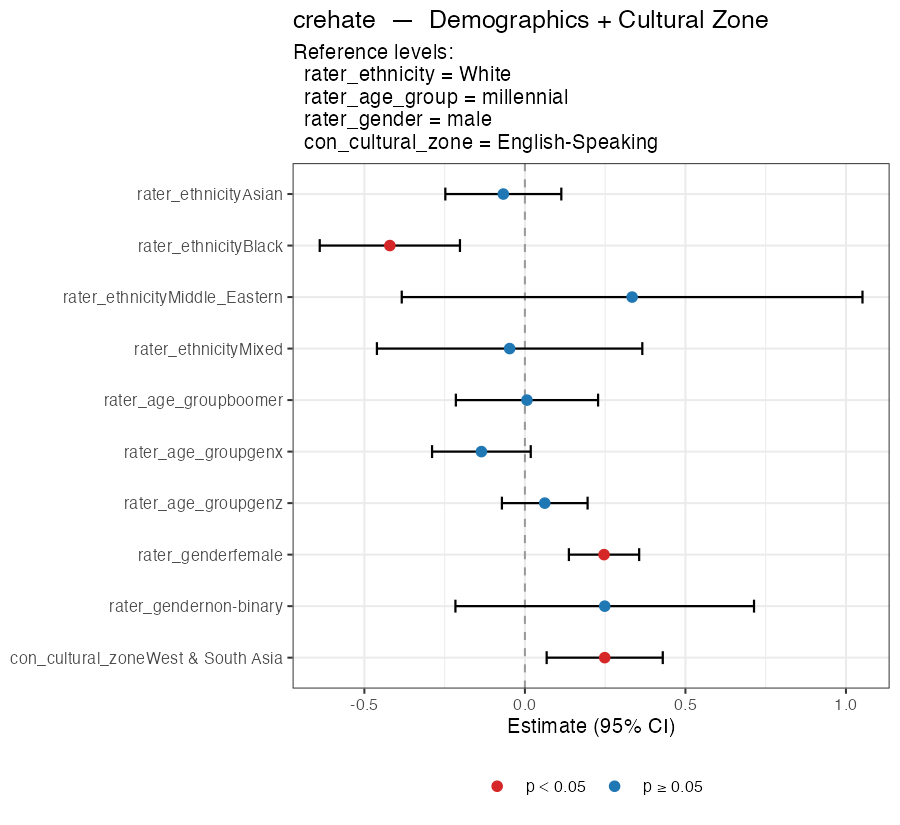}
        \caption{CREHate -- Demographics + Cultural Zone}
    \end{subfigure}
    \\[2ex]
    % ── Severity ──
    \begin{subfigure}[t]{0.32\textwidth}
        \includegraphics[width=\linewidth]{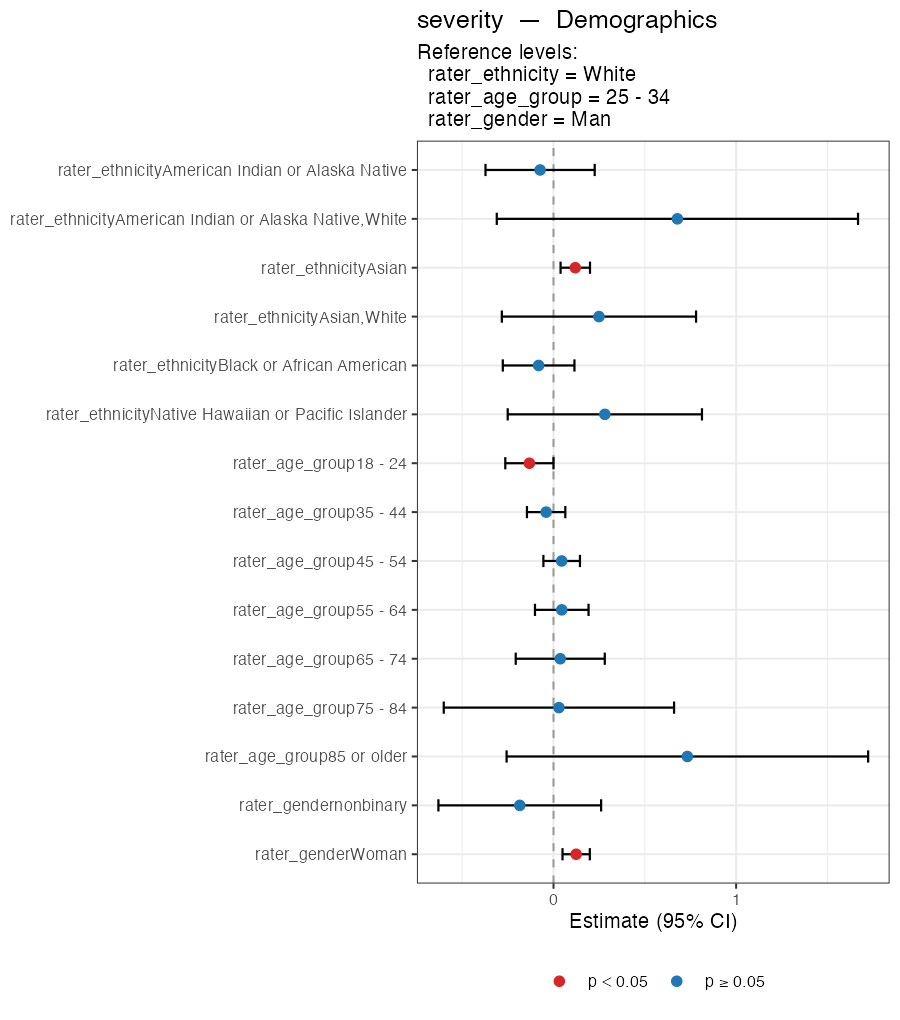}
        \caption{Severity -- Demographics}
    \end{subfigure}\hfill
    \begin{subfigure}[t]{0.32\textwidth}
        \includegraphics[width=\linewidth]{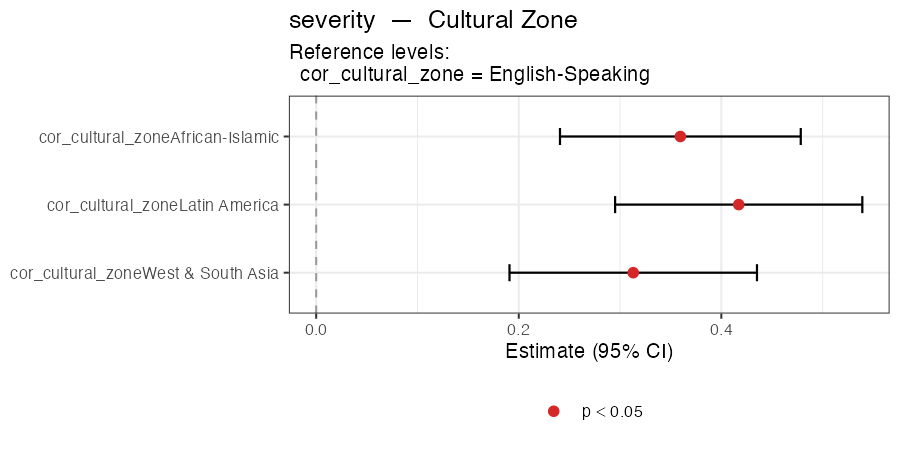}
        \caption{Severity -- Cultural Zone}
    \end{subfigure}\hfill
    \begin{subfigure}[t]{0.32\textwidth}
        \includegraphics[width=\linewidth]{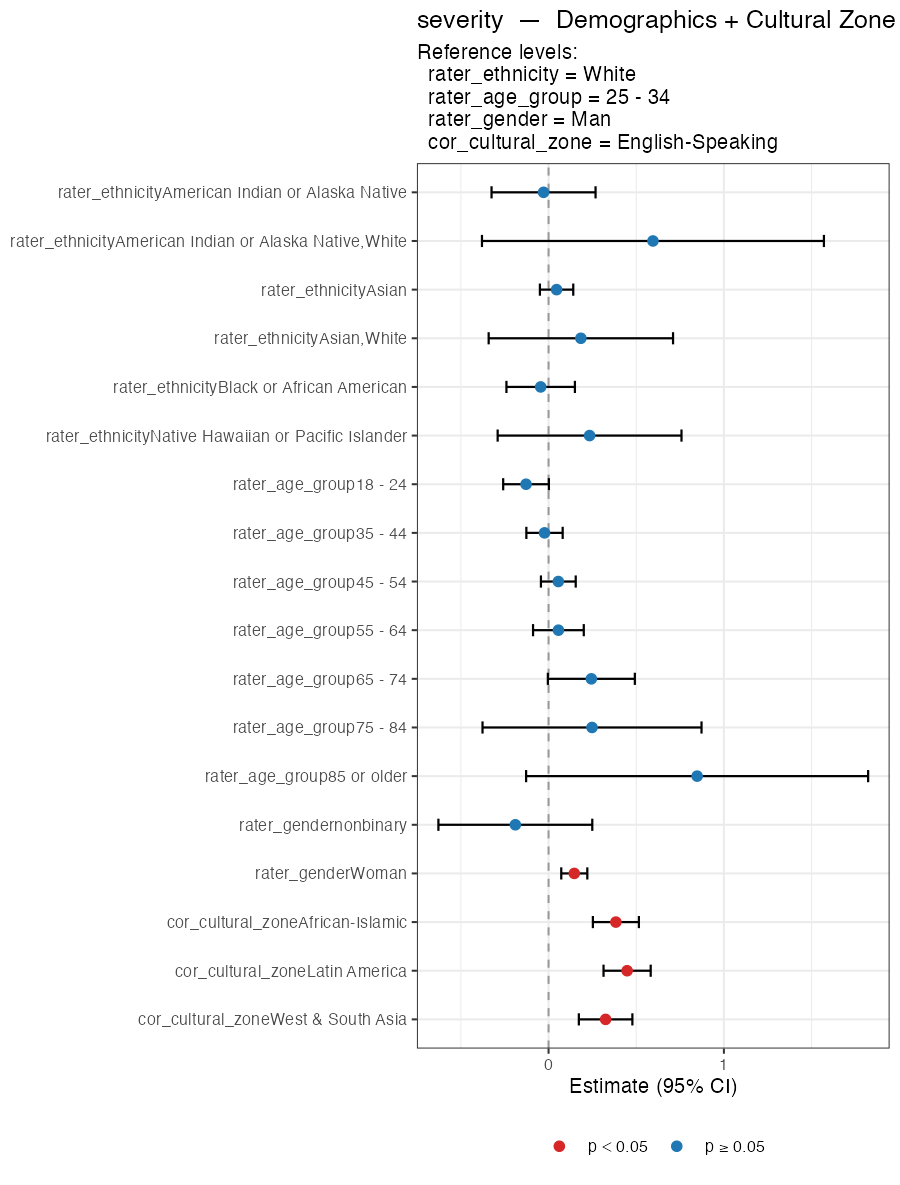}
        \caption{Severity -- Demographics + Cultural Zone}
    \end{subfigure}
    \\[2ex]
    \caption{Forest plots of fixed-effect coefficients (point estimates with 95\% CI) for the Demographics, Cultural Zone, and Demographics + Cultural Zone models (4/4). Red points: $p < 0.05$; blue points: $p \geq 0.05$. Reference levels are shown in each panel subtitle.}
    \label{fig:forest_plots_4}
\end{figure*}

Figures \ref{fig:forest_plots_1}, \ref{fig:forest_plots_2}, \ref{fig:forest_plots_3}, \ref{fig:forest_plots_4} display forest plots with fixed effect estimates from core multilevel models (Demographics, Cultural Zone, D+CZ) across datasets. Full model specifications and estimates available in Appendix \ref{app:lin_models}.

\clearpage

\clearpage

\clearpage

\clearpage
\section{Culturally sensitive items}

\subsection{Algorithm} \label{app:full_algo}

The full algorithm to identify culturally sensitive items is presented below in Algorithm \ref{alg:cultural_consensus}.

\begin{algorithm}[h]
   \caption{Cultural Sensitivity Scoring}
   \small
\begin{algorithmic}
   \STATE {\bfseries Input:} Dataset $\mathcal{D}$, quadrants $Q = \{I, II, III, IV\}$; unsafe rating threshold $\tau_{\text{vote}}$; number of raters threshold $\tau_n = 3$; gender, age, ethnicity homogeneity thresholds $\tau_{g,a,e}$; priors $\alpha = 1, \beta = 1$.
   \STATE {\bfseries Output:} Cultural Sensitivity scores $S_{i,q}$ for every item $i$ and quadrant $q$.
   
   \STATE \textit{// Phase 1: Aggregation \& Validity}
   \FOR{each item $i \in \mathcal{D}$}
       \FOR{each quadrant $q \in Q$}
           \STATE $n_{i,q} \leftarrow \text{Count}(\text{votes for } i \text{ in } q)$
           \STATE $k_{i,q} \leftarrow \text{Count}(\text{unsafe votes ($\leq \tau_{\text{vote}}$) for } i \text{ in } q)$
           
           \STATE \textit{// Calculate max homogeneity per demographic axis}
           \STATE $h^{\text{ethn}}_{i,q} \leftarrow \max(\%\text{Ethnicity groups in } q)$
           \STATE $h^{\text{gend}}_{i,q} \leftarrow \max(\%\text{Gender groups in } q)$
           \STATE $h^{\text{age}}_{i,q} \leftarrow \max(\%\text{Age groups in } q)$
           
           \STATE \textit{// Valid if populated AND diverse across all axes}
           \IF{$n_{i,q} \geq \tau_n$ \textbf{and} $h^{\text{ethn}}_{i,q} < \tau_e$ \textbf{and} $h^{\text{gend}}_{i,q} < \tau_g$ \textbf{and} $h^{\text{age}}_{i,q} < \tau_a$}
               \STATE $v_{i,q} \leftarrow 1$
           \ELSE
               \STATE $v_{i,q} \leftarrow 0$
           \ENDIF
       \ENDFOR
   \ENDFOR

   \STATE \textit{// Phase 2: Obtain the Probability}
   \FOR{each item $i \in \mathcal{D}$}
       \FOR{each quadrant $q \in Q$}
           \STATE $p'_{i,q} \leftarrow 1 - F_{\text{Beta}}(0.5; \alpha + k_{i,q}, \beta + n_{i,q} - k_{i,q})$
           \STATE $p_{i,q} \leftarrow p'_{i,q} \cdot v_{i,q}$ \COMMENT{Zero out if invalid}
       \ENDFOR
   \ENDFOR

   \STATE \textit{// Phase 3: Obtain the Sensitivity Score}
   \FOR{each item $i \in \mathcal{D}$}
       \FOR{each target $q \in Q$}
           \STATE $S_{i,q} \leftarrow p_{i,q}$
           \STATE $C_{opp} \leftarrow 0$
           
           \FOR{each opponent $r \in Q \setminus \{q\}$}
               \STATE \textit{// If opponent $r$ is invalid, $p_{i,r}=0$, so we multiply by 1.0}
               \STATE $S_{i,q} \leftarrow S_{i,q} \cdot (1 - p_{i,r})$
               \STATE $C_{opp} \leftarrow C_{opp} + v_{i,r}$
           \ENDFOR
           
           \STATE \textit{// Require at least one valid opponent}
           \IF{$C_{opp} == 0$}
               \STATE $S_{i,q} \leftarrow 0$
           \ENDIF
       \ENDFOR
   \ENDFOR
   
   \STATE \textbf{return} $S$
\end{algorithmic}
\label{alg:cultural_consensus}
\end{algorithm}
\clearpage

% for ICML:
% \subsection{Examples (CW: these examples may be offensive)} \label{app:cs_qual_ex}
% Tables \ref{tab:D3-qualitative-examples} (D3), \ref{tab:CREHate-qualitative-examples} (CREHate), \ref{tab:DICES-990-qualitative-examples} (DICES-990), \ref{tab:NLPositionality-qualitative-examples} (NLPositionality), \ref{tab:Severity-qualitative-examples} (Severity), \ref{tab:DIVE-qualitative-examples} (DIVE) display top $5$ items by probability of being culturally sensitive ($S_{iq}$).
% Tables \ref{tab:D3 (all quadrants valid)-qualitative-examples} (D3-all), \ref{tab:CREHate (all quadrants valid)-qualitative-examples} (CREHate-all) display top items when all quadrant votes present in the dataset are valid. The alignment with cultural quadrant values is not always obvious, showing potential for future qualitative research on quadrant-level disagreements and motivating the need for more representative quadrant data collection. Note that the examples were not cherry-picked, but selected using the $S_{iq}$ probability. See more examples at \href{https://asaakyan.github.io/culture-safety/}{\texttt{asaakyan.github.io/culture-safety}}.
% \input{examples}

% for arxiv:
\subsection{Examples} \label{app:cs_qual_ex}
Examples can be found at \href{https://asaakyan.github.io/culture-safety/}{\texttt{asaakyan.github.io/culture-safety}}. 
% Tables \ref{tab:D3-qualitative-examples} (D3), \ref{tab:CREHate-qualitative-examples} (CREHate), \ref{tab:DICES-990-qualitative-examples} (DICES-990), \ref{tab:NLPositionality-qualitative-examples} (NLPositionality), \ref{tab:Severity-qualitative-examples} (Severity), \ref{tab:DIVE-qualitative-examples} (DIVE) display top $5$ items by probability of being culturally sensitive ($S_{iq}$).
% Tables \ref{tab:D3 (all quadrants valid)-qualitative-examples} (D3-all), \ref{tab:CREHate (all quadrants valid)-qualitative-examples} (CREHate-all) display top items when all quadrant votes present in the dataset are valid. 
The alignment with cultural quadrant values is not always obvious, showing potential for future qualitative research on quadrant-level disagreements and motivating the need for more representative quadrant data collection.

% \clearpage
\subsection{Threshold Sensitivity Analysis} \label{app:thresh_sens}

We conduct a threshold sensitivity analysis to see how the rate of culturally sensitive items would change if we adjust $S_{iq}$ (the joint posterior probability that, among valid quadrants, only quadrant $q$ rated the item as unsafe) and $\tau_{\text{majority}}$ (the threshold on $\theta_{iq}$ used in $H_{iq} = P(\theta_{iq} > \tau_{\text{majority}})$). 

\begin{figure}[h!]
    \centering
    \includegraphics[width=0.7\linewidth]{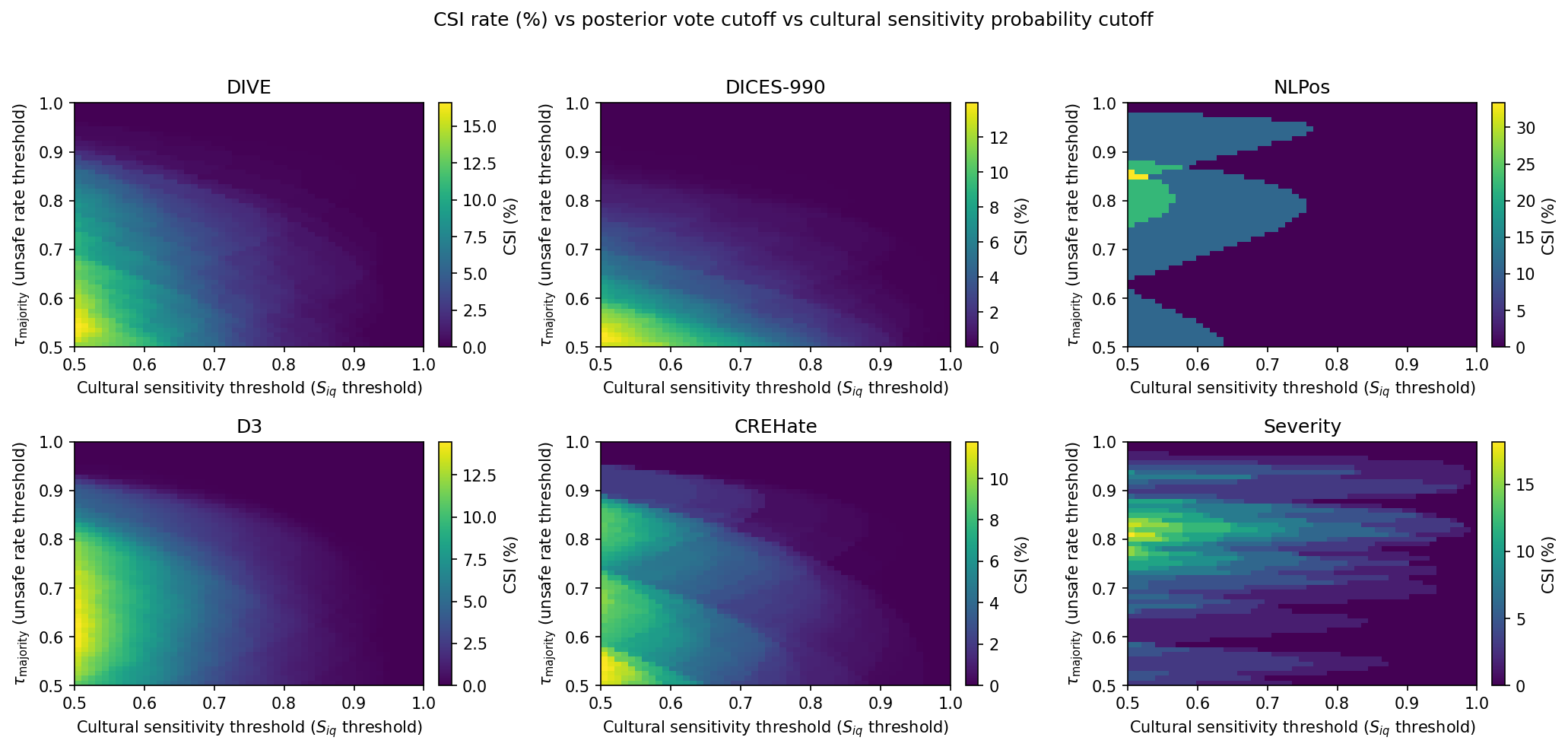}
    \caption{Sensitivity of culturally sensitive item rate to two thresholds: $S_{iq}$ (the joint posterior probability that, among valid quadrants, only quadrant $q$ rated the item as unsafe) and $\tau_{\text{majority}}$ (the threshold on $\theta_{iq}$ used in $H_{iq} = P(\theta_{iq} > \tau_{\text{majority}})$).}
    \label{fig:cs_thresh_sens}
\end{figure}

Fixing $\tau_{\text{majority}} = 0.5$ (the criterion used in the main text), Table \ref{tab:csi-aggregate} shows how the average Culturally Sensitive Item rate would differ as we increase the threshold for $S_{iq}$. Though our main results rely on the ``more likely than not'' standard, even when the threshold is increased to 0.7, most datasets display a non-trivial rate of culturally sensitive items (see the breakdown in Table \ref{tab:csi-per-dataset}). Heatmaps showing sensitivity to both thresholds can be found in Figure \ref{fig:cs_thresh_sens}.

% ================================================================
% Table 2 — Aggregate CSI (%) mean ± std across datasets
% ================================================================
\begin{table}[h]
\centering
\small
\caption{%
  CSI (\%) aggregated across all six datasets as a function of the $S_{iq}$
  threshold ($\tau_{\text{majority}} = 0.5$).%
}
\label{tab:csi-aggregate}
\begin{tabular}{S[table-format=1.2] S[table-format=2.2] @{\,\(\pm\)\,} S[table-format=1.2]}
  \toprule
  {$S_{iq}$ threshold} & \multicolumn{2}{c}{CSI (\%)\quad Mean $\pm$ Std} \\
  \midrule
  0.5  & 10.53 & 3.87 \\
  0.6  & 8.02 & 3.72 \\
  0.7  & 3.43 & 2.74 \\
  0.8  & 1.39 & 1.46 \\
  0.9  & 0.41 & 0.49 \\
  0.95 & 0.09 & 0.13 \\
  0.99 & 0.00 & 0.00 \\
  \bottomrule
\end{tabular}
\end{table}

% ================================================================
% Table 1 — CSI (%) per dataset across S_iq thresholds
% ================================================================
\begin{table}[h]
\centering
\small
\caption{%
  Culturally Sensitive Item rate (CSI, \%) per dataset as a function of the cultural sensitivity threshold $S_{iq}$, with
  $\tau_{\text{majority}} = 0.5$ as the threshold for considering a quadrant's judgment to be unsafe.
}
\label{tab:csi-per-dataset}
\sisetup{round-mode=places, round-precision=2}
\begin{tabular}{l S[table-format=2.2] S[table-format=2.2] S[table-format=2.2] S[table-format=2.2] S[table-format=2.2] S[table-format=2.2] S[table-format=2.2]}
  \toprule
  & \multicolumn{7}{c}{$S_{iq}$ threshold} \\
  \cmidrule(lr){2-8}
  Dataset & {0.5} & {0.6} & {0.7} & {0.8} & {0.9} & {0.95} & {0.99} \\
  \midrule
  DIVE         & 13.87 & 10.82 & 5.64 & 1.80 & 0.45 & 0.00 & 0.00 \\
  DICES-990    & 13.13 & 10.71 & 7.47 & 3.94 & 1.31 & 0.30 & 0.00 \\
  NLPos        & 11.11 & 11.11 & 0.00 & 0.00 & 0.00 & 0.00 & 0.00 \\
  D3           & 10.89 & 7.23 & 3.37 & 1.48 & 0.43 & 0.07 & 0.00 \\
  CREHate      & 11.14 & 6.72 & 2.56 & 1.09 & 0.26 & 0.19 & 0.00 \\
  Severity     & 3.03 & 1.52 & 1.52 & 0.00 & 0.00 & 0.00 & 0.00 \\
  \midrule
  \textit{Mean} & {10.53} & {8.02} & {3.43} & {1.39} & {0.41} & {0.09} & {0.00} \\
  \bottomrule
\end{tabular}
\end{table}

\clearpage
\section{Classifier Experiments}

\subsection{Quadrant-level prediction} \label{app:q-level-dataset}

\begin{figure}[h]
    \centering    \includegraphics[width=0.99\linewidth]{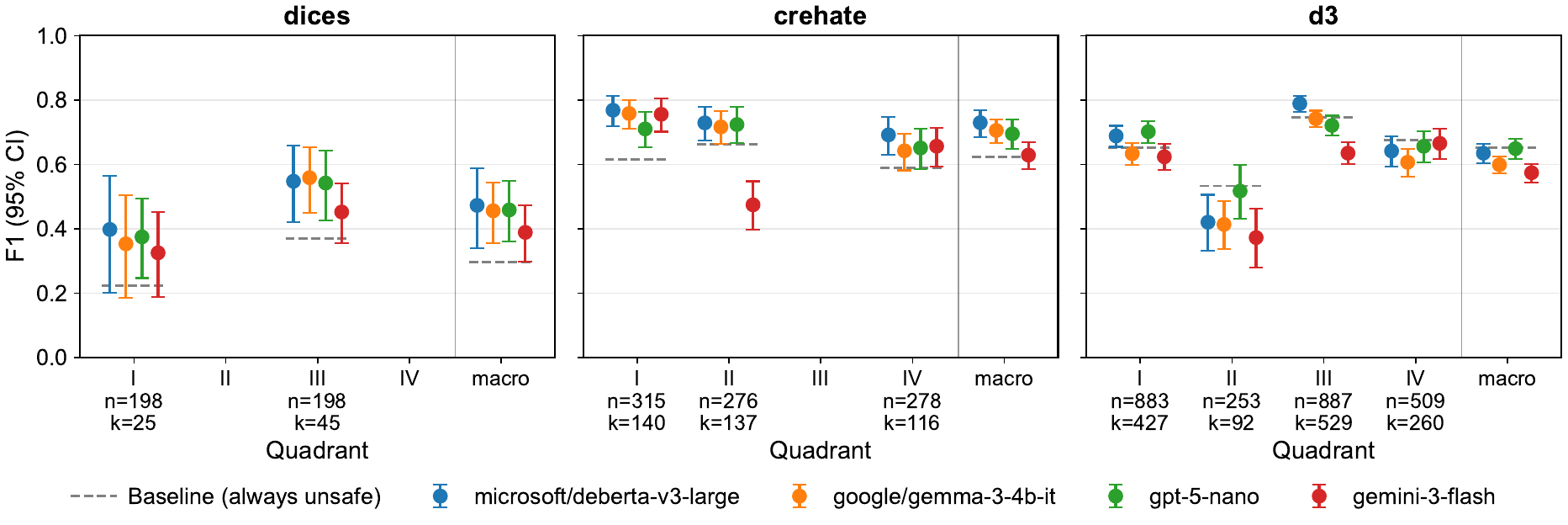}
    \caption{Model performance by quadrant and dataset. 95\% CIs obtained via hierarchical bootstrap on random seed and item level.}
    \label{fig:q-level-detail}
\end{figure}

As described in Section \ref{sec:cs_score}, each item can be associated with its estimated probability of each of the four quadrants rating it as harmful ($H_{iq})$. We assemble a dataset where each item is associated with a binarized quadrant-level safety rating (e.g., $1$ if Quadrant I is more likely to rate it as unsafe i.e. $H_{iI} > 0.5$, 0 for Quadrant II if $H_{iII} < 0.5$, etc.). We combine the data from D3, CREHate, and DICES-990 and split it into training, validation, and testing sets. Crucially, no item is present in multiple sets. Table \ref{tab:q-level-train} shows the statistics for the train set, Table \ref{tab:q-level-val} for the validation set, and Table \ref{tab:q-level-test} for the test set. Figure \ref{fig:q-level-detail} shows the detailed breakdown of model performance among quadrants by dataset.

\begin{table}[h!]
\centering
\small
\begin{tabular}{lrrrrrrrrrrrr}
\toprule
 & \multicolumn{3}{c}{Label I} & \multicolumn{3}{c}{Label II} & \multicolumn{3}{c}{Label III} & \multicolumn{3}{c}{Label IV} \\
 & 0 & 1 & \% unsafe & 0 & 1 & \% unsafe & 0 & 1 & \% unsafe & 0 & 1 & \% unsafe \\
Dataset &  &  &  &  &  &  &  &  &  &  &  &  \\
\midrule
crehate & 562 & 449 & 44.4 & 449 & 440 & 49.5 & -- & -- & -- & 515 & 375 & 42.1 \\
d3 & 1457 & 1360 & 48.3 & 509 & 298 & 36.9 & 1139 & 1694 & 59.8 & 794 & 835 & 51.3 \\
dices & 552 & 82 & 12.9 & -- & -- & -- & 491 & 143 & 22.6 & -- & -- & -- \\
\bottomrule
\end{tabular}
\caption{Train set 0/1 label distribution on the test set across datasets and quadrants. \textit{-- = no valid annotations for that quadrant in that dataset.}}
\label{tab:q-level-train}
\end{table}

\begin{table}[h!]
\centering
\small
\begin{tabular}{lrrrrrrrrrrrr}
\toprule
 & \multicolumn{3}{c}{Label I} & \multicolumn{3}{c}{Label II} & \multicolumn{3}{c}{Label III} & \multicolumn{3}{c}{Label IV} \\
 & 0 & 1 & \% unsafe & 0 & 1 & \% unsafe & 0 & 1 & \% unsafe & 0 & 1 & \% unsafe \\
Dataset &  &  &  &  &  &  &  &  &  &  &  &  \\
\midrule
crehate & 138 & 114 & 45.2 & 112 & 110 & 49.5 & -- & -- & -- & 128 & 93 & 42.1 \\
d3 & 364 & 342 & 48.4 & 128 & 77 & 37.6 & 287 & 423 & 59.6 & 202 & 207 & 50.6 \\
dices & 138 & 20 & 12.7 & -- & -- & -- & 123 & 35 & 22.2 & -- & -- & -- \\
\bottomrule
\end{tabular}
\caption{Validation set 0/1 label distribution on the test set across datasets and quadrants. \textit{-- = no valid annotations for that quadrant in that dataset.}}
\label{tab:q-level-val}
\end{table}

\begin{table}[h!]
\centering
\small
\begin{tabular}{lrrrrrrrrrrrr}
\toprule
 & \multicolumn{3}{c}{Label I} & \multicolumn{3}{c}{Label II} & \multicolumn{3}{c}{Label III} & \multicolumn{3}{c}{Label IV} \\
 & 0 & 1 & \% unsafe & 0 & 1 & \% unsafe & 0 & 1 & \% unsafe & 0 & 1 & \% unsafe \\
Dataset &  &  &  &  &  &  &  &  &  &  &  &  \\
\midrule
crehate & 175 & 140 & 44.4 & 139 & 137 & 49.6 & -- & -- & -- & 162 & 116 & 41.7 \\
d3 & 456 & 427 & 48.4 & 161 & 92 & 36.4 & 358 & 529 & 59.6 & 249 & 260 & 51.1 \\
dices & 173 & 25 & 12.6 & -- & -- & -- & 153 & 45 & 22.7 & -- & -- & -- \\
\bottomrule
\end{tabular}
\caption{Test set 0/1 label distribution on the test set across datasets and quadrants. \textit{-- = no valid annotations for that quadrant in that dataset.}}
\label{tab:q-level-test}
\end{table}

% \clearpage
\subsection{Culturally sensitive item identification} \label{app:sens-item-id}

Table \ref{tab:label_distribution_safe_vs_unsafe} shows the label distribution across train, validation, test sets for the safe-vs-unsafe task; Table \ref{tab:label_distribution_sensitive} for the safe-vs-sensitive task. Note the number of items is the same between tasks to enable a more fair comparison between them.

\begin{table}[h]
\centering
\small
\caption{0/1 label distribution for the safety classification task (safe vs.\ unsafe) across datasets and splits. Label 1 = unanimously unsafe; label 0 = unanimously safe.}
\label{tab:label_distribution_safe_vs_unsafe}
\begin{tabular}{lrrrrrrrrr}
\toprule
 & \multicolumn{3}{c}{Train} & \multicolumn{3}{c}{Val} & \multicolumn{3}{c}{Test} \\
 & 0 & 1 & \% unsafe & 0 & 1 & \% unsafe & 0 & 1 & \% unsafe \\
Dataset &  &  &  &  &  &  &  &  &  \\
\midrule
crehate & 111 & 111 & 50.0 & 28 & 28 & 50.0 & 35 & 35 & 50.0 \\
d3 & 310 & 310 & 50.0 & 78 & 78 & 50.0 & 97 & 97 & 50.0 \\
dices & 57 & 57 & 50.0 & 15 & 15 & 50.0 & 18 & 18 & 50.0 \\
\bottomrule
\end{tabular}
\end{table}

\begin{table}[h]
\centering
\small
\caption{0/1 label distribution for the sensitive item identification task (safe vs.\ culturally sensitive) across datasets and splits. Label 1 = culturally sensitive; label 0 = unanimously safe.}
\label{tab:label_distribution_sensitive}
\begin{tabular}{lrrrrrrrrr}
\toprule
 & \multicolumn{3}{c}{Train} & \multicolumn{3}{c}{Val} & \multicolumn{3}{c}{Test} \\
 & 0 & 1 & \% sensitive & 0 & 1 & \% sensitive & 0 & 1 & \% sensitive \\
Dataset &  &  &  &  &  &  &  &  &  \\
\midrule
crehate & 111 & 111 & 50.0 & 28 & 28 & 50.0 & 35 & 35 & 50.0 \\
d3 & 310 & 310 & 50.0 & 78 & 78 & 50.0 & 97 & 97 & 50.0 \\
dices & 57 & 57 & 50.0 & 15 & 15 & 50.0 & 18 & 18 & 50.0 \\
\bottomrule
\end{tabular}
\end{table}

\begin{figure}[h]
    \centering
    \includegraphics[width=0.8\linewidth]{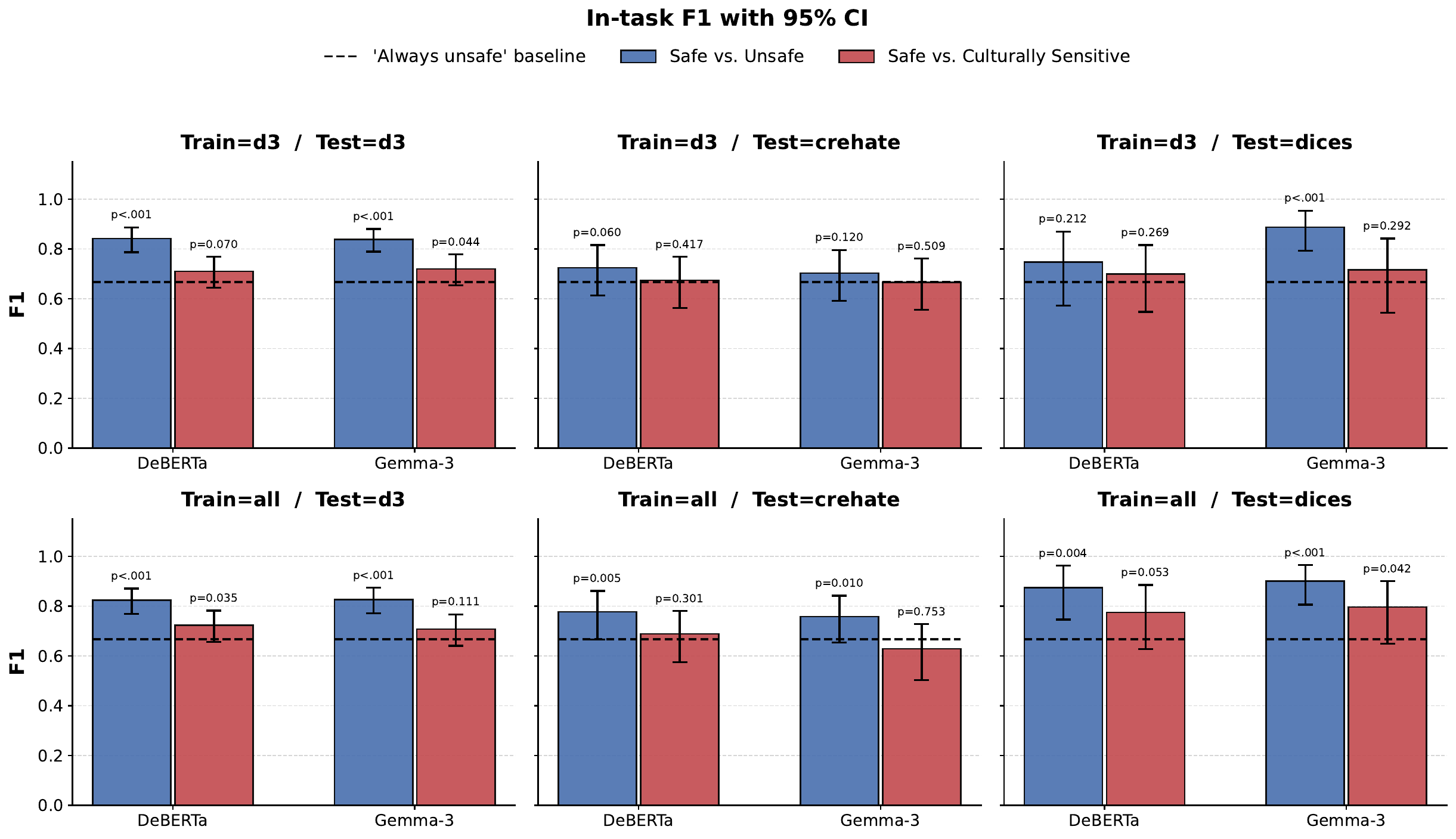}
    \caption{Cross-dataset generalization for safe-vs-unsafe and safe-vs-sensitive tasks. 95\% CIs obtained with hierarchical bootstrap on the item and random seed level. p-values indicate whether difference from the ``Always Unsafe'' baseline is significant.}
    \label{fig:sens-item-cross-dataset}
\end{figure}

\begin{figure}[h]
    \centering
    \includegraphics[width=0.7\linewidth]{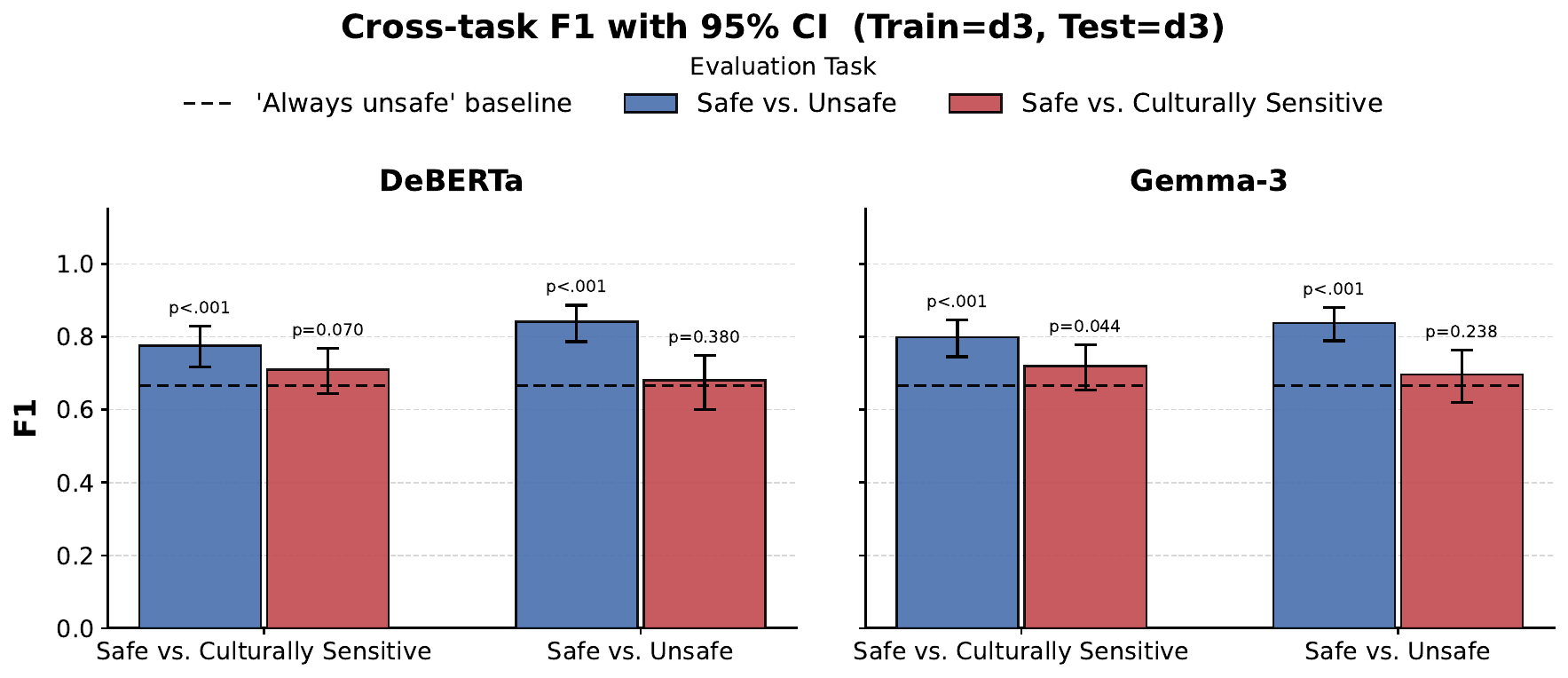}
    \caption{Cross-task generalization for safe-vs-unsafe and safe-vs-sensitive tasks on D3. 95\% CIs obtained with hierarchical bootstrap on the item and random seed level. p-values indicate whether difference from the ``Always Unsafe'' baseline is significant.}
    \label{fig:sens-item-cross-task}
\end{figure}

\paragraph{Cross-dataset and cross-task generalization.} Figure \ref{fig:sens-item-cross-dataset} shows limited cross-dataset generalization: training on D3 does not generalize to CREHate and DICES-990, neither for safe-vs-unsafe nor for safe-vs-sensitive classification, with the exception of Gemma on the safe-vs-unsafe task in DICES-990. Training on all datasets together does not yield conclusive improvements. Figure \ref{fig:sens-item-cross-task} shows that models trained on the safe-vs-unsafe task do not generalize to the safe-vs-sensitive task. However, models trained on the safe-vs-sensitive task do achieve above-random performance on the safe-vs-unsafe task. This suggests that training to detect culturally sensitive items teaches a richer notion of unsafety than training on unanimously unsafe content alone.

\clearpage
\section{Language Model Hyperparameters}

\subsection{Fine-tuning Hyperparameters} \label{app:d3exp}

We used the following hyperparameters for fine-tuning the models:

\begin{itemize}
    \item \textbf{DeBERTa-Large}
    \begin{itemize}
        \item Model version \texttt{deberta-v3-large}
        \item Fine-tuning type: full parameter
        \item Batch size: 16
        \item Learning rate: 2e-5
        \item Epochs: 5
        \item weight decay 0.01
    \end{itemize}
     \item \textbf{Gemma-3-4B}
    \begin{itemize}
        \item Model version \texttt{gemma-3-4b-it}
        \item Fine-tuning type: full parameter
        \item Batch size: 8
        \item Learning rate: 2e-5
        \item Epochs: 5
        \item weight decay 0.01
    \end{itemize}
\end{itemize}

All open model fine-tuning was ran on 4 A100 40GB NVIDIA GPUs using the transformers library \citep{wolf2020transformers}. Each model was trained with $10$ random seeds.

\subsection{Reasoning Model Hyperparameters}

\begin{itemize}
    \item \textbf{GPT-5 Nano}
    \begin{itemize}
        \item Model version \texttt{gpt-5-nano-2025-08-07}
        \item Effort: low
        \item Verbosity: low
    \end{itemize}
     \item \textbf{Gemini-3 Flash}
    \begin{itemize}
        \item Model version \texttt{gemini-3-flash-preview}
        \item Effort: low
    \end{itemize}
\end{itemize}

\subsection{Prompt for Reasoning Models} \label{app:llmj_prompt}

\begin{figure}[htbp]
    \centering
    \includegraphics[width=0.6\linewidth]{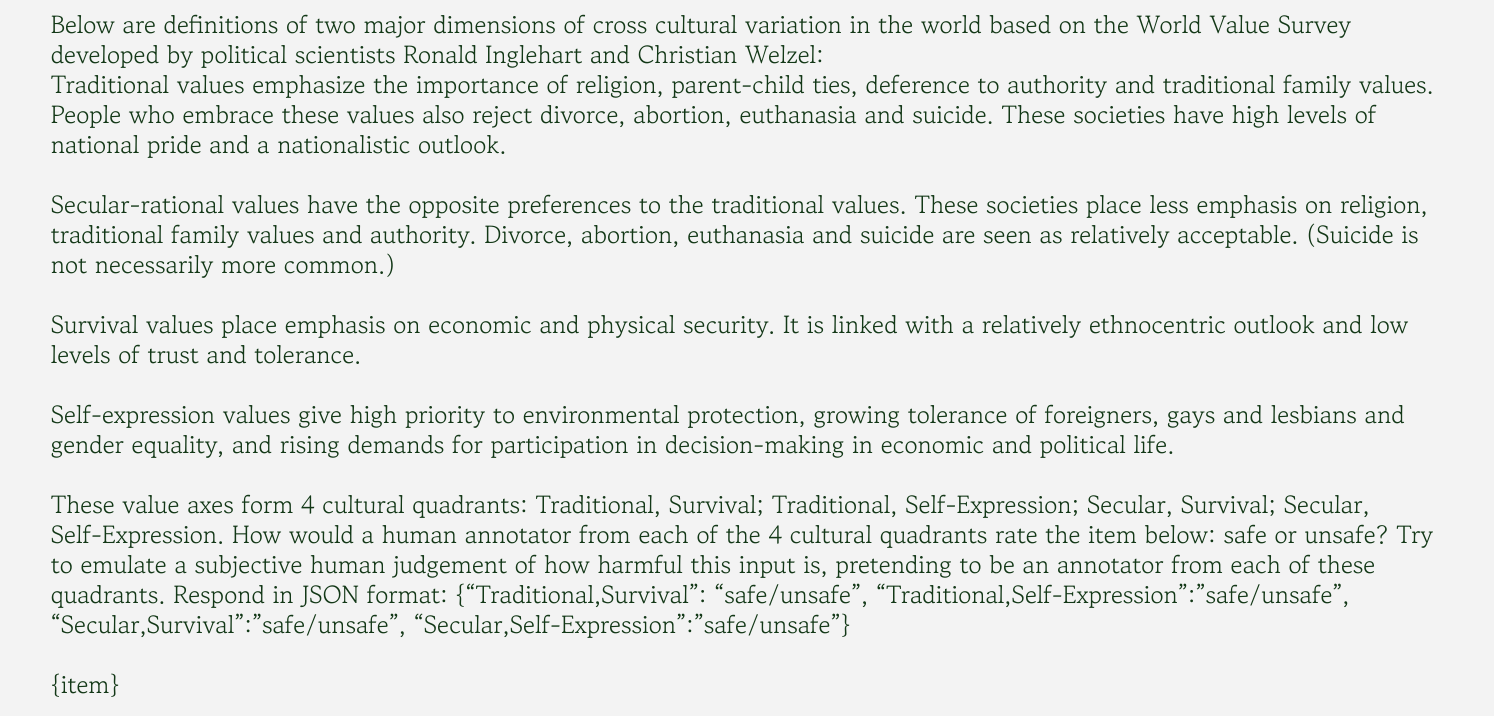}
    \caption{Prompt for the reasoning LLM-as-a-Judge model to emulate judgments from the 4 cultural quadrants (Definitions of the values taken directly from www.worldvaluessurvey.org).}
    \label{fig:prompt}
\end{figure}

The prompt for the reasoning models is provided in Figure \ref{fig:prompt}. 

\clearpage
\section{Hierarchical Linear Model Details} \label{app:lin_models}

% \input{tex/all_models}
% ==============================================================================
% CultureSafe — Linear-Model Results (all datasets)
%
% Include this file from your main Overleaf document:
%   \input{LinearModels_cr/tex/all_models}
%
% Prerequisites in your main preamble:
%   \usepackage{booktabs}
%   \usepackage{longtable}
% ==============================================================================

% ------------------------------------------------------------------------------
% Overview table — every model at a glance
% ------------------------------------------------------------------------------
\begin{longtable}{lllll}
\caption{Overview of all mixed-effects models fitted per dataset.
         Each row corresponds to one regression table included below.}
\label{tab:all_models_overview} \\
\toprule
\textbf{Dataset} & \textbf{Model} & \textbf{Specification} & \textbf{Family} & \textbf{Table} \\
\midrule
\endfirsthead
\toprule
\textbf{Dataset} & \textbf{Model} & \textbf{Specification} & \textbf{Family} & \textbf{Table} \\
\midrule
\endhead
\midrule
\multicolumn{5}{r}{\emph{continued on next page}} \\
\endfoot
\bottomrule
\endlastfoot

% -- CREHate ------------------------------------------------------------------
CREHate & Null                 & Intercept + RE                                & GLMER (binomial) & \ref{tab:crehate_null}              \\
CREHate & Demographics         & + ethnicity, age, gender                      & GLMER (binomial) & \ref{tab:crehate_demog}             \\
CREHate & Culture              & + cultural zone (nationality)              & GLMER (binomial) & \ref{tab:crehate_cult}              \\
CREHate & Culture (Trad)       & + trad values                                 & GLMER (binomial) & \ref{tab:crehate_cult_vals_trad}    \\
CREHate & Culture (Surv)       & + surv values                                 & GLMER (binomial) & \ref{tab:crehate_cult_vals_surv}    \\
CREHate & Culture (Trad+Surv)  & + trad + surv values                          & GLMER (binomial) & \ref{tab:crehate_cult_vals_add}     \\
CREHate & Culture (Trad*Surv)  & + trad $\times$ surv values                   & GLMER (binomial) & \ref{tab:crehate_cult_vals_inter}   \\
CREHate & Demog + Cult         & + demographics + culture (additive)           & GLMER (binomial) & \ref{tab:crehate_demog_cult}        \\
CREHate & Demog + Cult (Vals)  & + demographics + values (additive)            & GLMER (binomial) & \ref{tab:crehate_demog_cult_vals}   \\
CREHate & Demog $\times$ Cult  & + demographics $\times$ culture (interaction) & GLMER (binomial) & \ref{tab:crehate_demog_x_cult}      \\
CREHate & Demog $\times$ Cult (Vals) & + demographics $\times$ values (interaction)  & GLMER (binomial) & \ref{tab:crehate_demog_x_cult_vals} \\
CREHate & Culture (Quadrants)             & + cultural quadrants (Inglehart-Welzel)         & GLMER (binomial) & \ref{tab:crehate_cult_quadrants}         \\
CREHate & Demog + Cult (Quadrants)        & + demographics + quadrants (additive)           & GLMER (binomial) & \ref{tab:crehate_demog_cult_quadrants}   \\
% CREHate & Demog $\times$ Cult (Quadrants) & + demographics $\times$ quadrants (interaction) & GLMER (binomial) & \ref{tab:crehate_demog_x_cult_quadrants} \\
\midrule

% -- D3 -----------------------------------------------------------------------
D3 & Null                 & Intercept + RE                                & LMM              & \ref{tab:d3_null}                   \\
D3 & Demographics         & + age, gender                                 & LMM              & \ref{tab:d3_demog}                  \\
D3 & Culture              & + cultural zone (residence)                & LMM              & \ref{tab:d3_cult}                   \\
D3 & Culture (Trad)       & + trad values                                 & LMM              & \ref{tab:d3_cult_vals_trad}         \\
D3 & Culture (Surv)       & + surv values                                 & LMM              & \ref{tab:d3_cult_vals_surv}         \\
D3 & Culture (Trad+Surv)  & + trad + surv values                          & LMM              & \ref{tab:d3_cult_vals_add}          \\
D3 & Culture (Trad*Surv)  & + trad $\times$ surv values                   & LMM              & \ref{tab:d3_cult_vals_inter}        \\
D3 & Demog + Cult         & + demographics + culture (additive)           & LMM              & \ref{tab:d3_demog_cult}             \\
D3 & Demog + Cult (Vals)  & + demographics + values (additive)            & LMM              & \ref{tab:d3_demog_cult_vals}        \\
D3 & Demog $\times$ Cult  & + demographics $\times$ culture (interaction) & LMM              & \ref{tab:d3_demog_x_cult}           \\
D3 & Demog $\times$ Cult (Vals) & + demographics $\times$ values (interaction)  & LMM              & \ref{tab:d3_demog_x_cult_vals}      \\
D3 & Culture (Quadrants)             & + cultural quadrants (Inglehart-Welzel)         & LMM              & \ref{tab:d3_cult_quadrants}         \\
D3 & Demog + Cult (Quadrants)        & + demographics + quadrants (additive)           & LMM              & \ref{tab:d3_demog_cult_quadrants}   \\
% D3 & Demog $\times$ Cult (Quadrants) & + demographics $\times$ quadrants (interaction) & LMM              & \ref{tab:d3_demog_x_cult_quadrants} \\
\midrule

% -- CultFrames ---------------------------------------------------------------
CultFrames & Null                 & Intercept + controls + RE                     & GLMER (binomial) & \ref{tab:cultframes_null}               \\
CultFrames & Demographics         & + age, gender                                 & GLMER (binomial) & \ref{tab:cultframes_demog}              \\
CultFrames & Culture              & + cultural zone (birth)                    & GLMER (binomial) & \ref{tab:cultframes_cult}               \\
CultFrames & Culture (Trad)       & + trad values                                 & GLMER (binomial) & \ref{tab:cultframes_cult_vals_trad}     \\
CultFrames & Culture (Surv)       & + surv values                                 & GLMER (binomial) & \ref{tab:cultframes_cult_vals_surv}     \\
CultFrames & Culture (Trad+Surv)  & + trad + surv values                          & GLMER (binomial) & \ref{tab:cultframes_cult_vals_add}      \\
CultFrames & Culture (Trad*Surv)  & + trad $\times$ surv values                   & GLMER (binomial) & \ref{tab:cultframes_cult_vals_inter}    \\
CultFrames & Demog + Cult         & + demographics + culture (additive)           & GLMER (binomial) & \ref{tab:cultframes_demog_cult}         \\
CultFrames & Demog + Cult (Vals)  & + demographics + values (additive)            & GLMER (binomial) & \ref{tab:cultframes_demog_cult_vals}    \\
CultFrames & Demog $\times$ Cult  & + demographics $\times$ culture (interaction) & GLMER (binomial) & \ref{tab:cultframes_demog_x_cult}       \\
CultFrames & Demog $\times$ Cult (Vals) & + demographics $\times$ values (interaction)  & GLMER (binomial) & \ref{tab:cultframes_demog_x_cult_vals}  \\
CultFrames & Culture (Quadrants)             & + cultural quadrants (Inglehart-Welzel)         & GLMER (binomial) & \ref{tab:cultframes_cult_quadrants}         \\
CultFrames & Demog + Cult (Quadrants)        & + demographics + quadrants (additive)           & GLMER (binomial) & \ref{tab:cultframes_demog_cult_quadrants}   \\
% CultFrames & Demog $\times$ Cult (Quadrants) & + demographics $\times$ quadrants (interaction) & GLMER (binomial) & \ref{tab:cultframes_demog_x_cult_quadrants} \\
\midrule

% -- DICES Expert -------------------------------------------------------------
DICES Expert & Null                 & Intercept + controls + RE                     & GLMER (binomial) & \ref{tab:dices_expert_null}             \\
DICES Expert & Demographics         & + ethnicity, age, gender                      & GLMER (binomial) & \ref{tab:dices_expert_demog}            \\
DICES Expert & Culture              & + cultural zone (locale)                   & GLMER (binomial) & \ref{tab:dices_expert_cult}             \\
DICES Expert & Culture (Trad)       & + trad values                                 & GLMER (binomial) & \ref{tab:dices_expert_cult_vals_trad}   \\
DICES Expert & Culture (Surv)       & + surv values                                 & GLMER (binomial) & \ref{tab:dices_expert_cult_vals_surv}   \\
DICES Expert & Culture (Trad+Surv)  & + trad + surv values                          & GLMER (binomial) & \ref{tab:dices_expert_cult_vals_add}    \\
DICES Expert & Culture (Trad*Surv)  & + trad $\times$ surv values                   & GLMER (binomial) & \ref{tab:dices_expert_cult_vals_inter}  \\
DICES Expert & Demog + Cult         & + demographics + culture (additive)           & GLMER (binomial) & \ref{tab:dices_expert_demog_cult}       \\
DICES Expert & Demog + Cult (Vals)  & + demographics + values (additive)            & GLMER (binomial) & \ref{tab:dices_expert_demog_cult_vals}  \\
DICES Expert & Demog $\times$ Cult  & + demographics $\times$ culture (interaction) & GLMER (binomial) & \ref{tab:dices_expert_demog_x_cult}     \\
DICES Expert & Demog $\times$ Cult (Vals) & + demographics $\times$ values (interaction)  & GLMER (binomial) & \ref{tab:dices_expert_demog_x_cult_vals}\\
DICES Expert & Culture (Quadrants)             & + cultural quadrants (Inglehart-Welzel)         & GLMER (binomial) & \ref{tab:dices_expert_cult_quadrants}         \\
DICES Expert & Demog + Cult (Quadrants)        & + demographics + quadrants (additive)           & GLMER (binomial) & \ref{tab:dices_expert_demog_cult_quadrants}   \\
% DICES Expert & Demog $\times$ Cult (Quadrants) & + demographics $\times$ quadrants (interaction) & GLMER (binomial) & \ref{tab:dices_expert_demog_x_cult_quadrants} \\
\midrule

% -- DIVE ---------------------------------------------------------------------
DIVE & Null                 & Intercept + RE                                & LMM              & \ref{tab:dive_null}                 \\
DIVE & Demographics         & + ethnicity, age, gender                      & LMM              & \ref{tab:dive_demog}                \\
DIVE & Culture              & + cultural zone (birth)                    & LMM              & \ref{tab:dive_cult}                 \\
DIVE & Culture (Trad)       & + trad values                                 & LMM              & \ref{tab:dive_cult_vals_trad}       \\
DIVE & Culture (Surv)       & + surv values                                 & LMM              & \ref{tab:dive_cult_vals_surv}       \\
DIVE & Culture (Trad+Surv)  & + trad + surv values                          & LMM              & \ref{tab:dive_cult_vals_add}        \\
DIVE & Culture (Trad*Surv)  & + trad $\times$ surv values                   & LMM              & \ref{tab:dive_cult_vals_inter}      \\
DIVE & Demog + Cult         & + demographics + culture (additive)           & LMM              & \ref{tab:dive_demog_cult}           \\
DIVE & Demog + Cult (Vals)  & + demographics + values (additive)            & LMM              & \ref{tab:dive_demog_cult_vals}      \\
DIVE & Demog $\times$ Cult  & + demographics $\times$ culture (interaction) & LMM              & \ref{tab:dive_demog_x_cult}         \\
DIVE & Demog $\times$ Cult (Vals) & + demographics $\times$ values (interaction)  & LMM              & \ref{tab:dive_demog_x_cult_vals}    \\
DIVE & Culture (Quadrants)             & + cultural quadrants (Inglehart-Welzel)         & LMM              & \ref{tab:dive_cult_quadrants}         \\
DIVE & Demog + Cult (Quadrants)        & + demographics + quadrants (additive)           & LMM              & \ref{tab:dive_demog_cult_quadrants}   \\
% DIVE & Demog $\times$ Cult (Quadrants) & + demographics $\times$ quadrants (interaction) & LMM              & \ref{tab:dive_demog_x_cult_quadrants} \\
\midrule

% -- Severity --------------------------------------------------------------------
Severity & Null                 & Intercept + RE                                & LMM              & \ref{tab:severity_null}                 \\
Severity & Demographics         & + ethnicity, age, gender                      & LMM              & \ref{tab:severity_demog}                \\
Severity & Culture              & + cultural zone (residence)                & LMM              & \ref{tab:severity_cult}                 \\
Severity & Culture (Trad)       & + trad values                                 & LMM              & \ref{tab:severity_cult_vals_trad}       \\
Severity & Culture (Surv)       & + surv values                                 & LMM              & \ref{tab:severity_cult_vals_surv}       \\
Severity & Culture (Trad+Surv)  & + trad + surv values                          & LMM              & \ref{tab:severity_cult_vals_add}        \\
Severity & Culture (Trad*Surv)  & + trad $\times$ surv values                   & LMM              & \ref{tab:severity_cult_vals_inter}      \\
Severity & Demog + Cult         & + demographics + culture (additive)           & LMM              & \ref{tab:severity_demog_cult}           \\
Severity & Demog + Cult (Vals)  & + demographics + values (additive)            & LMM              & \ref{tab:severity_demog_cult_vals}      \\
Severity & Demog $\times$ Cult  & + demographics $\times$ culture (interaction) & LMM              & \ref{tab:severity_demog_x_cult}         \\
Severity & Demog $\times$ Cult (Vals) & + demographics $\times$ values (interaction)  & LMM              & \ref{tab:severity_demog_x_cult_vals}    \\
Severity & Culture (Quadrants)             & + cultural quadrants (Inglehart-Welzel)         & LMM              & \ref{tab:severity_cult_quadrants}         \\
Severity & Demog + Cult (Quadrants)        & + demographics + quadrants (additive)           & LMM              & \ref{tab:severity_demog_cult_quadrants}   \\
% Severity & Demog $\times$ Cult (Quadrants) & + demographics $\times$ quadrants (interaction) & LMM              & \ref{tab:severity_demog_x_cult_quadrants} \\
\midrule

% -- NLPositionality ----------------------------------------------------------
NLPositionality & Null                 & Intercept + RE                                & GLMER (binomial) & \ref{tab:nlpos_null}                 \\
NLPositionality & Demographics         & + ethnicity, age, gender                      & GLMER (binomial) & \ref{tab:nlpos_demog}                \\
NLPositionality & Culture              & + cultural zone                            & GLMER (binomial) & \ref{tab:nlpos_cult}                 \\
NLPositionality & Culture (Trad)       & + trad values                                 & GLMER (binomial) & \ref{tab:nlpos_cult_vals_trad}       \\
NLPositionality & Culture (Surv)       & + surv values                                 & GLMER (binomial) & \ref{tab:nlpos_cult_vals_surv}       \\
NLPositionality & Culture (Trad+Surv)  & + trad + surv values                          & GLMER (binomial) & \ref{tab:nlpos_cult_vals_add}        \\
NLPositionality & Culture (Trad*Surv)  & + trad $\times$ surv values                   & GLMER (binomial) & \ref{tab:nlpos_cult_vals_inter}      \\
NLPositionality & Demog + Cult         & + demographics + culture (additive)           & GLMER (binomial) & \ref{tab:nlpos_demog_cult}           \\
NLPositionality & Demog + Cult (Vals)  & + demographics + values (additive)            & GLMER (binomial) & \ref{tab:nlpos_demog_cult_vals}      \\
NLPositionality & Demog $\times$ Cult  & + demographics $\times$ culture (interaction) & GLMER (binomial) & \ref{tab:nlpos_demog_x_cult}         \\
NLPositionality & Demog $\times$ Cult (Vals) & + demographics $\times$ values (interaction)  & GLMER (binomial) & \ref{tab:nlpos_demog_x_cult_vals}    \\
NLPositionality & Culture (Quadrants)             & + cultural quadrants (Inglehart-Welzel)         & GLMER (binomial) & \ref{tab:nlpos_cult_quadrants}         \\
NLPositionality & Demog + Cult (Quadrants)        & + demographics + quadrants (additive)           & GLMER (binomial) & \ref{tab:nlpos_demog_cult_quadrants}   \\
% NLPositionality & Demog $\times$ Cult (Quadrants) & + demographics $\times$ quadrants (interaction) & GLMER (binomial) & \ref{tab:nlpos_demog_x_cult_quadrants} \\
\midrule

% -- PRISM --------------------------------------------------------------------
PRISM & Null                 & Intercept + controls + RE                     & LMM              & \ref{tab:prism_null}                 \\
PRISM & Demographics         & + ethnicity, age, gender                      & LMM              & \ref{tab:prism_demog}                \\
PRISM & Culture              & + cultural zone (birth)                    & LMM              & \ref{tab:prism_cult}                 \\
PRISM & Culture (Trad)       & + trad values                                 & LMM              & \ref{tab:prism_cult_vals_trad}       \\
PRISM & Culture (Surv)       & + surv values                                 & LMM              & \ref{tab:prism_cult_vals_surv}       \\
PRISM & Culture (Trad+Surv)  & + trad + surv values                          & LMM              & \ref{tab:prism_cult_vals_add}        \\
PRISM & Culture (Trad*Surv)  & + trad $\times$ surv values                   & LMM              & \ref{tab:prism_cult_vals_inter}      \\
PRISM & Demog + Cult         & + demographics + culture (additive)           & LMM              & \ref{tab:prism_demog_cult}           \\
PRISM & Demog + Cult (Vals)  & + demographics + values (additive)            & LMM              & \ref{tab:prism_demog_cult_vals}      \\
PRISM & Demog $\times$ Cult  & + demographics $\times$ culture (interaction) & LMM              & \ref{tab:prism_demog_x_cult}         \\
PRISM & Demog $\times$ Cult (Vals) & + demographics $\times$ values (interaction)  & LMM              & \ref{tab:prism_demog_x_cult_vals}    \\
PRISM & Culture (Quadrants)             & + cultural quadrants (Inglehart-Welzel)         & LMM              & \ref{tab:prism_cult_quadrants}         \\
PRISM & Demog + Cult (Quadrants)        & + demographics + quadrants (additive)           & LMM              & \ref{tab:prism_demog_cult_quadrants}   \\
% PRISM & Demog $\times$ Cult (Quadrants) & + demographics $\times$ quadrants (interaction) & LMM              & \ref{tab:prism_demog_x_cult_quadrants} \\

\end{longtable}

\clearpage

% ============================================================================
% Individual model tables
% ============================================================================

% -- CREHate ------------------------------------------------------------------
\begin{table}[htbp]
\begin{center}
\adjustbox{max width=\textwidth}{%
\begin{tabular}{l c}
\toprule
 & \shortstack[l]{harmful \textasciitilde{} \\ (1 \textbar{} rater\_id) + (1 \textbar{} item\_id)} \\
\midrule
(Intercept)                & $-0.41 \; [-0.53; -0.30]^{*}$ \\
\midrule
$R^2_m$ (theoretical)      & $0$                           \\
$R^2_c$ (theoretical)      & $0.59$                        \\
AIC                        & $39042.36$                    \\
BIC                        & $39068.06$                    \\
Log Likelihood             & $-19518.18$                   \\
Num. obs.                  & $38727$                       \\
Num. groups: item\_id      & $1580$                        \\
Num. groups: rater\_id     & $1037$                        \\
Var: item\_id (Intercept)  & $4.29$                        \\
Var: rater\_id (Intercept) & $0.53$                        \\
\bottomrule
\multicolumn{2}{l}{\parbox[t]{\textwidth}{\scriptsize{$^*$ 0 outside the confidence interval.}}}
\end{tabular}%
}
\caption{CREHate: Null model (GLMER, binomial)}
\label{tab:crehate_null}
\end{center}
\end{table}
\begin{table}[htbp]
\begin{center}
\adjustbox{max width=\textwidth}{%
\begin{tabular}{l c}
\toprule
 & \shortstack[l]{harmful \textasciitilde{} ethnicity + age + gender + \\ (1 \textbar{} rater\_id) + (1 \textbar{} item\_id)} \\
\midrule
(Intercept)                     & $-0.51 \; [-0.66; -0.36]^{*}$ \\
rater\_ethnicityAsian           & $0.08 \; [-0.07;  0.23]$      \\
rater\_ethnicityBlack           & $-0.22 \; [-0.38; -0.06]^{*}$ \\
rater\_ethnicityMiddle\_Eastern & $0.36 \; [-0.36;  1.08]$      \\
rater\_ethnicityMixed           & $0.02 \; [-0.40;  0.43]$      \\
rater\_age\_groupboomer         & $0.01 \; [-0.21;  0.24]$      \\
rater\_age\_groupgenx           & $-0.14 \; [-0.30;  0.01]$     \\
rater\_age\_groupgenz           & $0.08 \; [-0.06;  0.21]$      \\
rater\_genderfemale             & $0.24 \; [ 0.13;  0.35]^{*}$  \\
rater\_gendernon-binary         & $0.23 \; [-0.23;  0.70]$      \\
\midrule
$R^2_m$ (theoretical)           & $0.00$                        \\
$R^2_c$ (theoretical)           & $0.59$                        \\
AIC                             & $39026.74$                    \\
BIC                             & $39129.51$                    \\
Log Likelihood                  & $-19501.37$                   \\
Num. obs.                       & $38727$                       \\
Num. groups: item\_id           & $1580$                        \\
Num. groups: rater\_id          & $1037$                        \\
Var: item\_id (Intercept)       & $4.29$                        \\
Var: rater\_id (Intercept)      & $0.50$                        \\
\bottomrule
\multicolumn{2}{l}{\parbox[t]{\textwidth}{\scriptsize{$^*$ 0 outside the confidence interval. Reference levels: rater\_ethnicity = White; rater\_age\_group = millennial; rater\_gender = male.}}}
\end{tabular}%
}
\caption{CREHate: Demographics model (GLMER, binomial)}
\label{tab:crehate_demog}
\end{center}
\end{table}
\begin{table}[htbp]
\begin{center}
\adjustbox{max width=\textwidth}{%
\begin{tabular}{l c}
\toprule
 & \shortstack[l]{harmful \textasciitilde{} con\_cult\_cluster + \\ (1 \textbar{} rater\_id) + (1 \textbar{} item\_id)} \\
\midrule
(Intercept)                          & $-0.44 \; [-0.56; -0.31]^{*}$ \\
con\_cultural\_zoneWest \& South Asia & $0.08 \; [-0.04;  0.20]$      \\
\midrule
$R^2_m$ (theoretical)                & $0$                           \\
$R^2_c$ (theoretical)                & $0.59$                        \\
AIC                                  & $39042.74$                    \\
BIC                                  & $39077.00$                    \\
Log Likelihood                       & $-19517.37$                   \\
Num. obs.                            & $38727$                       \\
Num. groups: item\_id                & $1580$                        \\
Num. groups: rater\_id               & $1037$                        \\
Var: item\_id (Intercept)            & $4.29$                        \\
Var: rater\_id (Intercept)           & $0.52$                        \\
\bottomrule
\multicolumn{2}{l}{\parbox[t]{\textwidth}{\scriptsize{$^*$ 0 outside the confidence interval. Reference levels: con\_cultural\_zone = English-Speaking.}}}
\end{tabular}%
}
\caption{CREHate: Culture model (GLMER, binomial)}
\label{tab:crehate_cult}
\end{center}
\end{table}
\begin{table}[htbp]
\begin{center}
\adjustbox{max width=\textwidth}{%
\begin{tabular}{l c}
\toprule
 & \shortstack[l]{harmful \textasciitilde{} TradAgg + \\ (1 \textbar{} rater\_id) + (1 \textbar{} item\_id)} \\
\midrule
(Intercept)                & $-0.46 \; [-0.59; -0.34]^{*}$ \\
nat\_TradAgg               & $0.12 \; [ 0.01;  0.24]^{*}$  \\
\midrule
$R^2_m$ (theoretical)      & $0$                           \\
$R^2_c$ (theoretical)      & $0.59$                        \\
AIC                        & $39040.21$                    \\
BIC                        & $39074.47$                    \\
Log Likelihood             & $-19516.11$                   \\
Num. obs.                  & $38727$                       \\
Num. groups: item\_id      & $1580$                        \\
Num. groups: rater\_id     & $1037$                        \\
Var: item\_id (Intercept)  & $4.29$                        \\
Var: rater\_id (Intercept) & $0.52$                        \\
\bottomrule
\multicolumn{2}{l}{\parbox[t]{\textwidth}{\scriptsize{$^*$ 0 outside the confidence interval.}}}
\end{tabular}%
}
\caption{CREHate: Culture values (TradAgg) (GLMER, binomial)}
\label{tab:crehate_cult_vals_trad}
\end{center}
\end{table}
\begin{table}[htbp]
\begin{center}
\adjustbox{max width=\textwidth}{%
\begin{tabular}{l c}
\toprule
 & \shortstack[l]{harmful \textasciitilde{} SurvSAgg + \\ (1 \textbar{} rater\_id) + (1 \textbar{} item\_id)} \\
\midrule
(Intercept)                & $-0.43 \; [-0.57; -0.29]^{*}$ \\
nat\_SurvSAgg              & $0.01 \; [-0.05;  0.06]$      \\
\midrule
$R^2_m$ (theoretical)      & $0$                           \\
$R^2_c$ (theoretical)      & $0.59$                        \\
AIC                        & $39044.26$                    \\
BIC                        & $39078.51$                    \\
Log Likelihood             & $-19518.13$                   \\
Num. obs.                  & $38727$                       \\
Num. groups: item\_id      & $1580$                        \\
Num. groups: rater\_id     & $1037$                        \\
Var: item\_id (Intercept)  & $4.29$                        \\
Var: rater\_id (Intercept) & $0.53$                        \\
\bottomrule
\multicolumn{2}{l}{\parbox[t]{\textwidth}{\scriptsize{$^*$ 0 outside the confidence interval.}}}
\end{tabular}%
}
\caption{CREHate: Culture values (SurvSAgg) (GLMER, binomial)}
\label{tab:crehate_cult_vals_surv}
\end{center}
\end{table}
\begin{table}[htbp]
\begin{center}
\adjustbox{max width=\textwidth}{%
\begin{tabular}{l c}
\toprule
 & \shortstack[l]{harmful \textasciitilde{} TradAgg + SurvSAgg + \\ (1 \textbar{} rater\_id) + (1 \textbar{} item\_id)} \\
\midrule
(Intercept)                & $-0.35 \; [-0.50; -0.21]^{*}$ \\
nat\_TradAgg               & $0.45 \; [ 0.21;  0.69]^{*}$  \\
nat\_SurvSAgg              & $-0.17 \; [-0.28; -0.06]^{*}$ \\
\midrule
$R^2_m$ (theoretical)      & $0.00$                        \\
$R^2_c$ (theoretical)      & $0.59$                        \\
AIC                        & $39033.23$                    \\
BIC                        & $39076.05$                    \\
Log Likelihood             & $-19511.62$                   \\
Num. obs.                  & $38727$                       \\
Num. groups: item\_id      & $1580$                        \\
Num. groups: rater\_id     & $1037$                        \\
Var: item\_id (Intercept)  & $4.29$                        \\
Var: rater\_id (Intercept) & $0.51$                        \\
\bottomrule
\multicolumn{2}{l}{\parbox[t]{\textwidth}{\scriptsize{$^*$ 0 outside the confidence interval.}}}
\end{tabular}%
}
\caption{CREHate: Culture values (TradAgg + SurvSAgg) (GLMER, binomial)}
\label{tab:crehate_cult_vals_add}
\end{center}
\end{table}
\begin{table}[htbp]
\begin{center}
\adjustbox{max width=\textwidth}{%
\begin{tabular}{l c}
\toprule
 & \shortstack[l]{harmful \textasciitilde{} TradAgg * SurvSAgg + \\ (1 \textbar{} rater\_id) + (1 \textbar{} item\_id)} \\
\midrule
(Intercept)                & $-0.36 \; [-0.51; -0.20]^{*}$ \\
nat\_TradAgg               & $0.44 \; [-0.09;  0.96]$      \\
nat\_SurvSAgg              & $-0.17 \; [-0.28; -0.06]^{*}$ \\
nat\_TradAgg:nat\_SurvSAgg & $0.01 \; [-0.23;  0.25]$      \\
\midrule
$R^2_m$ (theoretical)      & $0.00$                        \\
$R^2_c$ (theoretical)      & $0.59$                        \\
AIC                        & $39035.23$                    \\
BIC                        & $39086.61$                    \\
Log Likelihood             & $-19511.61$                   \\
Num. obs.                  & $38727$                       \\
Num. groups: item\_id      & $1580$                        \\
Num. groups: rater\_id     & $1037$                        \\
Var: item\_id (Intercept)  & $4.29$                        \\
Var: rater\_id (Intercept) & $0.51$                        \\
\bottomrule
\multicolumn{2}{l}{\parbox[t]{\textwidth}{\scriptsize{$^*$ 0 outside the confidence interval.}}}
\end{tabular}%
}
\caption{CREHate: Culture values (TradAgg * SurvSAgg) (GLMER, binomial)}
\label{tab:crehate_cult_vals_inter}
\end{center}
\end{table}
\begin{table}[htbp]
\begin{center}
\adjustbox{max width=\textwidth}{%
\begin{tabular}{l c}
\toprule
 & \shortstack[l]{harmful \textasciitilde{} ethnicity + age + gender + \\ con\_cult\_cluster + \\ (1 \textbar{} rater\_id) + (1 \textbar{} item\_id)} \\
\midrule
(Intercept)                          & $-0.52 \; [-0.67; -0.37]^{*}$ \\
rater\_ethnicityAsian                & $-0.07 \; [-0.25;  0.11]$     \\
rater\_ethnicityBlack                & $-0.42 \; [-0.64; -0.20]^{*}$ \\
rater\_ethnicityMiddle\_Eastern      & $0.33 \; [-0.38;  1.05]$      \\
rater\_ethnicityMixed                & $-0.05 \; [-0.46;  0.37]$     \\
rater\_age\_groupboomer              & $0.01 \; [-0.22;  0.23]$      \\
rater\_age\_groupgenx                & $-0.14 \; [-0.29;  0.02]$     \\
rater\_age\_groupgenz                & $0.06 \; [-0.07;  0.20]$      \\
rater\_genderfemale                  & $0.25 \; [ 0.14;  0.36]^{*}$  \\
rater\_gendernon-binary              & $0.25 \; [-0.22;  0.71]$      \\
con\_cultural\_zoneWest \& South Asia & $0.25 \; [ 0.07;  0.43]^{*}$  \\
\midrule
$R^2_m$ (theoretical)                & $0.00$                        \\
$R^2_c$ (theoretical)                & $0.59$                        \\
AIC                                  & $39021.61$                    \\
BIC                                  & $39132.95$                    \\
Log Likelihood                       & $-19497.81$                   \\
Num. obs.                            & $38727$                       \\
Num. groups: item\_id                & $1580$                        \\
Num. groups: rater\_id               & $1037$                        \\
Var: item\_id (Intercept)            & $4.29$                        \\
Var: rater\_id (Intercept)           & $0.49$                        \\
\bottomrule
\multicolumn{2}{l}{\parbox[t]{\textwidth}{\scriptsize{$^*$ 0 outside the confidence interval. Reference levels: rater\_ethnicity = White; rater\_age\_group = millennial; rater\_gender = male; con\_cultural\_zone = English-Speaking.}}}
\end{tabular}%
}
\caption{CREHate: Demographics + Culture additive (GLMER, binomial)}
\label{tab:crehate_demog_cult}
\end{center}
\end{table}
\begin{table}[htbp]
\begin{center}
\adjustbox{max width=\textwidth}{%
\begin{tabular}{l c}
\toprule
 & \shortstack[l]{harmful \textasciitilde{} ethnicity + age + gender + \\ nat\_TradAgg + nat\_SurvSAgg + \\ (1 \textbar{} rater\_id) + (1 \textbar{} item\_id)} \\
\midrule
(Intercept)                     & $-0.42 \; [-0.64; -0.20]^{*}$ \\
rater\_ethnicityAsian           & $0.02 \; [-0.16;  0.19]$      \\
rater\_ethnicityBlack           & $-0.20 \; [-0.41;  0.01]$     \\
rater\_ethnicityMiddle\_Eastern & $0.41 \; [-0.31;  1.13]$      \\
rater\_ethnicityMixed           & $0.01 \; [-0.40;  0.42]$      \\
rater\_age\_groupboomer         & $0.01 \; [-0.21;  0.23]$      \\
rater\_age\_groupgenx           & $-0.15 \; [-0.30;  0.00]$     \\
rater\_age\_groupgenz           & $0.07 \; [-0.06;  0.20]$      \\
rater\_genderfemale             & $0.23 \; [ 0.12;  0.34]^{*}$  \\
rater\_gendernon-binary         & $0.23 \; [-0.23;  0.70]$      \\
nat\_TradAgg                    & $0.36 \; [ 0.10;  0.61]^{*}$  \\
nat\_SurvSAgg                   & $-0.16 \; [-0.28; -0.03]^{*}$ \\
\midrule
$R^2_m$ (theoretical)           & $0.00$                        \\
$R^2_c$ (theoretical)           & $0.59$                        \\
AIC                             & $39023.12$                    \\
BIC                             & $39143.02$                    \\
Log Likelihood                  & $-19497.56$                   \\
Num. obs.                       & $38727$                       \\
Num. groups: item\_id           & $1580$                        \\
Num. groups: rater\_id          & $1037$                        \\
Var: item\_id (Intercept)       & $4.29$                        \\
Var: rater\_id (Intercept)      & $0.49$                        \\
\bottomrule
\multicolumn{2}{l}{\parbox[t]{\textwidth}{\scriptsize{$^*$ 0 outside the confidence interval. Reference levels: rater\_ethnicity = White; rater\_age\_group = millennial; rater\_gender = male.}}}
\end{tabular}%
}
\caption{CREHate: Demographics + selected culture-values additive (add) (GLMER, binomial)}
\label{tab:crehate_demog_cult_vals}
\end{center}
\end{table}
\begin{table}[htbp]
\begin{center}
\adjustbox{max width=\textwidth}{%
\begin{tabular}{l c}
\toprule
 & \shortstack[l]{harmful \textasciitilde{} (ethnicity + age + gender) * \\ con\_cult\_cluster + \\ (1 \textbar{} rater\_id) + (1 \textbar{} item\_id)} \\
\midrule
(Intercept)                                                          & $-0.54 \; [-0.69; -0.38]^{*}$ \\
rater\_ethnicityAsian                                                & $-0.14 \; [-0.36;  0.08]$     \\
rater\_ethnicityBlack                                                & $-0.36 \; [-0.75;  0.03]$     \\
rater\_ethnicityMiddle\_Eastern                                      & $0.26 \; [-0.53;  1.05]$      \\
rater\_ethnicityMixed                                                & $0.14 \; [-0.35;  0.63]$      \\
rater\_age\_groupboomer                                              & $-0.04 \; [-0.28;  0.20]$     \\
rater\_age\_groupgenx                                                & $-0.14 \; [-0.31;  0.02]$     \\
\end{tabular}%
}
\caption{CREHate: Demographics x Culture interaction (GLMER, binomial)}
\label{tab:crehate_demog_x_cult}
\end{center}
\end{table}
{\small
\begin{longtable}{l c}
\caption{CREHate: Demographics x selected culture-values interaction (add) (GLMER, binomial)}\label{tab:crehate_demog_x_cult_vals} \\
\toprule
& \shortstack[l]{harmful \textasciitilde{} (ethnicity + age + gender) * \\ (nat\_TradAgg + nat\_SurvSAgg) + \\ (1 \textbar{} rater\_id) + (1 \textbar{} item\_id)} \\
\midrule
\endfirsthead
\multicolumn{2}{l}{\emph{CREHate (continued)}} \\
\toprule
& \shortstack[l]{harmful \textasciitilde{} (ethnicity + age + gender) * \\ (nat\_TradAgg + nat\_SurvSAgg) + \\ (1 \textbar{} rater\_id) + (1 \textbar{} item\_id)} \\
\midrule
\endhead
\midrule
\multicolumn{2}{r}{\emph{continued on next page}} \\
\endfoot
\bottomrule
\multicolumn{2}{l}{\parbox[t]{\linewidth}{\scriptsize{$^*$ 0 outside the confidence interval. Reference levels: rater\_ethnicity = White; rater\_age\_group = millennial; rater\_gender = male.}}}
\endlastfoot
(Intercept)                                   & $-0.87 \; [-1.27; -0.46]^{*}$ \\
rater\_ethnicityAsian                         & $0.53 \; [ 0.13;  0.93]^{*}$  \\
rater\_ethnicityBlack                         & $0.57 \; [-0.10;  1.25]$      \\
rater\_ethnicityMiddle\_Eastern               & $0.74 \; [-1.52;  2.99]$      \\
rater\_ethnicityMixed                         & $-0.03 \; [-1.07;  1.01]$     \\
rater\_age\_groupboomer                       & $0.31 \; [-0.22;  0.84]$      \\
rater\_age\_groupgenx                         & $0.03 \; [-0.44;  0.51]$      \\
rater\_age\_groupgenz                         & $0.10 \; [-0.11;  0.32]$      \\
\end{longtable}
}
\begin{table}[htbp]
\begin{center}
\adjustbox{max width=\textwidth}{%
\begin{tabular}{l c}
\toprule
 & \shortstack[l]{harmful \textasciitilde{} nat\_quadrant + \\ (1 \textbar{} rater\_id) + (1 \textbar{} item\_id)} \\
\midrule
(Intercept)                & $-0.44 \; [-0.56; -0.31]^{*}$ \\
nat\_quadrantII            & $0.28 \; [ 0.11;  0.44]^{*}$  \\
nat\_quadrantIV            & $-0.06 \; [-0.20;  0.09]$     \\
\midrule
$R^2_m$ (theoretical)      & $0.00$                        \\
$R^2_c$ (theoretical)      & $0.59$                        \\
AIC                        & $39034.28$                    \\
BIC                        & $39077.10$                    \\
Log Likelihood             & $-19512.14$                   \\
Num. obs.                  & $38727$                       \\
Num. groups: item\_id      & $1580$                        \\
Num. groups: rater\_id     & $1037$                        \\
Var: item\_id (Intercept)  & $4.29$                        \\
Var: rater\_id (Intercept) & $0.51$                        \\
\bottomrule
\multicolumn{2}{l}{\parbox[t]{\textwidth}{\scriptsize{$^*$ 0 outside the confidence interval. Reference levels: nat\_quadrant = I.}}}
\end{tabular}%
}
\caption{CREHate: Culture quadrants (GLMER, binomial)}
\label{tab:crehate_cult_quadrants}
\end{center}
\end{table}
\begin{table}[htbp]
\begin{center}
\adjustbox{max width=\textwidth}{%
\begin{tabular}{l c}
\toprule
 & \shortstack[l]{harmful \textasciitilde{} ethnicity + age + gender + \\ nat\_quadrant + \\ (1 \textbar{} rater\_id) + (1 \textbar{} item\_id)} \\
\midrule
(Intercept)                     & $-0.52 \; [-0.66; -0.37]^{*}$ \\
rater\_ethnicityAsian           & $-0.14 \; [-0.35;  0.07]$     \\
rater\_ethnicityBlack           & $-0.33 \; [-0.59; -0.08]^{*}$ \\
rater\_ethnicityMiddle\_Eastern & $0.35 \; [-0.37;  1.06]$      \\
rater\_ethnicityMixed           & $-0.02 \; [-0.43;  0.40]$     \\
rater\_age\_groupboomer         & $0.00 \; [-0.22;  0.22]$      \\
rater\_age\_groupgenx           & $-0.14 \; [-0.29;  0.01]$     \\
rater\_age\_groupgenz           & $0.07 \; [-0.07;  0.20]$      \\
rater\_genderfemale             & $0.24 \; [ 0.13;  0.35]^{*}$  \\
rater\_gendernon-binary         & $0.26 \; [-0.20;  0.73]$      \\
nat\_quadrantII                 & $0.36 \; [ 0.11;  0.60]^{*}$  \\
nat\_quadrantIV                 & $0.14 \; [-0.10;  0.38]$      \\
\midrule
$R^2_m$ (theoretical)           & $0.00$                        \\
$R^2_c$ (theoretical)           & $0.59$                        \\
AIC                             & $39021.91$                    \\
BIC                             & $39141.81$                    \\
Log Likelihood                  & $-19496.96$                   \\
Num. obs.                       & $38727$                       \\
Num. groups: item\_id           & $1580$                        \\
Num. groups: rater\_id          & $1037$                        \\
Var: item\_id (Intercept)       & $4.29$                        \\
Var: rater\_id (Intercept)      & $0.49$                        \\
\bottomrule
\multicolumn{2}{l}{\parbox[t]{\textwidth}{\scriptsize{$^*$ 0 outside the confidence interval. Reference levels: rater\_ethnicity = White; rater\_age\_group = millennial; rater\_gender = male; nat\_quadrant = I.}}}
\end{tabular}%
}
\caption{CREHate: Demographics + Quadrants additive (GLMER, binomial)}
\label{tab:crehate_demog_cult_quadrants}
\end{center}
\end{table}

\clearpage % <--- ADD THIS TO EMPTY THE FLOAT QUEUE

% -- D3 -----------------------------------------------------------------------
\begin{table}[htbp]
\begin{center}
\adjustbox{max width=\textwidth}{%
\begin{tabular}{l c}
\toprule
 & \shortstack[l]{harmful \textasciitilde{} \\ (1 \textbar{} rater\_id) + (1 \textbar{} topic/subtopic/item\_id)} \\
\midrule
(Intercept)                              & $1.01 \; [0.50; 1.53]^{*}$ \\
\midrule
$R^2_m$ (theoretical)                    & $0$                        \\
$R^2_c$ (theoretical)                    & $0.44$                     \\
AIC                                      & $429107.78$                \\
BIC                                      & $429166.85$                \\
Log Likelihood                           & $-214547.89$               \\
Num. obs.                                & $139184$                   \\
Num. groups: item\_id:subtopic:topic     & $4590$                     \\
Num. groups: rater\_id                   & $4075$                     \\
Num. groups: subtopic:topic              & $11$                       \\
Num. groups: topic                       & $3$                        \\
Var: item\_id:subtopic:topic (Intercept) & $0.22$                     \\
Var: rater\_id (Intercept)               & $0.43$                     \\
Var: subtopic:topic (Intercept)          & $0.01$                     \\
Var: topic (Intercept)                   & $0.20$                     \\
Var: Residual                            & $1.11$                     \\
\bottomrule
\multicolumn{2}{l}{\parbox[t]{\textwidth}{\scriptsize{$^*$ 0 outside the confidence interval.}}}
\end{tabular}%
}
\caption{D3: Null model (LMM)}
\label{tab:d3_null}
\end{center}
\end{table}
\begin{table}[htbp]
\begin{center}
\adjustbox{max width=\textwidth}{%
\begin{tabular}{l c}
\toprule
 & \shortstack[l]{harmful \textasciitilde{} age + gender + \\ (1 \textbar{} rater\_id) + (1 \textbar{} topic/subtopic/item\_id)} \\
\midrule
(Intercept)                              & $1.06 \; [ 0.55;  1.58]^{*}$  \\
rater\_age\_group30-50                   & $-0.12 \; [-0.16; -0.07]^{*}$ \\
rater\_age\_group50+                     & $-0.12 \; [-0.17; -0.06]^{*}$ \\
rater\_genderOther                       & $-0.21 \; [-0.42;  0.01]$     \\
rater\_genderWoman                       & $0.03 \; [-0.02;  0.07]$      \\
\midrule
$R^2_m$ (theoretical)                    & $0.00$                        \\
$R^2_c$ (theoretical)                    & $0.44$                        \\
AIC                                      & $429101.26$                   \\
BIC                                      & $429199.70$                   \\
Log Likelihood                           & $-214540.63$                  \\
Num. obs.                                & $139184$                      \\
Num. groups: item\_id:subtopic:topic     & $4590$                        \\
Num. groups: rater\_id                   & $4075$                        \\
Num. groups: subtopic:topic              & $11$                          \\
Num. groups: topic                       & $3$                           \\
Var: item\_id:subtopic:topic (Intercept) & $0.22$                        \\
Var: rater\_id (Intercept)               & $0.43$                        \\
Var: subtopic:topic (Intercept)          & $0.01$                        \\
Var: topic (Intercept)                   & $0.20$                        \\
Var: Residual                            & $1.11$                        \\
\bottomrule
\multicolumn{2}{l}{\parbox[t]{\textwidth}{\scriptsize{$^*$ 0 outside the confidence interval. Reference levels: rater\_age\_group = 18-30; rater\_gender = Man.}}}
\end{tabular}%
}
\caption{D3: Demographics model (LMM)}
\label{tab:d3_demog}
\end{center}
\end{table}
\begin{table}[htbp]
\begin{center}
\adjustbox{max width=\textwidth}{%
\begin{tabular}{l c}
\toprule
 & \shortstack[l]{harmful \textasciitilde{} cor\_cult\_cluster + \\ (1 \textbar{} rater\_id) + (1 \textbar{} topic/subtopic/item\_id)} \\
\midrule
(Intercept)                              & $0.89 \; [ 0.37;  1.41]^{*}$  \\
cor\_cultural\_zoneAfrican-Islamic       & $0.30 \; [ 0.24;  0.36]^{*}$  \\
cor\_cultural\_zoneConfucian             & $-0.19 \; [-0.27; -0.11]^{*}$ \\
cor\_cultural\_zoneLatin America         & $0.29 \; [ 0.22;  0.36]^{*}$  \\
cor\_cultural\_zoneProtestant Europe     & $0.16 \; [ 0.07;  0.25]^{*}$  \\
cor\_cultural\_zoneWest \& South Asia     & $0.15 \; [ 0.09;  0.21]^{*}$  \\
\midrule
$R^2_m$ (theoretical)                    & $0.01$                        \\
$R^2_c$ (theoretical)                    & $0.44$                        \\
AIC                                      & $428936.57$                   \\
BIC                                      & $429044.85$                   \\
Log Likelihood                           & $-214457.29$                  \\
Num. obs.                                & $139184$                      \\
Num. groups: item\_id:subtopic:topic     & $4590$                        \\
Num. groups: rater\_id                   & $4075$                        \\
Num. groups: subtopic:topic              & $11$                          \\
Num. groups: topic                       & $3$                           \\
Var: item\_id:subtopic:topic (Intercept) & $0.22$                        \\
Var: rater\_id (Intercept)               & $0.41$                        \\
Var: subtopic:topic (Intercept)          & $0.01$                        \\
Var: topic (Intercept)                   & $0.20$                        \\
Var: Residual                            & $1.11$                        \\
\bottomrule
\multicolumn{2}{l}{\parbox[t]{\textwidth}{\scriptsize{$^*$ 0 outside the confidence interval. Reference levels: cor\_cultural\_zone = English-Speaking.}}}
\end{tabular}%
}
\caption{D3: Culture model (LMM)}
\label{tab:d3_cult}
\end{center}
\end{table}
\begin{table}[htbp]
\begin{center}
\adjustbox{max width=\textwidth}{%
\begin{tabular}{l c}
\toprule
 & \shortstack[l]{harmful \textasciitilde{} TradAgg + \\ (1 \textbar{} rater\_id) + (1 \textbar{} topic/subtopic/item\_id)} \\
\midrule
(Intercept)                              & $1.01 \; [ 0.49;  1.52]^{*}$  \\
cor\_TradAgg                             & $-0.11 \; [-0.13; -0.09]^{*}$ \\
\midrule
$R^2_m$ (theoretical)                    & $0.00$                        \\
$R^2_c$ (theoretical)                    & $0.44$                        \\
AIC                                      & $429027.93$                   \\
BIC                                      & $429096.84$                   \\
Log Likelihood                           & $-214506.97$                  \\
Num. obs.                                & $139184$                      \\
Num. groups: item\_id:subtopic:topic     & $4590$                        \\
Num. groups: rater\_id                   & $4075$                        \\
Num. groups: subtopic:topic              & $11$                          \\
Num. groups: topic                       & $3$                           \\
Var: item\_id:subtopic:topic (Intercept) & $0.22$                        \\
Var: rater\_id (Intercept)               & $0.42$                        \\
Var: subtopic:topic (Intercept)          & $0.01$                        \\
Var: topic (Intercept)                   & $0.20$                        \\
Var: Residual                            & $1.11$                        \\
\bottomrule
\multicolumn{2}{l}{\parbox[t]{\textwidth}{\scriptsize{$^*$ 0 outside the confidence interval.}}}
\end{tabular}%
}
\caption{D3: Culture values (TradAgg) (LMM)}
\label{tab:d3_cult_vals_trad}
\end{center}
\end{table}
\begin{table}[htbp]
\begin{center}
\adjustbox{max width=\textwidth}{%
\begin{tabular}{l c}
\toprule
 & \shortstack[l]{harmful \textasciitilde{} SurvSAgg + \\ (1 \textbar{} rater\_id) + (1 \textbar{} topic/subtopic/item\_id)} \\
\midrule
(Intercept)                              & $1.06 \; [ 0.55;  1.58]^{*}$  \\
cor\_SurvSAgg                            & $-0.07 \; [-0.09; -0.06]^{*}$ \\
\midrule
$R^2_m$ (theoretical)                    & $0.01$                        \\
$R^2_c$ (theoretical)                    & $0.44$                        \\
AIC                                      & $429011.57$                   \\
BIC                                      & $429080.48$                   \\
Log Likelihood                           & $-214498.79$                  \\
Num. obs.                                & $139184$                      \\
Num. groups: item\_id:subtopic:topic     & $4590$                        \\
Num. groups: rater\_id                   & $4075$                        \\
Num. groups: subtopic:topic              & $11$                          \\
Num. groups: topic                       & $3$                           \\
Var: item\_id:subtopic:topic (Intercept) & $0.22$                        \\
Var: rater\_id (Intercept)               & $0.42$                        \\
Var: subtopic:topic (Intercept)          & $0.01$                        \\
Var: topic (Intercept)                   & $0.20$                        \\
Var: Residual                            & $1.11$                        \\
\bottomrule
\multicolumn{2}{l}{\parbox[t]{\textwidth}{\scriptsize{$^*$ 0 outside the confidence interval.}}}
\end{tabular}%
}
\caption{D3: Culture values (SurvSAgg) (LMM)}
\label{tab:d3_cult_vals_surv}
\end{center}
\end{table}
\begin{table}[htbp]
\begin{center}
\adjustbox{max width=\textwidth}{%
\begin{tabular}{l c}
\toprule
 & \shortstack[l]{harmful \textasciitilde{} TradAgg + SurvSAgg + \\ (1 \textbar{} rater\_id) + (1 \textbar{} topic/subtopic/item\_id)} \\
\midrule
(Intercept)                              & $1.05 \; [ 0.53;  1.56]^{*}$  \\
cor\_TradAgg                             & $-0.04 \; [-0.08; -0.00]^{*}$ \\
cor\_SurvSAgg                            & $-0.05 \; [-0.08; -0.03]^{*}$ \\
\midrule
$R^2_m$ (theoretical)                    & $0.01$                        \\
$R^2_c$ (theoretical)                    & $0.44$                        \\
AIC                                      & $429014.90$                   \\
BIC                                      & $429093.65$                   \\
Log Likelihood                           & $-214499.45$                  \\
Num. obs.                                & $139184$                      \\
Num. groups: item\_id:subtopic:topic     & $4590$                        \\
Num. groups: rater\_id                   & $4075$                        \\
Num. groups: subtopic:topic              & $11$                          \\
Num. groups: topic                       & $3$                           \\
Var: item\_id:subtopic:topic (Intercept) & $0.22$                        \\
Var: rater\_id (Intercept)               & $0.42$                        \\
Var: subtopic:topic (Intercept)          & $0.01$                        \\
Var: topic (Intercept)                   & $0.20$                        \\
Var: Residual                            & $1.11$                        \\
\bottomrule
\multicolumn{2}{l}{\parbox[t]{\textwidth}{\scriptsize{$^*$ 0 outside the confidence interval.}}}
\end{tabular}%
}
\caption{D3: Culture values (TradAgg + SurvSAgg) (LMM)}
\label{tab:d3_cult_vals_add}
\end{center}
\end{table}
\begin{table}[htbp]
\begin{center}
\adjustbox{max width=\textwidth}{%
\begin{tabular}{l c}
\toprule
 & \shortstack[l]{harmful \textasciitilde{} TradAgg * SurvSAgg + \\ (1 \textbar{} rater\_id) + (1 \textbar{} topic/subtopic/item\_id)} \\
\midrule
(Intercept)                              & $1.01 \; [ 0.49;  1.52]^{*}$  \\
cor\_TradAgg                             & $-0.04 \; [-0.08; -0.01]^{*}$ \\
cor\_SurvSAgg                            & $-0.06 \; [-0.09; -0.04]^{*}$ \\
cor\_TradAgg:cor\_SurvSAgg               & $0.05 \; [ 0.02;  0.07]^{*}$  \\
\midrule
$R^2_m$ (theoretical)                    & $0.01$                        \\
$R^2_c$ (theoretical)                    & $0.44$                        \\
AIC                                      & $429009.04$                   \\
BIC                                      & $429097.63$                   \\
Log Likelihood                           & $-214495.52$                  \\
Num. obs.                                & $139184$                      \\
Num. groups: item\_id:subtopic:topic     & $4590$                        \\
Num. groups: rater\_id                   & $4075$                        \\
Num. groups: subtopic:topic              & $11$                          \\
Num. groups: topic                       & $3$                           \\
Var: item\_id:subtopic:topic (Intercept) & $0.22$                        \\
Var: rater\_id (Intercept)               & $0.42$                        \\
Var: subtopic:topic (Intercept)          & $0.01$                        \\
Var: topic (Intercept)                   & $0.20$                        \\
Var: Residual                            & $1.11$                        \\
\bottomrule
\multicolumn{2}{l}{\parbox[t]{\textwidth}{\scriptsize{$^*$ 0 outside the confidence interval.}}}
\end{tabular}%
}
\caption{D3: Culture values (TradAgg * SurvSAgg) (LMM)}
\label{tab:d3_cult_vals_inter}
\end{center}
\end{table}
\begin{table}[htbp]
\begin{center}
\adjustbox{max width=\textwidth}{%
\begin{tabular}{l c}
\toprule
 & \shortstack[l]{harmful \textasciitilde{} age + gender + \\ cor\_cult\_cluster + \\ (1 \textbar{} rater\_id) + (1 \textbar{} topic/subtopic/item\_id)} \\
\midrule
(Intercept)                              & $0.88 \; [ 0.36;  1.40]^{*}$  \\
rater\_age\_group30-50                   & $-0.05 \; [-0.10; -0.00]^{*}$ \\
rater\_age\_group50+                     & $-0.03 \; [-0.09;  0.03]$     \\
rater\_genderOther                       & $-0.16 \; [-0.37;  0.05]$     \\
rater\_genderWoman                       & $0.07 \; [ 0.02;  0.11]^{*}$  \\
cor\_cultural\_zoneAfrican-Islamic       & $0.30 \; [ 0.24;  0.36]^{*}$  \\
cor\_cultural\_zoneConfucian             & $-0.18 \; [-0.26; -0.09]^{*}$ \\
\end{tabular}%
}
\caption{D3: Demographics + Culture additive (LMM)}
\label{tab:d3_demog_cult}
\end{center}
\end{table}
\begin{table}[htbp]
\begin{center}
\adjustbox{max width=\textwidth}{%
\begin{tabular}{l c}
\toprule
 & \shortstack[l]{harmful \textasciitilde{} age + gender + \\ cor\_TradAgg * cor\_SurvSAgg + \\ (1 \textbar{} rater\_id) + (1 \textbar{} topic/subtopic/item\_id)} \\
\midrule
(Intercept)                              & $1.02 \; [ 0.51;  1.54]^{*}$  \\
rater\_age\_group30-50                   & $-0.08 \; [-0.13; -0.04]^{*}$ \\
rater\_age\_group50+                     & $-0.07 \; [-0.13; -0.01]^{*}$ \\
rater\_genderOther                       & $-0.15 \; [-0.36;  0.06]$     \\
rater\_genderWoman                       & $0.06 \; [ 0.02;  0.10]^{*}$  \\
cor\_TradAgg                             & $-0.03 \; [-0.07;  0.01]$     \\
cor\_SurvSAgg                            & $-0.07 \; [-0.09; -0.04]^{*}$ \\
\end{tabular}%
}
\caption{D3: Demographics + selected culture-values additive (inter) (LMM)}
\label{tab:d3_demog_cult_vals}
\end{center}
\end{table}
{\small
\begin{longtable}{l c}
\caption{D3: Demographics x Culture interaction (LMM)}\label{tab:d3_demog_x_cult} \\
\toprule
& \shortstack[l]{harmful \textasciitilde{} (age + gender) * \\ cor\_cult\_cluster + \\ (1 \textbar{} rater\_id) + (1 \textbar{} topic/subtopic/item\_id)} \\
\midrule
\endfirsthead
\multicolumn{2}{l}{\emph{D3 (continued)}} \\
\toprule
& \shortstack[l]{harmful \textasciitilde{} (age + gender) * \\ cor\_cult\_cluster + \\ (1 \textbar{} rater\_id) + (1 \textbar{} topic/subtopic/item\_id)} \\
\midrule
\endhead
\midrule
\multicolumn{2}{r}{\emph{continued on next page}} \\
\endfoot
\bottomrule
\multicolumn{2}{l}{\parbox[t]{\linewidth}{\scriptsize{$^*$ 0 outside the confidence interval. Reference levels: rater\_age\_group = 18-30; rater\_gender = Man; cor\_cultural\_zone = English-Speaking.}}}
\endlastfoot
(Intercept)                                                 & $0.83 \; [ 0.31;  1.36]^{*}$  \\
rater\_age\_group30-50                                      & $-0.20 \; [-0.28; -0.12]^{*}$ \\
rater\_age\_group50+                                        & $-0.11 \; [-0.20; -0.02]^{*}$ \\
rater\_genderOther                                          & $0.23 \; [-0.11;  0.57]$      \\
rater\_genderWoman                                          & $0.26 \; [ 0.19;  0.33]^{*}$  \\
cor\_cultural\_zoneAfrican-Islamic                          & $0.32 \; [ 0.22;  0.42]^{*}$  \\
cor\_cultural\_zoneConfucian                                & $0.16 \; [-0.06;  0.38]$      \\
cor\_cultural\_zoneLatin America                            & $0.29 \; [ 0.16;  0.41]^{*}$  \\
\end{longtable}
}
{\small
\begin{longtable}{l c}
\caption{D3: Demographics x selected culture-values interaction (inter) (LMM)}\label{tab:d3_demog_x_cult_vals} \\
\toprule
& \shortstack[l]{harmful \textasciitilde{} (age + gender) * \\ (cor\_TradAgg * cor\_SurvSAgg) + \\ (1 \textbar{} rater\_id) + (1 \textbar{} topic/subtopic/item\_id)} \\
\midrule
\endfirsthead
\multicolumn{2}{l}{\emph{D3 (continued)}} \\
\toprule
& \shortstack[l]{harmful \textasciitilde{} (age + gender) * \\ (cor\_TradAgg * cor\_SurvSAgg) + \\ (1 \textbar{} rater\_id) + (1 \textbar{} topic/subtopic/item\_id)} \\
\midrule
\endhead
\midrule
\multicolumn{2}{r}{\emph{continued on next page}} \\
\endfoot
\bottomrule
\multicolumn{2}{l}{\parbox[t]{\linewidth}{\scriptsize{$^*$ 0 outside the confidence interval. Reference levels: rater\_age\_group = 18-30; rater\_gender = Man.}}}
\endlastfoot
(Intercept)                                       & $1.14 \; [ 0.62;  1.66]^{*}$  \\
rater\_age\_group30-50                            & $-0.12 \; [-0.20; -0.05]^{*}$ \\
rater\_age\_group50+                              & $-0.21 \; [-0.31; -0.12]^{*}$ \\
rater\_genderOther                                & $-0.54 \; [-1.00; -0.09]^{*}$ \\
rater\_genderWoman                                & $-0.10 \; [-0.16; -0.03]^{*}$ \\
cor\_TradAgg                                      & $0.06 \; [-0.01;  0.13]$      \\
cor\_SurvSAgg                                     & $-0.11 \; [-0.16; -0.07]^{*}$ \\
cor\_TradAgg:cor\_SurvSAgg                        & $-0.04 \; [-0.08;  0.00]$     \\
\end{longtable}
}
\begin{table}[htbp]
\begin{center}
\adjustbox{max width=\textwidth}{%
\begin{tabular}{l c}
\toprule
 & \shortstack[l]{harmful \textasciitilde{} cor\_quadrant + \\ (1 \textbar{} rater\_id) + (1 \textbar{} topic/subtopic/item\_id)} \\
\midrule
(Intercept)                              & $0.92 \; [ 0.40;  1.44]^{*}$  \\
cor\_quadrantII                          & $-0.24 \; [-0.32; -0.17]^{*}$ \\
cor\_quadrantIII                         & $0.27 \; [ 0.22;  0.31]^{*}$  \\
cor\_quadrantIV                          & $0.12 \; [ 0.06;  0.19]^{*}$  \\
\midrule
$R^2_m$ (theoretical)                    & $0.01$                        \\
$R^2_c$ (theoretical)                    & $0.44$                        \\
AIC                                      & $428914.64$                   \\
BIC                                      & $429003.24$                   \\
Log Likelihood                           & $-214448.32$                  \\
Num. obs.                                & $139184$                      \\
Num. groups: item\_id:subtopic:topic     & $4590$                        \\
Num. groups: rater\_id                   & $4075$                        \\
Num. groups: subtopic:topic              & $11$                          \\
Num. groups: topic                       & $3$                           \\
Var: item\_id:subtopic:topic (Intercept) & $0.22$                        \\
Var: rater\_id (Intercept)               & $0.41$                        \\
Var: subtopic:topic (Intercept)          & $0.01$                        \\
Var: topic (Intercept)                   & $0.20$                        \\
Var: Residual                            & $1.11$                        \\
\bottomrule
\multicolumn{2}{l}{\parbox[t]{\textwidth}{\scriptsize{$^*$ 0 outside the confidence interval. Reference levels: cor\_quadrant = I.}}}
\end{tabular}%
}
\caption{D3: Culture quadrants (LMM)}
\label{tab:d3_cult_quadrants}
\end{center}
\end{table}
\begin{table}[htbp]
\begin{center}
\adjustbox{max width=\textwidth}{%
\begin{tabular}{l c}
\toprule
 & \shortstack[l]{harmful \textasciitilde{} age + gender + \\ cor\_quadrant + \\ (1 \textbar{} rater\_id) + (1 \textbar{} topic/subtopic/item\_id)} \\
\midrule
(Intercept)                              & $0.91 \; [ 0.39;  1.43]^{*}$  \\
rater\_age\_group30-50                   & $-0.05 \; [-0.10; -0.00]^{*}$ \\
rater\_age\_group50+                     & $-0.04 \; [-0.10;  0.01]$     \\
rater\_genderOther                       & $-0.14 \; [-0.35;  0.07]$     \\
rater\_genderWoman                       & $0.07 \; [ 0.03;  0.11]^{*}$  \\
cor\_quadrantII                          & $-0.23 \; [-0.31; -0.16]^{*}$ \\
cor\_quadrantIII                         & $0.27 \; [ 0.22;  0.31]^{*}$  \\
\end{tabular}%
}
\caption{D3: Demographics + Quadrants additive (LMM)}
\label{tab:d3_demog_cult_quadrants}
\end{center}
\end{table}

\clearpage % <--- ADD THIS TO EMPTY THE FLOAT QUEUE

% -- CultFrames ---------------------------------------------------------------
\begin{table}[htbp]
\begin{center}
\adjustbox{max width=\textwidth}{%
\begin{tabular}{l c}
\toprule
 & \shortstack[l]{harmful \textasciitilde{} category + model\_name + \\ (1 \textbar{} rater\_id) + (1 \textbar{} item\_id)} \\
\midrule
(Intercept)                & $-3.37 \; [-3.68; -3.06]^{*}$ \\
categoryetiquette          & $-0.01 \; [-0.21;  0.19]$     \\
categoryfamily             & $-0.02 \; [-0.27;  0.23]$     \\
categorygreetings          & $0.21 \; [-0.03;  0.45]$      \\
categoryreligion           & $-0.05 \; [-0.29;  0.20]$     \\
model\_namegpt-image       & $0.49 \; [ 0.27;  0.71]^{*}$  \\
model\_nameimagegen3       & $0.73 \; [ 0.50;  0.97]^{*}$  \\
model\_nameSD35            & $0.66 \; [ 0.44;  0.88]^{*}$  \\
\midrule
$R^2_m$ (theoretical)      & $0.01$                        \\
$R^2_c$ (theoretical)      & $0.52$                        \\
AIC                        & $6595.11$                     \\
BIC                        & $6667.14$                     \\
Log Likelihood             & $-3287.56$                    \\
Num. obs.                  & $9922$                        \\
Num. groups: item\_id      & $3577$                        \\
Num. groups: rater\_id     & $379$                         \\
Var: item\_id (Intercept)  & $0.74$                        \\
Var: rater\_id (Intercept) & $2.68$                        \\
\bottomrule
\multicolumn{2}{l}{\parbox[t]{\textwidth}{\scriptsize{$^*$ 0 outside the confidence interval. Reference levels: category = dates-of-significance; model\_name = flux.}}}
\end{tabular}%
}
\caption{CultFrames: Null model (GLMER, binomial)}
\label{tab:cultframes_null}
\end{center}
\end{table}
\begin{table}[htbp]
\begin{center}
\adjustbox{max width=\textwidth}{%
\begin{tabular}{l c}
\toprule
 & \shortstack[l]{harmful \textasciitilde{} age + gender + \\ category + model\_name + \\ (1 \textbar{} rater\_id) + (1 \textbar{} item\_id)} \\
\midrule
(Intercept)                & $-3.67 \; [-4.10; -3.23]^{*}$ \\
rater\_age\_groupboomer    & $0.71 \; [-0.77;  2.18]$      \\
rater\_age\_groupgenx      & $0.28 \; [-0.28;  0.85]$      \\
rater\_age\_groupgenz      & $0.04 \; [-0.40;  0.48]$      \\
rater\_genderFemale        & $0.46 \; [ 0.06;  0.86]^{*}$  \\
categoryetiquette          & $-0.01 \; [-0.21;  0.19]$     \\
categoryfamily             & $-0.02 \; [-0.27;  0.23]$     \\
categorygreetings          & $0.22 \; [-0.02;  0.46]$      \\
categoryreligion           & $-0.04 \; [-0.29;  0.20]$     \\
model\_namegpt-image       & $0.49 \; [ 0.27;  0.71]^{*}$  \\
model\_nameimagegen3       & $0.73 \; [ 0.50;  0.97]^{*}$  \\
model\_nameSD35            & $0.66 \; [ 0.44;  0.88]^{*}$  \\
\midrule
$R^2_m$ (theoretical)      & $0.02$                        \\
$R^2_c$ (theoretical)      & $0.51$                        \\
AIC                        & $6596.07$                     \\
BIC                        & $6696.90$                     \\
Log Likelihood             & $-3284.03$                    \\
Num. obs.                  & $9922$                        \\
Num. groups: item\_id      & $3577$                        \\
Num. groups: rater\_id     & $379$                         \\
Var: item\_id (Intercept)  & $0.74$                        \\
Var: rater\_id (Intercept) & $2.56$                        \\
\bottomrule
\multicolumn{2}{l}{\parbox[t]{\textwidth}{\scriptsize{$^*$ 0 outside the confidence interval. Reference levels: rater\_age\_group = millennial; rater\_gender = Male; category = dates-of-significance; model\_name = flux.}}}
\end{tabular}%
}
\caption{CultFrames: Demographics model (GLMER, binomial)}
\label{tab:cultframes_demog}
\end{center}
\end{table}
\begin{table}[htbp]
\begin{center}
\adjustbox{max width=\textwidth}{%
\begin{tabular}{l c}
\toprule
 & \shortstack[l]{harmful \textasciitilde{} cob\_cult\_cluster + \\ category + model\_name + \\ (1 \textbar{} rater\_id) + (1 \textbar{} item\_id)} \\
\midrule
(Intercept)                          & $-4.20 \; [-4.92; -3.48]^{*}$ \\
cob\_cultural\_zoneAfrican-Islamic   & $2.55 \; [ 1.63;  3.47]^{*}$  \\
cob\_cultural\_zoneCatholic Europe   & $0.85 \; [-0.11;  1.82]$      \\
cob\_cultural\_zoneConfucian         & $1.61 \; [ 0.82;  2.40]^{*}$  \\
cob\_cultural\_zoneLatin America     & $0.66 \; [-0.27;  1.60]$      \\
cob\_cultural\_zoneProtestant Europe & $0.70 \; [-0.16;  1.56]$      \\
cob\_cultural\_zoneWest \& South Asia & $0.37 \; [-0.38;  1.12]$      \\
\end{tabular}%
}
\caption{CultFrames: Culture model (GLMER, binomial)}
\label{tab:cultframes_cult}
\end{center}
\end{table}
\begin{table}[htbp]
\begin{center}
\adjustbox{max width=\textwidth}{%
\begin{tabular}{l c}
\toprule
 & \shortstack[l]{harmful \textasciitilde{} cob\_TradAgg + \\ category + model\_name + \\ (1 \textbar{} rater\_id) + (1 \textbar{} item\_id)} \\
\midrule
(Intercept)                & $-3.42 \; [-3.74; -3.10]^{*}$ \\
cob\_TradAgg               & $0.22 \; [-0.09;  0.53]$      \\
categoryetiquette          & $-0.01 \; [-0.21;  0.19]$     \\
categoryfamily             & $-0.03 \; [-0.28;  0.22]$     \\
categorygreetings          & $0.21 \; [-0.03;  0.45]$      \\
categoryreligion           & $-0.04 \; [-0.29;  0.20]$     \\
model\_namegpt-image       & $0.49 \; [ 0.27;  0.71]^{*}$  \\
model\_nameimagegen3       & $0.73 \; [ 0.50;  0.97]^{*}$  \\
model\_nameSD35            & $0.66 \; [ 0.45;  0.88]^{*}$  \\
\midrule
$R^2_m$ (theoretical)      & $0.02$                        \\
$R^2_c$ (theoretical)      & $0.52$                        \\
AIC                        & $6595.14$                     \\
BIC                        & $6674.37$                     \\
Log Likelihood             & $-3286.57$                    \\
Num. obs.                  & $9922$                        \\
Num. groups: item\_id      & $3577$                        \\
Num. groups: rater\_id     & $379$                         \\
Var: item\_id (Intercept)  & $0.74$                        \\
Var: rater\_id (Intercept) & $2.68$                        \\
\bottomrule
\multicolumn{2}{l}{\parbox[t]{\textwidth}{\scriptsize{$^*$ 0 outside the confidence interval. Reference levels: category = dates-of-significance; model\_name = flux.}}}
\end{tabular}%
}
\caption{CultFrames: Culture values (TradAgg) (GLMER, binomial)}
\label{tab:cultframes_cult_vals_trad}
\end{center}
\end{table}
\begin{table}[htbp]
\begin{center}
\adjustbox{max width=\textwidth}{%
\begin{tabular}{l c}
\toprule
 & \shortstack[l]{harmful \textasciitilde{} cob\_SurvSAgg + \\ category + model\_name + \\ (1 \textbar{} rater\_id) + (1 \textbar{} item\_id)} \\
\midrule
(Intercept)                & $-3.25 \; [-3.58; -2.93]^{*}$ \\
cob\_SurvSAgg              & $-0.23 \; [-0.43; -0.02]^{*}$ \\
categoryetiquette          & $-0.01 \; [-0.21;  0.19]$     \\
categoryfamily             & $-0.02 \; [-0.27;  0.23]$     \\
categorygreetings          & $0.21 \; [-0.03;  0.45]$      \\
categoryreligion           & $-0.05 \; [-0.29;  0.20]$     \\
model\_namegpt-image       & $0.49 \; [ 0.27;  0.71]^{*}$  \\
model\_nameimagegen3       & $0.73 \; [ 0.50;  0.97]^{*}$  \\
model\_nameSD35            & $0.66 \; [ 0.44;  0.88]^{*}$  \\
\midrule
$R^2_m$ (theoretical)      & $0.02$                        \\
$R^2_c$ (theoretical)      & $0.51$                        \\
AIC                        & $6592.40$                     \\
BIC                        & $6671.63$                     \\
Log Likelihood             & $-3285.20$                    \\
Num. obs.                  & $9922$                        \\
Num. groups: item\_id      & $3577$                        \\
Num. groups: rater\_id     & $379$                         \\
Var: item\_id (Intercept)  & $0.74$                        \\
Var: rater\_id (Intercept) & $2.60$                        \\
\bottomrule
\multicolumn{2}{l}{\parbox[t]{\textwidth}{\scriptsize{$^*$ 0 outside the confidence interval. Reference levels: category = dates-of-significance; model\_name = flux.}}}
\end{tabular}%
}
\caption{CultFrames: Culture values (SurvSAgg) (GLMER, binomial)}
\label{tab:cultframes_cult_vals_surv}
\end{center}
\end{table}
\input{tex/cultframes_m_cult_vals_add}
\input{tex/cultframes_m_cult_vals_inter}
\begin{table}[htbp]
\begin{center}
\adjustbox{max width=\textwidth}{%
\begin{tabular}{l c}
\toprule
 & \shortstack[l]{harmful \textasciitilde{} age + gender + \\ cob\_cult\_cluster + category + model\_name + \\ (1 \textbar{} rater\_id) + (1 \textbar{} item\_id)} \\
\midrule
(Intercept)                          & $-4.47 \; [-5.24; -3.70]^{*}$ \\
rater\_age\_groupboomer              & $1.05 \; [-0.34;  2.43]$      \\
rater\_age\_groupgenx                & $0.17 \; [-0.37;  0.70]$      \\
rater\_age\_groupgenz                & $0.03 \; [-0.38;  0.44]$      \\
rater\_genderFemale                  & $0.38 \; [-0.00;  0.76]$      \\
cob\_cultural\_zoneAfrican-Islamic   & $2.59 \; [ 1.68;  3.50]^{*}$  \\
cob\_cultural\_zoneCatholic Europe   & $0.95 \; [ 0.00;  1.90]^{*}$  \\
\end{tabular}%
}
\caption{CultFrames: Demographics + Culture additive (GLMER, binomial)}
\label{tab:cultframes_demog_cult}
\end{center}
\end{table}
\input{tex/cultframes_m_demog_cult_vals}
\input{tex/cultframes_m_demog_x_cult}
\input{tex/cultframes_m_demog_x_cult_vals}
\input{tex/cultframes_m_cult_quadrants}
\input{tex/cultframes_m_demog_cult_quadrants}
\input{tex/cultframes_m_demog_x_cult_quadrants}

\clearpage % <--- ADD THIS TO EMPTY THE FLOAT QUEUE

% -- DICES Expert -------------------------------------------------------------
\input{tex/dices_expert_m_null}
\input{tex/dices_expert_m_demog}
\input{tex/dices_expert_m_cult}
\input{tex/dices_expert_m_cult_vals_trad}
\input{tex/dices_expert_m_cult_vals_surv}
\input{tex/dices_expert_m_cult_vals_add}
\input{tex/dices_expert_m_cult_vals_inter}
\input{tex/dices_expert_m_demog_cult}
\input{tex/dices_expert_m_demog_cult_vals}
\input{tex/dices_expert_m_demog_x_cult}
\input{tex/dices_expert_m_demog_x_cult_vals}
\input{tex/dices_expert_m_cult_quadrants}
\input{tex/dices_expert_m_demog_cult_quadrants}

\clearpage % <--- ADD THIS TO EMPTY THE FLOAT QUEUE

% -- DIVE ---------------------------------------------------------------------
\input{tex/dive_m_null}
\input{tex/dive_m_demog}
\input{tex/dive_m_cult}
\input{tex/dive_m_cult_vals_trad}
\input{tex/dive_m_cult_vals_surv}
\input{tex/dive_m_cult_vals_add}
\input{tex/dive_m_cult_vals_inter}
\input{tex/dive_m_demog_cult}
\input{tex/dive_m_demog_cult_vals}
\input{tex/dive_m_demog_x_cult}
\input{tex/dive_m_demog_x_cult_vals}
\input{tex/dive_m_cult_quadrants}
\input{tex/dive_m_demog_cult_quadrants}

\clearpage

% -- Severity --------------------------------------------------------------------
\input{tex/severity_m_null}
\input{tex/severity_m_demog}
\input{tex/severity_m_cult}
\input{tex/severity_m_cult_vals_trad}
\input{tex/severity_m_cult_vals_surv}
\input{tex/severity_m_cult_vals_add}
\input{tex/severity_m_cult_vals_inter}
\input{tex/severity_m_demog_cult}
\input{tex/severity_m_demog_cult_vals}
\input{tex/severity_m_demog_x_cult}
\input{tex/severity_m_demog_x_cult_vals}
\input{tex/severity_m_cult_quadrants}
\input{tex/severity_m_demog_cult_quadrants}

\clearpage

% -- NLPositionality ----------------------------------------------------------
\input{tex/nlpos_m_null}
\input{tex/nlpos_m_demog}
\input{tex/nlpos_m_cult}
\input{tex/nlpos_m_cult_vals_trad}
\input{tex/nlpos_m_cult_vals_surv}
\input{tex/nlpos_m_cult_vals_add}
\input{tex/nlpos_m_cult_vals_inter}
\input{tex/nlpos_m_demog_cult}
\input{tex/nlpos_m_demog_cult_vals}
\input{tex/nlpos_m_demog_x_cult}
\input{tex/nlpos_m_demog_x_cult_vals}
\input{tex/nlpos_m_cult_quadrants}
\input{tex/nlpos_m_demog_cult_quadrants}

\clearpage

% -- PRISM --------------------------------------------------------------------
\input{tex/prism_m_null}
\input{tex/prism_m_demog}
\input{tex/prism_m_cult}
\input{tex/prism_m_cult_vals_trad}
\input{tex/prism_m_cult_vals_surv}
\input{tex/prism_m_cult_vals_add}
\input{tex/prism_m_cult_vals_inter}
\input{tex/prism_m_demog_cult}
\input{tex/prism_m_demog_cult_vals}
\input{tex/prism_m_demog_x_cult}
\input{tex/prism_m_demog_x_cult_vals}
\input{tex/prism_m_cult_quadrants}
\input{tex/prism_m_demog_cult_quadrants}

\clearpage

%%%%%%%%%%%%%%%%%%%%%%%%%%%%%%%%%%%%%%%%%%%%%%%%%%%%%%%%%%%%%%%%%%%%%%%%%%%%%%%
%%%%%%%%%%%%%%%%%%%%%%%%%%%%%%%%%%%%%%%%%%%%%%%%%%%%%%%%%%%%%%%%%%%%%%%%%%%%%%%

\end{document}